\documentclass[12pt]{article}
\pdfoutput=1
\DeclareMathAlphabet{\scr}{U}{rsfs}{m}{n}
\usepackage{latexsym}
\usepackage[mathscr]{eucal}
\usepackage{amsfonts}
\usepackage{amscd}
\usepackage{cite}
\usepackage{amssymb}
\usepackage[centertags]{amsmath}
\usepackage{dsfont}
\usepackage{yfonts}
\usepackage{enumerate}
\usepackage{graphicx}
\usepackage{feynmf}

\newcommand{\cleqn}{\setcounter{equation}{0}}
\newcommand{\newc}{\newcommand}
\newc{\be}{\begin{equation}}
\newc{\ee}{\end{equation}}
\newc{\bea}{\begin{eqnarray}}
\newc{\eea}{\end{eqnarray}}
\newc{\ben}{\begin{equation*}}
\newc{\een}{\end{equation*}}
\newc{\bean}{\begin{eqnarray*}}
\newc{\eean}{\end{eqnarray*}}
\newc{\ol}{\overline}
\newc{\wt}{\widetilde}
\newc{\bs}{\boldsymbol}
\newc{\m}{\mathcal}
\newc{\la}{\lambda}
\newc{\lra}{\longrightarrow}
\newc{\vp}{\varphi}
\newc{\ti}{\tilde}

\begin{document}

\title{\hfill ~\\[-30mm]
       \hfill\mbox{\small \begin{tabular}{l}
                                   SISSA 16/2010/EP \\[1mm]
                                   SHEP-10-08
                          \end{tabular}}\\[15mm]
       \textbf{A SUSY GUT of Flavour with ${\bs{S_4 \times SU(5)}}$ to NLO} }
\date{}
\author{\\Claudia Hagedorn\footnote{E-mail: {\tt hagedorn@sissa.it}}\\[2mm]
  \emph{\small{}SISSA and INFN-Sezione di Trieste},\\
  \emph{\small via Beirut 2-4, I-34014 Trieste, Italy}\\[2mm]
\\Stephen F. King\footnote{E-mail: {\tt king@soton.ac.uk}}~~and
        Christoph Luhn\footnote{E-mail: {\tt christoph.luhn@soton.ac.uk}}\\[2mm]
  \emph{\small{}School of Physics and Astronomy, University of Southampton,}\\
  \emph{\small Southampton, SO17 1BJ, United Kingdom}}

\maketitle

\begin{abstract}
\noindent
We construct a Supersymmetric (SUSY) Grand Unified Theory (GUT) of Flavour
based on $S_4\times SU(5)$, together with an additional (global or local) Abelian symmetry,
and study it to next-to-leading order (NLO) accuracy.
The model includes a successful description of quark and lepton masses and
mixing angles at leading order (LO) incorporating the Gatto-Sartori-Tonin (GST) relation
and the Georgi-Jarlskog (GJ) relations. 
We study the vacuum alignment arising from $F$-terms to NLO 
and such corrections are shown to have a negligible effect on the results for fermion masses and mixings
achieved at LO. Tri-bimaximal (TB) mixing in the neutrino sector is predicted 
very accurately up to NLO corrections of order 0.1\%.
Including charged lepton mixing corrections implies small deviations from TB mixing described by
a precise sum rule, accurately maximal atmospheric mixing and a reactor mixing angle close to three degrees.
\end{abstract}
\thispagestyle{empty}
\vfill
\newpage
\setcounter{page}{1}


\section{Introduction}
\cleqn


A long standing quest of theories of particle
physics beyond the Standard Model (SM) is to
formulate a theory of quark and charged lepton
masses and quark mixings.
In recent years, this quest has been extended to include the
neutrino masses and lepton mixing as a result of tremendous experimental advances and discoveries
in neutrino physics. Indeed, perhaps
the greatest advance in particle physics over the past dozen years has
been the measurement of neutrino masses and mixing involving two large
mixing angles associated with atmospheric and solar neutrino oscillation
experiments, while the remaining mixing angle, although unmeasured, is
constrained by reactor neutrino oscillation
experiments to be relatively small.
The empirical observation of TB lepton mixing \cite{HPS}
contrasts sharply with the smallness of
quark mixing, and this observation, together with the
smallness of neutrino masses, provides new and tantalising clues
in the search for the origin of quark and lepton flavour
in terms of a theory of flavour that would supersede the SM.

TB lepton mixing in particular hints at a spontaneously broken family symmetry $G_f$
which might underpin a flavour theory of all quarks and leptons, but which might only
reveal itself in the neutrino sector.
What is the nature of such a family symmetry?
In the (diagonal) charged lepton mass basis,
it has been shown that the neutrino mass matrix leading to TB mixing is invariant
under (off-diagonal) transformations $S$ and $U$ which constitute the Klein
group\cite{Lam}.\footnote{In a different basis $S$ and $U$ could as
  well be represented by diagonal matrices.}
The observed neutrino flavour symmetry corresponding to the two generators $S$ and $U$ of the Klein group
may arise either directly or indirectly from certain classes of discrete groups \cite{King:2009ap}.
Several models have been constructed that account for the
structure of leptonic mixings, e.g. \cite{S3-L,Dn-L,A4-L,S4-L,delta54-L}, while other models
extend the underlying family symmetry to provide
a description of the complete fermionic structure
\cite{S3-LQ,Dn-LQ,Q6-LQ,A4-LQ,A4-LQ:Morisi,A4-LQ:King,doubleA4-LQ,S4-LQ:Hagedorn,S4-group,S4-LQ,Mohapatra:2003tw,King:2009mk,A4-SU5,Tpr-SU5,Z7Z3-LQ,T7-LQ,delta27-LQ:King,SO(3)-LQ:King,SU(3)-LQ:Ross}.\footnote{See
  \cite{Reviews} for review papers with more extensive references.}
If the neutrino flavour symmetry arises directly from the
family symmetry \cite{King:2009ap} then this implies that the family symmetry
should contain the generators $S$ and $U$ so that they can be preserved in the
neutrino sector at LO.
The smallest group that contains the generators
$S$ and $U$ together with a (diagonal) phase matrix $T$ is $S_4$ \cite{Lam} and the
models found in
\cite{S4-L,S4-LQ,S4-LQ:Hagedorn,S4-group,Mohapatra:2003tw}
are based on $S_4$.
The fact that it is possible to construct direct models based on the family
symmetry $A_4$ (generated by $S$ and $T$ only) is owed to the required absence of
family symmetry breaking fields (flavons)
 in the representations ${\bf 1'}$ and ${\bf 1''}$ of $A_4$. In such $A_4$ models
the symmetry associated with the generator $U$ arises accidentally at LO
\cite{A4-L}.

Despite the plethora of models, there are surprisingly few which successfully
combine a discrete family symmetry containing triplet representations (necessary to
account for TB mixing) together with a GUT. Examples are
the $A_4\times SU(5)$ models \cite{A4-SU5}, the $T'\times SU(5)$ model \cite{Tpr-SU5}, the $A_4\times SO(10)$ models \cite{A4-LQ:Morisi}, 
the $S_4\times SO(10)$ models \cite{Mohapatra:2003tw},
the $PSL(2,7) \times SO(10)$ model \cite{King:2009mk}, and the $\Delta_{27}\times SO(10)$ models
\cite{delta27-LQ:King}. The possible combination $S_4\times SU(5)$ stands out
in the sense that it combines the minimal GUT with the minimal choice of
family symmetry, which contains the generators $S$ and $U$. 

In this paper we construct a SUSY GUT of
Flavour based on $S_4\times SU(5)$ in which the ${\bf \ol{5}}$ matter fields
of $SU(5)$ are assigned to a triplet of $S_4$, while the
${\bf 10}$ matter fields are in a doublet plus a (trivial) singlet of $S_4$.
The operators are also controlled by an additional $U(1)$ symmetry
which segregates different types of flavons into different (charge) sectors at LO,
e.g. flavons, whose vacuum expectation values
(VEVs) preserve the generators $S$ and $U$, only couple to neutrinos at LO. Furthermore, the $U(1)$ symmetry
 controls the amount of flavon contamination
between different sectors beyond LO.
We shall show that the model predicts
TB neutrino mixing very accurately up to corrections of order 0.1\% at the GUT scale.
In order to do so, we specify the complete effective theory, valid just
below the GUT scale, and perform a full operator analysis of all relevant terms including several flavons.
Furthermore, we make an exhaustive study of vacuum alignment to NLO arising from the $F$-terms of driving fields.
These fields are, similar to the flavons, gauge singlets which only transform non-trivially under $S_4 \times U(1)$.
The model leads to a successful description of quark and
charged lepton masses and quark mixing angles, including
the GST relation between down and strange
quark masses and the Cabibbo angle $\theta^q_{12}$ \cite{gst}, and the GJ
relations between charged lepton and down quark masses \cite{GJ},
with bottom-tau Yukawa unification. The GJ factor is also
responsible for the (left-handed) charged lepton mixing angle $\theta^e_{12}$ being
$\theta^e_{12}\approx \theta^q_{12}/3$.
Including corrections due to non-zero mixing in the charged lepton sector
induces deviations from TB
lepton mixing expressed in a lepton mixing sum rule \cite{King:2005bj}
with a reactor mixing angle of order $\theta^q_{12}/(3\sqrt{2})$.
Since $\theta^e_{13}\approx0$ and $\theta^e_{23} \approx 0$,
maximal atmospheric mixing holds to good precision at the GUT scale.
We note that in the realisation of the model we discuss in detail, small
  and moderate values of $\tan\beta$, the ratio of the VEVs of the two
  electroweak Higgs doublets present in the Minimal Supersymmetric Standard
  Model (MSSM), are preferred because the hierarchy among the top and the
  bottom quark mass is accounted for by the family symmetry.
 Since our main concern in this work is the explanation of fermion masses and mixings, we 
leave aside the problem of constructing a GUT Higgs (super-)potential ensuring the correct breaking
of the gauge group $SU(5)$ to the SM.

We remark that an $S_4\times SU(5)$ model has also been proposed in \cite{S4-LQ}, in  which, 
 however, NLO corrections as well as the vacuum alignment of the flavons are not studied in detail.
 By contrast in the different $S_4\times SU(5)$ model proposed here 
 the LO predictions are robust against the NLO corrections which are explicitly calculated and shown to be small.
Furthermore, the alignment of the flavon VEVs is a natural result of the flavon superpotential. The
latter is thoroughly investigated to NLO.

The layout of the remainder of the paper is as follows:
in section 2 we define the SUSY $S_4 \times SU(5)$ model for a general class of $U(1)$ charges
and discuss the results for fermion masses and mixings at LO. In section 3 we
perform an operator analysis of all relevant terms including several flavon
fields. In this context, we introduce the notion of desired, dangerous,
marginal and irrelevant operators. We find 26 possible $U(1)$ charge assignments
which neither lead to dangerous nor to marginal operators. Section 4
contains a study of the vacuum alignment which justifies the alignments
assumed in previous sections. On the basis of the results of the analysis
of higher-order terms disturbing this alignment and of the possibility to
correlate the VEVs of different flavons we choose the actual $U(1)$
charges. In section 5 we discuss the NLO corrections to Yukawa couplings and to the
flavon superpotential for a particular choice of $U(1)$ charges and show that 
all corrections induced to fermion masses and mixings are small.
 Section 6 concludes the paper. The first three appendices contain
the group theory of $S_4$, an example of messengers generating the operators
giving rise to the GJ and the GST relations, and the list of dangerous
and marginal operators with less than four flavons contributing to the fermion mass matrices, according to
the classification introduced in section 3.  Appendix~D is dedicated to a
  discussion of how to ensure that the family symmetry is broken in the SUSY
  limit and how to (further) reduce the number of free parameters among the
  flavon VEVs introducing additional driving fields and
using couplings with positive mass dimension.


\section{\label{sec:model} The $\bs{S_4\times SU(5)}$ model and LO results}
\cleqn


In this section we present the model and discuss the LO result for fermion masses and mixings.
In table~\ref{model} we show the superfield charge assignments of
our SUSY GUT of Flavour based on $S_4\times SU(5)$.
For convenience the group theory of $S_4$ is summarised in appendix~A.
The ${\bf \ol{5}}$ matter fields $F$ of $SU(5)$ are assigned to a triplet of
$S_4$, while the ten-dimensional matter fields are assigned to a doublet $T$
plus the trivial singlet $T_3$ of $S_4$. The right-handed neutrinos $N$ are
taken to be a triplet of $S_4$, analogous to the $A_4$ see-saw models in
\cite{A4-L}, however there are some differences in the neutrino
sector, as discussed below.
The GUT Higgs fields $H^{}_{5}$, $H_{\ol{5}}$ and
$H_{\overline{45}}$ are all singlets under the family symmetry
$S_4$.\footnote{The $SU(5)$ symmetry might be broken by an additional ${\bf 24}$
  Higgs field which can be rendered irrelevant for the Yukawa operators by suitable
  charges under the $U(1)$ symmetry. The ${\bf 45}$ Higgs field which
  should be added due to anomaly cancellation may similarly decouple from
  the up quark sector. Therefore we disregard these Higgs fields in the following.
 Since the actual construction of a GUT Higgs (super-)potential is beyond the scope 
of this work, we do not specify further flavon fields which might be necessary in order
to allow relevant couplings in this (super-)potential, which were otherwise forbidden by
the $U(1)$ symmetry. An alternative possibility to achieve the breaking of the 
GUT symmetry might arise from appropriately chosen boundary conditions in an extra-dimensional
scenario, see \cite{SU5-EDs}. In this case also the problem related to the splitting of 
doublets and colour triplets is elegantly solved.
}
We note that these Higgs representations
each contain a Higgs doublet. The MSSM Higgs doublets
$H_{u}$ and $H_{d}$ then originate from, respectively,
$H^{}_{{5}}$ and one linear combination of the doublets in
$H_{\ol{5}}$ and $H_{\ol{45}}$. 
\footnote{
Again, it might be necessary to invoke the presence of further flavons to generate $S_4\times U(1)$
invariant couplings between $H_{\ol{5}}$ and $H_{\ol{45}}$ in order to introduce mixing among
their Higgs doublet components.
}
 The $H_{\overline{45}}$ component within
$H_{d}$ is responsible for the GJ relations between charged lepton and
down quark masses \cite{GJ}. Concerning the other (orthogonal) linear combination
we assume that it decouples from the low-energy theory by acquiring a GUT scale
mass just as the colour triplets, contained in the GUT Higgs fields, and other non-MSSM states
\cite{SU5DTsplit}. The (light) MSSM Higgs doublets $H_{u,d}$ acquire VEVs $v_{u,d}$ with $\tan
\beta = {v_u}/{v_d}$.

In addition, we introduce a number of flavon fields $\Phi_{\rho}^f$. An
important feature of the model is that different flavons couple to different
sectors of the theory at LO.
The flavons $\Phi_{\rho}^f$ are labelled both by the representation $\rho$ of
$S_4$ under which they transform (${\bf 1,2,3,3'}$) and by the fermion sector
$f$ to which they couple at LO, namely $u,d$ and $\nu$, where $d\sim e$ up to
the difference in the GJ factor.
Thus, for example, at LO, the flavon doublet $\Phi_2^u$ appears only
with $TT$, the flavon triplet $\Phi_3^d$ appears only with
$FT_3$, while the neutrino flavons $\Phi_\rho^{\nu}$ only appear with $NN$.
Notice that $\Phi_\rho^{\nu}$ consist of singlet, doublet and (primed) triplet
representations with vacuum alignments which preserve the generators
$S$ and $U$ contained in $S_4$ leading to TB neutrino mixing.

The segregation of the different flavons,
coupling to distinct sectors, at the LO level is
achieved through an additional $U(1)$ symmetry. For the time being, we assume this symmetry to be global
  in order to avoid constraints coming from the requirement of anomaly cancellation. The $U(1)$
 charges of the fields are expressed in terms of
three integers $x$, $y$ and $z$, as shown in table~\ref{model}. Note that
the Higgs fields $H^{}_{5}$ and $H_{\ol{5}}$ are taken to be neutral under this
symmetry.

The family symmetry $S_4$ is only broken spontaneously by flavon VEVs in
our model. On the other hand, the spontaneous breakdown of the global $U(1)$
symmetry leads to the appearance of a (very light) Goldstone boson unless the
$U(1)$ is also explicitly broken. For this reason
we assume a scenario in which the $U(1)$ symmetry is explicitly broken in the
hidden sector of the theory which is also responsible for SUSY breaking, so
that the soft terms do not respect  the $U(1)$ symmetry.
Then the would-be Goldstone boson will have a mass of the order of the soft
SUSY mass scale of around $1$ TeV.\footnote{We remark that in this context it would be interesting to investigate
 whether the family symmetry $S_4$ alone is sufficient to constrain the soft mass terms of sfermions in such a way
that all bounds associated with flavour changing neutral current and lepton flavour violating processes can be
satisfied without tuning the soft mass parameters.}
Alternatively one could gauge the $U(1)$ symmetry and add further
particles in order to cancel the anomalies. As the set of these additional
particles would depend on the explicit $U(1)$ charge assignments, we do not
follow this approach.

In our study, we disregard possible corrections to fermion
  masses and mixings which are due to deviations from the canonical
  normalisation of the K\"ahler potential. Such deviations arise in general, if subleading corrections
involving (several) flavons are taken into account. Studies of the possible effects of non-canonically normalised
kinetic terms on fermion masses and mixings can be found in, e.g., \cite{noncanonkin}.
\begin{table}
\begin{center}
$$
\begin{array}{|c||c|c|c|c|c|c|c|c|c|c|c|c|c|c|c|}\hline
\text{Field}\!\!\phantom{\Big|} & T_3 & T & F & N & H^{}_{5} & H_{\ol{5}} & H_{\ol{45}} &  \Phi^u_2 & \wt\Phi^u_2 & \Phi^d_3 & \wt\Phi^d_3 & \Phi^d_2  & \Phi^\nu_{3'} & \Phi^\nu_2 & \Phi^\nu_1 \\\hline
\!SU(5)\!\!\!\phantom{\Big|} & \bf 10 & \bf 10 & \bf \ol 5 & \bf 1 &\bf  5 &\bf \ol 5 &\bf \ol{45}
&\bf 1&\bf 1&\bf 1&\bf 1&\bf 1&\bf 1&\bf 1&\bf 1\\\hline
S_4\!\!\phantom{\Big|} & \bf 1&\bf 2&\bf 3&\bf 3&\bf 1&\bf 1&\bf 1&\bf 2&\bf 2&\bf 3&\bf 3&\bf
2&\bf 3'&\bf 2&\bf 1 \\ \hline
U(1)\!\!\phantom{\Big|} & 0&x&y&-y&0&0&z&\!-2x\!&0&\!-y\!&\!-x-y-2z\! &z &2y&2y&2y \\\hline
\end{array}
$$
\end{center}
\caption{\label{model}The symmetries and charges of the
superfields in the $SU(5)\times S_4 \times U(1)$ model. The $U(1)$ assignment depends on three integers $x$, $y$ and $z$.}
\end{table}

The lowest dimensional Yukawa operators invariant under the family symmetry $S_4 \times U(1)$, contributing to the
up quark mass matrix, are
(we omit order one coefficients in the following)
\be
 T_3T_3H^{}_{5} + \frac{1}{M} T T  \Phi^u_2 H^{}_{5}  +  \frac{1}{M^2} TT \Phi^u_2 \wt\Phi^u_2 H^{}_{5}  \; .
\label{upLO}
\ee
By $M$ we denote a generic messenger scale which is common for all higher-dimensional operators
we discuss. It is expected to be around the GUT scale. If not explicitly stated, we take into
account all possible independent $S_4$ contractions for each operator so that frequently
one given operator entails - depending on the vacuum alignment of the flavons -
a certain number of different contributions to the fermion mass matrices.

The LO operators giving
rise to masses for charged leptons and down quarks read
\be
\frac{1}{M} F T_3 \Phi^d_3 H_{\ol{5}} + \frac{1}{M^2} (F \wt\Phi^d_3)_1 ( T \Phi^d_2 )_1 H_{\ol{45}}
+ \frac{1}{M^3} (F \Phi^d_2 \Phi^d_2)_3 ( T \wt\Phi^d_3 )_3 H_{\ol{5}} \ ,
\label{downLO}
\ee
where $(\cdots)_1$ and $(\cdots)_3$ denote the contraction to an $S_4$
invariant ${\bf 1}$ and to the triplet ${\bf 3}$, respectively.
Note that there are other possible operator
contractions involving the same fields that we do not write down.
As a first step towards achieving the GJ and GST relations we have
assumed that the two contractions shown in Eq.~(\ref{downLO}) are the dominant ones among the various possible ones,
existing in a generic effective theory with a cutoff scale $M$.
One example of messengers which only give rise to these contractions 
 is discussed in detail in appendix B and shown diagrammatically in figure~\ref{diagram}.
The operator involving the Higgs field $H_{\ol{45}}$ must have the appropriate
form,
\be
\label{GJfact}
\frac{1}{M^2}(F_1 \wt\Phi^d_{3,1} + F_2 \wt\Phi^d_{3,3} + F_3 \wt\Phi^d_{3,2}) (T_1 \Phi^d_{2,2} + T_2 \Phi^d_{2,1})
 H_{\ol{45}} \ ,
\ee
to give rise to the GJ relations, $m_d = 3
m_e$ and $m_s = m_\mu/3$ and $m_b = m_\tau$, after insertion of the flavon VEVs. The third operator in
Eq.~(\ref{downLO}) leads to
\bea \nonumber
&& \!\!\!\frac{x_1}{M^3}  \, \Phi^d_{2,1} \Phi^d_{2,2} [F_1 (T_1 \wt\Phi^d_{3,2} +T_2 \wt\Phi^d_{3,3}) + F_2 (T_1 \wt\Phi^d_{3,1} +T_2 \wt\Phi^d_{3,2})
+ F_3 (T_1 \wt\Phi^d_{3,3} +T_2 \wt\Phi^d_{3,1})] H_{\ol{5}} \\[2mm] \nonumber
&\!\!\!+\!\!& \!\!\frac{x_2}{M^3}  [ (F_2 (\Phi^d_{2,2})^2\! + F_3 (\Phi^d_{2,1})^2)(T_1\wt\Phi^d_{3,2}+T_2\wt\Phi^d_{3,3})
+ (F_3 (\Phi^d_{2,2})^2 \! + F_1 (\Phi^d_{2,1})^2)(T_1\wt\Phi^d_{3,1}+T_2\wt\Phi^d_{3,2}) \\[1mm]
&& \!\phantom{x_2 [~\,} + (F_1 (\Phi^d_{2,2})^2\! + F_2
(\Phi^d_{2,1})^2)(T_1\wt\Phi^d_{3,3}+T_2\wt\Phi^d_{3,1}) ] H_{\ol{5}} \ ,
\label{GSTrel}
\eea
where the two coupling constants $x_1$ and $x_2$ indicate two independent
invariants. With an appropriate vacuum alignment of the flavon VEVs,
Eq.~(\ref{GSTrel}) gives rise to equal (12) and (21) entries in the mass
matrices  necessary to achieve the GST relation. 
%
%
\setlength{\unitlength}{1mm}
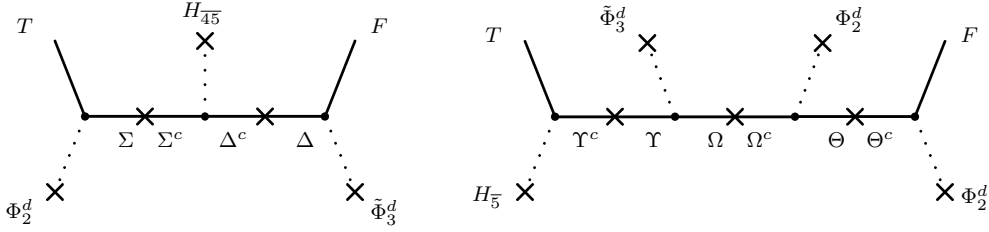
\begin{figure}[t]
\begin{center}
\begin{fmffile}{diagram}
\parbox{50mm}{
  \begin{fmfgraph*}(50,20)
    \fmfleft{i1,i2}
    \fmfright{o1,o2}
    \fmftop{o3}
    \fmf{dots}{v1,i1}
    \fmflabel{$\mbox{\scriptsize $\Phi^d_2$ \footnotesize}$}{i1}
    \fmfv{decor.shape=cross,decor.filled=full,decor.size=4thick}{i1}
    \fmf{plain}{v1,i2}
    \fmfv{decor.shape=circle,decor.filled=full,decor.size=1thick}{v1}
    \fmflabel{$\mbox{\scriptsize $T$ \footnotesize}$}{i2}
    \fmf{plain,label=$\mbox{\scriptsize $\phantom{\Sigma^c}\Sigma$ \footnotesize}$}{v1,v2}
    \fmfv{decor.shape=cross,decor.filled=full,decor.size=4thick}{v2}
    \fmf{plain,label=$\mbox{\scriptsize $\Sigma^c$ \footnotesize}$}{v2,v3}
    \fmfv{decor.shape=circle,decor.filled=full,decor.size=1thick}{v3}
    \fmf{dots,tension=0.0}{v3,o3}
    \fmflabel{$\mbox{\scriptsize $H_{\overline{45}}$ \footnotesize}$}{o3}
    \fmfv{decor.shape=cross,decor.filled=full,decor.size=4thick}{o3}
    \fmf{plain,label=$\mbox{\scriptsize $\Delta^c$ \footnotesize}$}{v3,v4}
    \fmfv{decor.shape=cross,decor.filled=full,decor.size=4thick}{v4}
    \fmf{plain,label=$\mbox{\scriptsize $\phantom{\Delta^c}\Delta$ \footnotesize}$}{v4,v5}
    \fmfv{decor.shape=circle,decor.filled=full,decor.size=1thick}{v5}
    \fmf{dots}{v5,o1}
    \fmflabel{$\mbox{\scriptsize $\tilde\Phi^d_3$ \footnotesize}$}{o1}
    \fmfv{decor.shape=cross,decor.filled=full,decor.size=4thick}{o1}
    \fmf{plain}{v5,o2}
    \fmflabel{$\mbox{\scriptsize $F$ \footnotesize}$}{o2}
   \end{fmfgraph*}
}
\hspace{0.3in}
\parbox{70mm}{
  \begin{fmfgraph*}(70,20)
    \fmfleft{i1,i2}
    \fmfright{o1,o2}
    \fmftop{o5,o3,o4,o6}
    \fmf{dots}{v1,i1}
    \fmflabel{$\mbox{\scriptsize $H_{\ol{5}}$ \footnotesize}$}{i1}
    \fmfv{decor.shape=cross,decor.filled=full,decor.size=4thick}{i1}
    \fmf{plain}{v1,i2}
    \fmfv{decor.shape=circle,decor.filled=full,decor.size=1thick}{v1}
    \fmflabel{$\mbox{\scriptsize $T$ \footnotesize}$}{i2}
    \fmf{plain,label=$\mbox{\scriptsize $\Upsilon^c$\footnotesize}$}{v1,v2}
    \fmfv{decor.shape=cross,decor.filled=full,decor.size=4thick}{v2}
    \fmf{plain,label=$\mbox{\scriptsize  $\phantom{\Upsilon^c}\Upsilon$ \footnotesize}$}{v2,v3}
    \fmfv{decor.shape=circle,decor.filled=full,decor.size=1thick}{v3}
    \fmf{dots,tension=0.0}{v3,o3}
    \fmflabel{$\mbox{\scriptsize $\tilde\Phi^{d}_3$ \footnotesize}$}{o3}
    \fmfv{decor.shape=cross,decor.filled=full,decor.size=4thick}{o3}
    \fmf{plain,label=$\mbox{\scriptsize $\phantom{\Omega^c}\Omega$ \footnotesize}$}{v3,v4}
    \fmfv{decor.shape=cross,decor.filled=full,decor.size=4thick}{v4}
    \fmf{plain,label=$\mbox{\scriptsize $\Omega^c$ \footnotesize}$}{v4,v5}
    \fmfv{decor.shape=circle,decor.filled=full,decor.size=1thick}{v5}
    \fmf{dots,tension=0.0}{v5,o4}
    \fmflabel{$\mbox{\scriptsize $\Phi^d_2$ \footnotesize}$}{o4}
    \fmfv{decor.shape=cross,decor.filled=full,decor.size=4thick}{o4}
    \fmf{plain,label=$\mbox{\scriptsize $\phantom{\Theta^c}\Theta$ \footnotesize}$}{v5,v6}
    \fmfv{decor.shape=cross,decor.filled=full,decor.size=4thick}{v6}
    \fmf{plain,label=$\mbox{\scriptsize $\Theta^c$ \footnotesize}$}{v6,v7}
    \fmfv{decor.shape=circle,decor.filled=full,decor.size=1thick}{v7}
    \fmf{dots}{v7,o1}
    \fmflabel{$\mbox{\scriptsize $\Phi^d_2$ \footnotesize}$}{o1}
    \fmfv{decor.shape=cross,decor.filled=full,decor.size=4thick}{o1}
    \fmf{plain}{v7,o2}
    \fmflabel{$\mbox{\scriptsize $F$ \footnotesize}$}{o2}
   \end{fmfgraph*}}
\end{fmffile}
\end{center}
\begin{center}
\caption{The Feynman diagram on the left shows how the contribution $(T
  \Phi^d_2)_1 (F\wt \Phi^d_3)_1 H_{\overline{45}}/M^2$ can arise in the
  context of a high energy completion. On the right we show the relevant
  diagram for generating the effective operator $(T \wt\Phi^d_3)_3(F \Phi^d_2
  \Phi^d_2)_3 H_{\ol{5}}/M^3$. In these diagrams scalars/fermions are
  displayed by dotted/solid lines. Crosses indicate a VEV for scalar components
   and mass insertions for fermions. See appendix B for details.\label{diagram}}
\end{center}
\end{figure}
%
%

In the neutrino sector the leading terms are
\be
y_D F N H^{}_{5} + \alpha N N \Phi^\nu_1 + \beta N N \Phi^\nu_2 + \gamma N N \Phi^\nu_{3'} \; .
\label{nuLO}
\ee

Using the vacuum alignment
\be
\langle \Phi^u_2 \rangle ~ = ~ \varphi^u_2 \begin{pmatrix} 0 \\ 1 \end{pmatrix}
\;\;\; \mbox{and} \;\;\;
\langle \wt\Phi^u_2 \rangle ~=~
\wt\varphi^u_2 \begin{pmatrix}0\\1\end{pmatrix} \ ,
\label{vacuum-up}
\ee
for the two flavons in doublet representations coupling to up quarks we see that
the up quark mass matrix $M_u$ is diagonal
\be \label{MuLO}
M_u \approx
\begin{pmatrix}
\varphi^u_2 \wt\varphi^u_2/M^2 & 0 & 0 \\
0 & \varphi^u_2/M & 0 \\
0 & 0 & 1
\end{pmatrix} \, v_u \ .
\ee
Taking
\be\label{orderphiu}
\varphi^u_2/M ~\approx~ \lambda^4 \;\;\; \mbox{and} \;\;\; \wt\varphi^u_2/M
~\approx~ \lambda^4 \ ,
\ee
with $\la \approx 0.22$ being the Wolfenstein parameter \cite{Wolfpara}, we obtain
the well-known mass hierarchy among the up quarks
\be
m_u:m_c:m_t \approx \la^8: \la^4:1 \; .
\ee

Similarly, we see using the alignment
\be
\langle \Phi^d_3 \rangle ~ = ~ \varphi^d_3 \begin{pmatrix}0\\1\\0\end{pmatrix}
\ , \qquad
\langle \wt\Phi^d_3 \rangle ~ =~
\wt\varphi^d_3 \begin{pmatrix}0\\-1\\1\end{pmatrix}
\ , \qquad
\langle \Phi^d_2 \rangle ~=~ \varphi^d_2 \begin{pmatrix}1\\0\end{pmatrix} \ ,
\label{vacuum-down}
\ee
that the down quark, $M_d$, and charged lepton mass matrix, $M_e$, are at LO of the form
(in the convention in which left-handed fields are on the left-hand side and right-handed fields on
the right-hand side of the mass matrix)
\be \label{MdLO}
M_d \approx
\begin{pmatrix}
0 & (\varphi^d_2)^2 \wt\varphi^d_3/M^3  & -(\varphi^d_2)^2 \wt\varphi^d_3/M^3   \\
-(\varphi^d_2)^2 \wt\varphi^d_3/M^3   & \varphi^d_2 \wt\varphi^d_3/M^2 &   -\varphi^d_2 \wt\varphi^d_3/M^2  + (\varphi^d_2)^2 \wt\varphi^d_3/M^3   \\
0  & 0 & \varphi^d_3/M
\end{pmatrix} v_d \ ,
\ee
and
\be \label{MeLO}
M_e \approx
\begin{pmatrix}
0 & -(\varphi^d_2)^2 \wt\varphi^d_3/M^3  & 0\\
(\varphi^d_2)^2 \wt\varphi^d_3/M^3   & -3 \varphi^d_2 \wt\varphi^d_3/M^2  & 0\\
-(\varphi^d_2)^2 \wt\varphi^d_3/M^3   & 3 \varphi^d_2 \wt\varphi^d_3/M^2  +
(\varphi^d_2)^2 \wt\varphi^d_3/M^3  & \varphi^d_3/M
\end{pmatrix} v_d \ .
\ee
$v_d$ denotes the VEV of the electroweak Higgs field $H_d$ which is in general a linear
combination of the doublet components of the GUT Higgs fields
$H_{\ol{5}}$ and $H_{\ol{45}}$. Assuming that
the angle associated with this mixing is of order one we can absorb the corresponding order one
factors into the other (not displayed) order one coefficients of each operator.
For
\be\label{orderphid}
\varphi^d_3/M ~\approx~ \lambda^{1+k} \ , \qquad
\wt\varphi^d_3/M ~\approx~ \lambda^{2+k} \ , \qquad
\varphi^d_2/M ~\approx~ \lambda \ ,
\ee
with $k=0$ or $k=1$, we find for the down quark and charged lepton mass hierarchy
\bea
&& m_d:m_s:m_b \approx  \, \la^4: \la^2: 1 \ , \label{mdLO}\\
&& m_e:m_\mu:m_\tau \approx (1/3) \, \la^4: 3 \, \la^2: 1 \ ,\label{meLO}
\eea
and for the mixing angles $\theta_{ij}^d$ and $\theta_{ij}^e$ of the left-handed fields
\bea
&& \theta_{12}^d \approx  \, \la \; , \;\; \theta_{13}^d \approx \la^3 \; , \;\;
\theta_{23}^d \approx  \la^2 \; , \label{thdLO}\\
&&  \theta_{12}^e \approx (1/3) \, \la \; , \;\; \theta_{13}^e \approx 0 \; , \;\;
\theta_{23}^e \approx 0 \; . \label{theLO}
\eea
The mass of the third generation of charged leptons and down quarks is at LO given by
\be
m_b \approx m_\tau \approx \la^{1+k} v_d \; .
\ee
As a consequence, the two possible choices of $k$ are equivalent to two different types of models: for $k=0$ we have
$m_b \approx m_\tau \approx 40/\tan\beta\, \rm{GeV}$ so that the value of
$\tan\beta$ is expected to be larger than 30, while for $k=1$ smaller values
of $\tan\beta$ in the range $5 \lesssim \tan\beta \lesssim 15$ are
preferred. Since the up quark mass matrix is diagonal, see Eq.~(\ref{MuLO}),
the Cabibbo angle has to be generated in the down quark sector, as one can see
from Eq.~(\ref{thdLO}). Also the two other quark mixing angles
$\theta_{13,23}^{q} \approx \theta_{13,23}^d$ turn out to be of the correct
order of magnitude. Furthermore, the model incorporates the GST relation
$\theta_{12}^q \approx \theta_{12}^d \approx \sqrt{m_d/m_s}$, see
Eqs.~(\ref{mdLO},\ref{thdLO}), arising from the equality of the
$(12)$ and $(21)$ elements as well as the vanishing of the $(11)$ element
in the down quark mass matrix $M_d$ at LO.
 Note that we can achieve the same LO results in the down
quark and charged lepton sector if we assume  $\langle \wt\Phi^d_3\rangle=
\wt\varphi^d_3 \, (0, \kappa, 1)^t$ with $|\kappa|=1$, $\kappa$ complex,
instead of using $\langle \wt\Phi^d_3\rangle\propto (0, -1, 1)^t$ as shown in
Eq.~(\ref{vacuum-down}).
For this reason we perform the study of marginal and dangerous Yukawa operators, which can be found
in the next section, assuming the alignment $\langle \wt\Phi^d_3\rangle\propto (0, \kappa, 1)^t$.
However, as one can see the flavon superpotential, discussed in section 4, only gives rise to the 
alignment $\langle \wt\Phi^d_3\rangle\propto (0, -1, 1)^t$. Thus, in the actual realisation, given in 
section 5, the latter alignment is used.

Finally, we display the LO results for the neutrino sector: the
neutrino Dirac mass matrix $M_D$ arising from Eq.~(\ref{nuLO}) has a very
simple form
\be
M_D = y_D \begin{pmatrix}
                     1 & 0 & 0\\
		     0 & 0 & 1\\
		     0 & 1 & 0
\end{pmatrix} v_u \; ,
\ee
while the right-handed neutrino mass matrix
\be
M_R = \begin{pmatrix}
                     \alpha \varphi^\nu_1 + 2 \gamma \varphi^\nu_{3'} & \beta \varphi^\nu_2 - \gamma \varphi^\nu_{3'}
		              & \beta \varphi^\nu_2 -\gamma \varphi^\nu_{3'}\\
		      \beta \varphi^\nu_2 - \gamma \varphi^\nu_{3'} & \beta \varphi^\nu_2 + 2 \gamma \varphi^\nu_{3'}
		              & \alpha \varphi^\nu_1 - \gamma \varphi^\nu_{3'}\\
		     \beta \varphi^\nu_2 -\gamma \varphi^\nu_{3'} &  \alpha \varphi^\nu_1 - \gamma \varphi^\nu_{3'}
		              & \beta \varphi^\nu_2 + 2 \gamma \varphi^\nu_{3'}
\end{pmatrix} \; ,
\ee
is the origin of TB mixing in this model if we use the vacuum alignment
\be
\langle\Phi^\nu_{3'} \rangle~=~ \varphi^\nu_{3'} \begin{pmatrix}1\\1\\1 \end{pmatrix}
\  , \qquad
\langle\Phi^\nu_2\rangle~=~ \varphi^\nu_2 \begin{pmatrix}1\\1 \end{pmatrix}
\  , \qquad
\langle \Phi^\nu_1 \rangle~=~ \varphi^\nu_1 \label{vacuum-nu}
\ee
(together with the information that these VEVs are all of the same order of
magnitude). Applying the type I see-saw formula
\be
m_\nu^{eff} = M_D M_R^{-1} M_D^t \ ,
\label{mnueffLO}
\ee
we find for the (complex) light neutrino masses
\be
m_1 = \frac{y_D^2 v_u^2}{\alpha \varphi^\nu_1 -\beta \varphi^\nu_2 + 3 \gamma \varphi^\nu_{3'}} \; , \;\;
m_2 = \frac{y_D^2 v_u^2}{\alpha \varphi^\nu_1 + 2 \beta \varphi^\nu_2} \; , \;\;
m_3 = \frac{y_D^2 v_u^2}{-\alpha \varphi^\nu_1 +\beta \varphi^\nu_2 + 3 \gamma \varphi^\nu_{3'}}
\; .
\ee
Due to the three different couplings $\alpha$, $\beta$, $\gamma$ the three light neutrino masses
are unrelated and any type of mass hierarchy can be accommodated. Especially, the former are not
constrained by a sum rule, as it is the case in the $A_4$ models \cite{A4-L}.

Concerning the approximate scale of the VEVs
of the flavons $\Phi^\nu_1$, $\Phi^\nu_2$ and $\Phi^\nu_{3'}$ we note that,
since they set the scale for right-handed neutrino masses, the physical neutrino masses
\be
m_i \sim 0.1 \ \mathrm{eV} \sim \frac{y_D^2v_u^2}{\varphi^\nu_{1,2,3'}} \ ,
\ee
imply that,
$\varphi^\nu_{1,2,3'} \sim 10^{13}$ GeV,
assuming $y_D\sim 0.3$ and $\tan \beta \sim 10$ for example.
Assuming the generic messenger scale $M$ to be of the order of the GUT scale,
$M \approx 10^{16}$ GeV, we see that $\varphi^\nu_{1,2,3'}$ fulfil
\be\label{orderphinu}
\varphi^\nu_{1} ~\approx~ \lambda^{4} M \ , \qquad
\varphi^\nu_{2} ~\approx~ \lambda^{4} M \ , \qquad
\varphi^\nu_{3'} ~\approx~ \lambda^{4} M \; .
\ee
The neutrino mixing stemming from Eq.~(\ref{mnueffLO}) is exactly TB mixing. However, it will be corrected by the
non-trivial $(12)$ mixing present in the charged lepton sector, see Eq.~(\ref{theLO}), so that the lepton
mixing angles at the high energy scale are given by \cite{King:2005bj},
\be
\label{thl_approx}
\sin^2 \theta^l_{23} \approx 1/2 \; , \;\; \sin^2 \theta^l_{12} \approx 1/3 + 2/9 \, \la \cos \delta^l \; , \;\;
\sin \theta^l_{13} \approx \la/(3 \sqrt{2}) \; ,
\ee
which incorporates the usual prediction for the reactor mixing angle, associated with the presence of the GJ factor and the $SU(5)$ context,
leading to the prediction $\theta^l_{13} \approx 3^{\circ}$ for
$\lambda \approx 0.22$, and, after eliminating $\lambda$, to the sum rule relation \cite{King:2005bj},
\be
\sin^2 \theta^l_{12} ~\approx~ \frac{1}{3}\left( 1+ 2 \sqrt{2} \sin \theta^l_{13} \cos \delta^l \right),
\ee
where $\delta^l$ is the leptonic Dirac CP phase.

The neutrino sector in the $S_4$ model above differs from that in the $A_4$ one \cite{A4-L}
by the presence of the doublet flavon $\Phi^\nu_2$ whose VEV structure preserves the generators $S$ and $U$.
 Note that, if the (irreducible) representations of $S_4$ are decomposed into those of its subgroup $A_4$, we find that
the doublet $\Phi^\nu_2$ decomposes  into the two non-trivial singlets ${\bf
  1'}$ and ${\bf 1''}$, see appendix A. In the $A_4$ model separate flavons in representations ${\bf
  1'}$ and ${\bf 1''}$, respectively, which, if allowed to appear with independent couplings, would violate the
symmetry associated with the generator $U$, have to be absent in order to achieve TB mixing \cite{A4-L}. This is why the $A_4$ model
accidentally preserves the generator $U$ in the neutrino sector, even though $U$ is not contained
in the group $A_4$ \cite{A4-L}. In the present model, both the generators
$S$ and $U$ are contained in $S_4$, and remain preserved in the neutrino sector at LO, so
that the neutrino flavour symmetry is reproduced in a more direct way.

The vacuum alignment of the flavons which has been only assumed in this
section will be discussed in more detail in section 4. We will show that the
alignment 
 (in which $\langle \wt\Phi^d_3\rangle\propto (0, -1, 1)^t$)
can be produced through $F$-terms of an additional set of gauge
singlet fields charged under the family symmetry $S_4 \times U(1)$. Regarding
the assumed sizes of the VEVs we find that these can be partly explained by
the superpotential which gives rise to correlations among the VEVs
and partly by introducing additional gauge singlets which allow couplings of
  positive mass dimension in the flavon superpotential, whose magnitude can be
appropriately chosen in order to reproduce the sizes of the VEVs.
 This issue is
  discussed in section~5.1 and appendix~D.


\section{Dangerous and marginal Yukawa operators}
\cleqn


After presenting the LO result which incorporates the prediction of TB mixing in the neutrino sector and
the successful accommodation of all charged fermion masses and quark mixings, we now discuss in more detail the role
of the additional $U(1)$ symmetry in forbidding all operators which would otherwise have a considerable effect on these
LO results. For example, as already remarked, beyond the LO we expect the
segregation of different flavons  $\Phi^{f}$ associated with a particular quark and lepton type $f=u,d,\nu$ to break down.

In order to identify operators which should be forbidden, we first classify them according to which
contributions they give to the fermion mass matrices. In this analysis we assume for the VEVs of the flavons
to have the LO form, as shown in Eqs.~(\ref{vacuum-up},\ref{vacuum-down},\ref{vacuum-nu})
 together with the generalised alignment of $\langle \wt\Phi^d_3\rangle$. However, as we will discuss
below, these VEVs receive in general corrections stemming from subleading terms present in the flavon superpotential.
 We fix the actual values of the $U(1)$ charges $x$, $y$ and $z$ on the
  basis of the results for the flavon superpotential. For this particular choice 
(and the specific alignment $\langle \wt\Phi^d_3\rangle\propto (0, -1, 1)^t$)
we discuss, in section 5, the subleading corrections induced through shifts in
  the flavon VEVs (and subleading operators), showing that their effects on
  fermion masses and mixings are negligible.

We can distinguish the following four types of Yukawa operators
\begin{itemize}
\item {\it desired} operators: These are the operators which - by definition of the $U(1)$ charges -
  are present at LO, see
  Eqs.~(\ref{upLO},\ref{downLO},\ref{nuLO}).
\item {\it dangerous} operators: These operators strongly perturb
the form of the mass matrices achieved at LO. In the case of charged fermions
their contribution is larger than the one stemming from the desired operators.
In the case of right-handed neutrinos, any contribution which is larger than or of the
same order of magnitude in $\la$ as the one coming from the desired operators has to be considered as
dangerous because TB mixing crucially depends on the form of the right-handed
neutrino mass matrix as well as on the fact that all entries of the latter
are of the same order of magnitude in $\la$.
\item {\it marginal} operators: These operators give contributions to
charged fermion mass matrices which are of the same order in $\la$ as the LO
contribution. Although not dangerous in the above sense, their presence has
a significant impact on the final result. For example, in the
case of the GST relation it might happen that such a marginal operator
contributes differently to the $(12)$ and the $(21)$ elements of the down
quark mass matrix $M_d$ so that the relation between the Cabibbo angle and the
masses of down and strange quark is lost.
\item {\it irrelevant} operators: These operators do not contribute to fermion
masses or mixings at LO in $\la$ and thus do not need to be
forbidden.
For phenomenology they are however not completely negligible, since they (can) give rise to
corrections to the LO result, e.g. they are responsible for deviations from
exact TB mixing in the neutrino sector.
\end{itemize}

According to this classification we wish to forbid all dangerous and all
marginal operators. Since the entries of the mass matrices $M_u$, $M_{d,e}$, $M_D$ and $M_{R}$
are of different order in $\la$, we list the
structures of the LO as well as the dangerous and the marginal
contributions for each sector separately. Note that we constrain ourselves in this study to
the case $k=1$, since it turns out that this choice reduces the number of dangerous and
marginal operators to a certain extent and thus facilitates the search for appropriate
$U(1)$ charge combinations $x$, $y$ and $z$, especially with small absolute
values. The value of $k$ is thus specified to $k=1$ for the rest of the paper.

In the up quark sector, Eq.~(\ref{MuLO}) tells us that the LO mass matrix has the form
\be
M_{u}^{\rm LO}\sim\left(
\begin{array}{ccc}
	\la^8 & 0 & 0\\
	0 & \la^4 & 0\\
	0 & 0 & 1
\end{array}
\right)  \ ,
\ee
so that we classify as dangerous (dang) all mass matrix entries which are equal or larger than
\be
M_{u}^{\rm dang} \gtrsim \left(
\begin{array}{ccc}
	\la^7 & \la^5 & \la^3\\
	\la^5 & \la^3 & \la\\
	\la^3 & \la & .
\end{array}
\right) \; .
\ee
Since the (33) entry of $M_u^{\rm LO}$ is $\mathcal{O}(1)$ any
corrections to this entry are irrelevant. The sizes of the other diagonal
entries are determined by the requirement of not having too large up and charm
quark masses, while the bounds on the off-diagonal elements originate from the
constraints on the quark mixing angles, $\theta_{12}^q \approx \la$,
$\theta_{23}^q \approx \la^2$ and $\theta_{13}^q \approx \la^3$, as well as
from achieving the correct mass hierarchy. Similarly, the
operators characterised as marginal (marg) give rise to entries in $M_u$ of the order
\be
M_{u}^{\rm marg} \sim \left(
\begin{array}{ccc}
	\la^8 & \la^6 & \la^4\\
	\la^6 & \la^4 & \la^2\\
	\la^4 & \la^2 & .
\end{array}
\right) \; .
\ee

Using Eqs.~(\ref{MdLO},\ref{MeLO}) we see that the LO of the entries in $M_{d,e}$ is
\be
M_{d,e}^{\mathrm{LO}} \sim \left(
\begin{array}{ccc}
	0 & \la^{5} & \la^{5}\\
	\la^{5} & \la^{4} & \la^{4}\\
	\la^{5} & \la^{4} & \la^{2}
\end{array}
\right) \; .
\ee
Thus, we classify operators as dangerous and marginal which lead to the
following mass matrix entries
\be
M_{d,e}^{\mathrm{dang}} \gtrsim \left(
\begin{array}{ccc}
	\la^{5} & \la^{4} & \la^{4}\\
	\la^{4} & \la^{3} & \la^{3}\\
	\la^{4} & \la^{3} & \la
\end{array}
\right)
\;\;\; \mbox{and} \;\;\;
M_{d,e}^{\mathrm{marg}} \sim \left(
\begin{array}{ccc}
	\la^{6} & \la^{5} & \la^{5}\\
	\la^{5} & \la^{4} & \la^{4}\\
	\la^{5} & \la^{4} & \la^{2}
\end{array}
\right) \; ,
\ee
respectively. Note that our results are based on the assumption that the mass
matrices are symmetric regarding  the order of magnitude in $\la$.
 We make this assumption although the off-diagonal elements in the third row
and column of $M_d$ and $M_e$, at LO, are non-zero only for one of the two matrices but not for
  both simultaneously.
Note further that
the (11) entry of $M_{d,e}$ vanishes at LO.  The constraint on this
entry to be smaller than $\la^5$ results from the requirement that the
determinant of $M_{d,e}$ should not exceed $\la^{12}$.

In the neutrino sector, all operators involving flavons, which contribute to
the neutrino Dirac mass matrix $M_D$,
\be
M_{D}^{\mathrm{LO}} \sim \left(
\begin{array}{ccc}
	1 & 0 & 0\\
	0 & 0 & 1\\
	0 & 1 & 0
\end{array}
\right) \; ,
\ee
can be classified as irrelevant, because the LO term, see
Eq.~(\ref{nuLO}), originates at the renormalisable level, i.e. does not
require the presence of any flavons. As already explained, since the form of
the LO result of $M_R$ is crucial to achieve TB neutrino mixing,
\be
M_{R}^{\mathrm{LO}} \sim \left(
\begin{array}{ccc}
	\la^{4} & \la^{4} & \la^{4}\\
	\la^{4} & \la^{4} & \la^{4}\\
	\la^{4} & \la^{4} & \la^{4}
\end{array}
\right) \ ,
\ee
any further contribution being of order $\la^4$ or larger is associated with a
dangerous operator
\be
M_{R}^{\mathrm{dang}} \gtrsim \left(
\begin{array}{ccc}
	\la^{4} & \la^{4} & \la^{4}\\
	\la^{4} & \la^{4} & \la^{4}\\
	\la^{4} & \la^{4} & \la^{4}
\end{array}
\right) \; .
\ee
All other operators contributing at the level $\lesssim \la^5$ are irrelevant.

Any operator comprising two superfields of the type $T_3$, $T$, $F$ and $N$ and an arbitrary number of
flavon fields that gives a dangerous or marginal contribution to a mass
matrix should be forbidden by the additional $U(1)$ symmetry. In the following, we first classify
all operators with up to three flavons according to the categories above, because
the LO result for fermion masses and mixings is generated by operators
with at maximum three flavons, see Eq.~(\ref{downLO}). The structures of the
resulting mass matrices determine the unwanted operators which are listed in
appendix~\ref{app-list}. Note that in this calculation we assumed the vacuum
alignment of the flavons as given in
Eqs.~(\ref{vacuum-up},\ref{vacuum-down},\ref{vacuum-nu}),  apart from the fact
that we allow $\langle\wt\Phi^d_3\rangle$ to be aligned as $(0, \kappa, 1)^t$ with
$|\kappa|=1$, $\kappa$ complex, instead of using $(0, -1, 1)^t$ as shown in
Eq.~(\ref{vacuum-down}).  The reason for this slightly generalised alignment of
$\langle\wt\Phi^d_3\rangle$ lies in the fact that keeping the relative
phase among the two non-vanishing entries of $\langle\wt\Phi^d_3\rangle$
arbitrary might leave us more freedom in the construction of the flavon
superpotential, from which the alignment of the flavons originates.
 Again, we emphasise that using the actual realisation of the flavon (super-)potential
presented in section 4, we arrive at the alignment $\langle \wt\Phi^d_3\rangle\propto (0, -1, 1)^t$.
However, an analysis of the Yukawa operators in the slightly more general framework
is still useful, because in any case the solutions found in this analysis can also be
applied to the specific alignment of $\langle \wt\Phi^d_3\rangle$ in which $\kappa$ is fixed to a certain
value. As we comment below, fixing $\kappa$ to $-1$ leads to some more possible sets of charges
$x$, $y$ and $z$, which however do not give rise to any feature not already revealed in the sets found
through the analysis using the generalised alignment of $\langle \wt\Phi^d_3\rangle$.
Apart from the unwanted operators the table in appendix \ref{app-list} also
shows the corresponding $\la$-suppression as well as the entries of
the mass matrices which are in conflict with the LO
setup.\footnote{Note that we only give one of the two entries $(ij)$ and
  $(ji)$ in the case of the symmetric or symmetrised terms $TTH^{}_{5}$,
  $T_3TH^{}_{5}$, $NN$.}
Entries for which the operator is marginal in the above sense are  marked with
square brackets, whereas in all other cases the operator is dangerous.
The three operators denoted with a prime ($43'$, $48'$, $54'$) differ from the
LO terms of the down quark sector in
Eq.~(\ref{downLO}) only by the exchange of $H_{\ol{5}}$ and
$H_{\ol{45}}$. All other terms given for the down quark sector must be forbidden for both
 Higgs fields, $H_{\ol{5}}$ as well as $H_{\ol{45}}$.

A complete scan over the parameters $x,y,z$ with $|x|,|y|,|z|\leq 5$ yields 43
different $U(1)$ symmetries which forbid all unwanted operators with up
to three flavon fields. Here we have identified the $U(1)$ symmetry related to
$(-x,-y,-z)$ with the one represented by $(x,y,z)$. Apart from this also dangerous or
marginal operators with more than three flavons should be forbidden. The dangerous
operators are
$$
TTH^{}_{5}(\Phi^d_2)^4/M^4  \ , \qquad
TTH^{}_{5}(\Phi^d_3)^3 \Phi^d_2/M^4 \ , \qquad
TTH^{}_{5}(\Phi^d_2)^3 \Phi^\nu_2/M^4 \ ,
$$
$$
TTH^{}_{5}(\Phi^d_2)^3 (\Phi^d_3)^2/M^5 \ , ~\quad
TTH^{}_{5}(\Phi^d_2)^7/M^7 \ , ~\quad
FTH_{\ol{5},\ol{45}} \Phi^d_3 (\Phi^d_2)^3/M^4 \ , ~\quad
NN(\Phi^d_2)^4/M^3 \ .
$$
As marginal operators we find
$$
TTH^{}_{5}(\Phi^d_3)^2 (\Phi^d_2)^2/M^4 \ , \qquad
TTH^{}_{5}(\Phi^d_2)^2 (\wt\Phi^d_3)^2/M^4 \ , \qquad
TTH^{}_{5}(\Phi^d_3)^2 \wt\Phi^d_3 \Phi^d_2/M^4 \ , \qquad
$$
$$
TTH^{}_{5} \Phi^d_3 (\Phi^d_2)^2 \Phi^\nu_{3'}/M^4 \ , \qquad
TTH^{}_{5} \Phi^d_3 \wt \Phi^d_3 (\Phi^d_2)^3/M^5 \ , \qquad
TTH^{}_{5}(\Phi^d_2)^4 \Phi^\nu_1/M^5 \ , \qquad
$$
$$
TTH^{}_{5} (\Phi^d_2)^6/M^6 \ , \qquad
FTH_{\ol{5},\ol{45}} \wt\Phi^d_3 (\Phi^d_2)^3/M^4 \ .
$$
Obviously, these must be removed as well,\footnote{We remark that the
  classification of these operators into dangerous and marginal does not depend
  on the relative phase introduced in the generalised alignment of
  $\langle\wt\Phi^d_3\rangle$.} so that we end up with 26 viable $U(1)$
symmetries listed in table~\ref{tab-U1s}.
\begin{table}
$$
\begin{array}{|c||c|c|c|c|c|c|c|c|c|c|c|c|c|}\hline
\#&1&2&3&4&5&6&7&8&9&10&11&12&13 \\\hline\hline
x&1&2&2&2&2&2&3&4&4&4&4&5&5 \\\hline
y&2&1&1&4&5&5&1&2&3&5&5&1&4\\\hline
z&5&4&5&5&1&4&5&5&5&2&3&4&1\\\hline  \multicolumn{14}{c}{} \\[-2mm] \hline
\#&14&15&16&17&18&19&20&21&22&23&24&25&26 \\\hline\hline
x&-1 &1&2 &2 &2 &2 &-3&3 &3 &4 &4 &4 &5 \\\hline
y&3&-4 &-5&-3&4 &5 &4 &-2&-2&-5&-1&-1&-2\\\hline
z&5 &5 &4 &5 &-5&-3&5 &4 &5 &3 &2 &5 &4\\\hline
\end{array}
$$
\caption{\label{tab-U1s}The 26 viable $U(1)$ symmetries defined by the
  parameters $x,y,z$ ($|x|,|y|,|z|\leq~5$) for the alignment $\langle\wt\Phi^d_3\rangle \propto (0, \kappa, 1)^t$ with $|\kappa|=1$,
$\kappa$ complex. Obviously, for each set of charges $(x,y,z)$ also the
set $(-x,-y,-z)$ is a viable candidate.}
\end{table}
This set will serve as a source of a candidate $U(1)$ symmetry which
eventually leads to a successful $S_4 \times SU(5)$ model.

Assuming the alignment of $\langle\wt\Phi^d_3\rangle$ to be the one as given
in Eq.~(\ref{vacuum-down}), we find that two operators among those classified
as dangerous or marginal become irrelevant, namely operators $\# 18$ and $\#
32$ in the table found in appendix \ref{app-list}.
Allowing these two operators, we find 18 additional solutions for the
$U(1)$ charges $x$, $y$ and $z$ as compared to the 43 mentioned above.
Including eventually the requirement to forbid the dangerous and marginal
operators with more than three flavons leaves us with 15 new sets $(x,y,z)$
that are added to the 26 $U(1)$ symmetries of table~\ref{tab-U1s}.
However, as we do not find any set with charges $x$, $y$ and $z$ with $|x|$,
$|y|$, $|z|<4$, these 15 new solutions are qualitatively not different from the
ones given in table~\ref{tab-U1s}, so that we do not consider them any further.
 Nevertheless in the subsequent sections 4 and 5 the alignment of $\langle \wt\Phi^d_3\rangle$ 
is fixed through the flavon superpotential to be proportional to $(0,-1,1)^t$.

Finally, we remark that the high energy completion we proposed in order to 
only generate the operators $(F\wt \Phi^d_3)_1 (T \Phi^d_2)_1 H_{\ol{45}}/M^2$
and $(F \Phi^d_2 \Phi^d_2)_3 (T \wt\Phi^d_3)_3 H_{\ol{5}}/M^3$  in the down quark
sector actually depends on the choice of the combination $x$, $y$ and $z$,
because in the calculation for generic charges $x$, $y$ and $z$ we implicitly
relied on the fact that all heavy fields appearing as messengers carry 
(different) charges under the $U(1)$ symmetry so that only the operators given
in appendix B are generated at the renormalisable level. This must be taken into account as an
additional constraint on the solutions presented in this section. We will
comment on this point in section 5 and appendix B. 


\section{Vacuum alignment}
\cleqn


The origin of the vacuum alignment is an integral part of a model of fermion
masses and mixings using a non-Abelian family symmetry.\footnote{There
    are also other possibilities to break a family symmetry, e.g.
through non-trivial boundary conditions in extra-dimensional models \cite{GfbreakED}.}
We first discuss in
section 4.1 how to achieve the vacuum alignment shown in
Eqs.~(\ref{vacuum-up},\ref{vacuum-down},\ref{vacuum-nu})  by introducing a new
set of fields, called driving fields in the following, from whose $F$-terms we
derive the alignment. 
 We actually show that in this case $\langle \wt\Phi^d_3\rangle \propto (0,-1,1)^t$
is the only solution, so that the parameter $\kappa$ in the generalised form of the
alignment of $\langle \wt\Phi^d_3\rangle$, used in the preceding section, is fixed 
to $\kappa=-1$.
The $U(1)$ charges of the driving fields are
given in terms of the three parameters $x$, $y$ and $z$ which have been
introduced in section 2. The additionally allowed operators of the flavon
superpotential beyond
those given in section 4.1 are then determined for all 26 sets of $U(1)$
charges $x$, $y$ and $z$ shown in table 2.
On the basis of this study we exclude all sets $(x,y,z)$
for which these additional operators strongly perturb the LO
vacuum alignment.
Focusing on the remaining four choices of $U(1)$ charges $x$, $y$
and $z$ for which no such operators arise if the LO results of the flavon VEVs
are used, we search for possibilities to (partly) correlate the flavon VEVs by
introducing further driving fields.
We eventually fully specify the values of the $U(1)$ charges $x$, $y$ and $z$
by choosing the possibility which allows for the largest number of
correlations among the scales of the various flavon VEVs. 
 This is explained in section 4.3 and in detail shown in appendix D. Furthermore,
we discuss in section 5.1 and appendix D that a
minimum of undetermined parameters
among the flavon VEVs can be reached, if driving fields are included which allow for
couplings with positive mass dimension.


\subsection{Flavon superpotential at LO}


 In our approach we generate
 the vacuum alignment through $F$-terms by coupling the flavons to
 driving fields. The latter are - similar to the flavons - gauge singlets and
transform in general in a non-trivial way under $S_4 \times U(1)$. We introduce
furthermore a $U(1)_R$ symmetry under which all driving fields carry
charge $+2$. In contrast to this, flavons and the GUT Higgs fields are uncharged under $U(1)_R$
and supermultiplets containing SM fields (or right-handed neutrinos) have $U(1)_R$ charge $+1$.
In this way, the driving fields can only appear linearly in the superpotential and in addition do not have direct
interactions with SM fermions (and right-handed neutrinos).
Under the assumption that the family symmetry $S_4 \times U(1)$ is broken at high energies,
a scale at which SUSY is not broken in the visible sector,
we can deduce the alignment of the flavon VEVs from the equations arising
from setting the $F$-terms of the driving fields to zero.
Table~\ref{tab-driving} gives a list of driving fields with which we
 generate the vacuum alignment in
Eqs.~(\ref{vacuum-up},\ref{vacuum-down},\ref{vacuum-nu}). The $U(1)$ charges
are expressed in terms of the parameters $x,y$ and $z$ so as to allow the
relevant superpotential operators which give rise to the desired alignments.
\begin{table}
$$
\begin{array}{|c|c|c|c|c|c|c|c|c|}\hline
\text{Driving field}\phantom{\Big|}\!\!&X^d_1 & Y^d_2 & Z^\nu_{3'} & Y^\nu_2 & \ol X^d_1 &
X^{\nu d}_{1'} & Y^{du}_2 & X^u_1 \\\hline
SU(5)\phantom{\Big|}\!\!&\bf 1&\bf 1&\bf 1&\bf 1&\bf 1&\bf 1&\bf 1&\bf 1\\\hline
S_4\phantom{\Big|}\!\!&\bf 1&\bf 2&\bf 3'&\bf 2&\bf 1&\bf 1'&\bf 2&\bf 1 \\\hline
U(1)\phantom{\Big|}\!\!&-2z&2y-2z&-4y&-4y&x+2y+z&x-y+2z&2x-z&2x \\\hline
\end{array}
$$
\caption{\label{tab-driving}The driving fields required for obtaining
  the vacuum alignment. All these fields carry charge $+2$ under $U(1)_R$.}
\end{table}
In the following we will discuss these terms in turn.
 Most of the alignments are achieved through renormalisable
operators with three fields in order not to introduce further mass scales.
In the case of non-renormalisable terms we suppress the operators by
appropriate powers of the generic messenger scale $M$. Note that in such a setup
with no superpotential couplings of positive mass dimension it is impossible to exclude the
trivial solution, i.e. a vacuum in which all flavon VEVs vanish. However,
having chosen the specific set of $U(1)$ charges
$x$, $y$ and $z$ we comment on this issue in section 5.1 and present
a way to enforce spontaneous family symmetry breaking in appendix D.

The driving field $X^d_1$, coupled to $\Phi^d_2$ through
\be
X^d_1 (\Phi^d_2)^2 ~=~
X^d_1 \Phi^d_{2,1}\Phi^d_{2,2}  \ ,
\ee
allows to align $\langle \Phi^d_2 \rangle$ either as
\be
\langle \Phi^d_2 \rangle ~\propto ~
\begin{pmatrix}   1\\0 \end{pmatrix}
~\text{or as}~
\begin{pmatrix}   0\\1 \end{pmatrix} \ .\label{d2-alignm}
\ee
In the following we choose the alignment in which the component $\Phi^d_{2,1}$
receives a non-zero VEV. Assuming Eq.~(\ref{orderphid}) to hold, the alignment of the
VEV of $\Phi^d_2$ is generated through an operator of the order $\la^2$.


Using the field $Y^d_2$ and the
alignment achieved for $\langle\Phi^d_2\rangle$ we align the VEV of
$\Phi^d_3$. In general we find three independent dimension-5 terms coming
from $Y^d_2 (\Phi^d_2)^2 (\Phi^d_3)^2/M^2$
\bea
&&\frac{1}{M^2}((\Phi^d_{3,1})^2 + 2 \Phi^d_{3,2}\Phi^d_{3,3})(Y^d_{2,1} (\Phi^d_{2,1})^2 + Y^d_{2,2} (\Phi^d_{2,2})^2)
\\ \nonumber
&+&\frac{1}{M^2}\Phi^d_{2,1}\Phi^d_{2,2}\left[Y^d_{2,1}((\Phi^d_{3,3})^2 + 2 \Phi^d_{3,1}\Phi^d_{3,2})+Y^d_{2,2}((\Phi^d_{3,2})^2 + 2 \Phi^d_{3,1}\Phi^d_{3,3})
\right]
\\ \nonumber
&+&\frac{1}{M^2}\left[
Y^d_{2,1} (\Phi^d_{2,2})^2 ((\Phi^d_{3,2})^2 + 2 \Phi^d_{3,1}\Phi^d_{3,3})+ Y^d_{2,2} (\Phi^d_{2,1})^2 ((\Phi^d_{3,3})^2 + 2 \Phi^d_{3,1}\Phi^d_{3,2})
\right] \ ,
\eea
which yield the following conditions
\be\label{condphid3}
(\Phi^d_{3,1})^2 + 2 \Phi^d_{3,2}\Phi^d_{3,3} = 0
~~~\mbox{and}~~~
(\Phi^d_{3,3})^2 + 2 \Phi^d_{3,1}\Phi^d_{3,2} = 0 \; ,
\ee
if the alignment of $\langle\Phi^d_2\rangle$ is plugged into the $F$-terms of $Y^d_{2,1}$ and $Y^d_{2,2}$. Eq.~(\ref{condphid3})
shows that $\langle\Phi^d_3\rangle$ has to be aligned as
\be
\langle \Phi^d_3 \rangle ~\propto~
\begin{pmatrix}   0\\1\\0 \end{pmatrix}
~\text{or as}~~
\frac{1}{3} \begin{pmatrix}   2\omega^p \\-1\\ 2\omega^{-p} \end{pmatrix} ,
~p=0,\pm 1 \ ,\label{d3-alignm}
\ee
with $\omega= e^{2\pi i/3}$. As before, we select the first of these
four possible alignments. Assuming the relative size of the VEVs $\varphi^d_2$
and $\varphi^d_3$ with respect to the messenger scale $M$ as given in
Eq.~(\ref{orderphid}) we find that the operators responsible for the alignment
of $\langle\Phi^d_3\rangle$ arise at the level $\la^6$.


The vacua of the fields $\Phi^\nu_1$,$\Phi^\nu_2$ and $\Phi^\nu_{3'}$, whose
alignments give rise to TB mixing in the neutrino sector, are driven by the
fields $Y^\nu_{2}$ and $Z^\nu_{3'}$. The part  of the superpotential
responsible for the correct alignment reads
\bea\label{nu-potential}
&&
\!\!\!\!\!\!\!\!
a^\nu_1 Y^\nu_2 \Phi^\nu_1 \Phi^\nu_2 +a^\nu_2 Y^\nu_2 (\Phi^\nu_2)^2 + a^\nu_3 Y^\nu_2 (\Phi^\nu_{3'})^2
+ b^\nu_1 Z^\nu_{3'} \Phi^\nu_1 \Phi^\nu_{3'} + b^\nu_2 Z^\nu_{3'} \Phi^\nu_2 \Phi^\nu_{3'} + b^\nu_3 Z^\nu_{3'} (\Phi^\nu_{3'})^2
\\ \nonumber
&&\!\!\!\!\!\!\!\!\!\!\!\!
 \;\;\,= a^\nu_1 \Phi^\nu_1 \left[ Y^\nu_{2,1}\Phi^\nu_{2,2}\!+Y^\nu_{2,2}\Phi^\nu_{2,1} \right]
+ a^\nu_2 \left[ Y^\nu_{2,1} (\Phi^\nu_{2,1})^2 \!+ Y^\nu_{2,2} (\Phi^\nu_{2,2})^2 \right]
\\ \nonumber
&&\!\!\!\!\!\!\!\!\!\!\!\!
 \;\;\, + \,a^\nu_3 \left[ Y^\nu_{2,1} ((\Phi^\nu_{3',3})^2 \!+ 2 \Phi^\nu_{3',1} \Phi^\nu_{3',2})
+Y^\nu_{2,2} ((\Phi^\nu_{3',2})^2 \!+ 2 \Phi^\nu_{3',1} \Phi^\nu_{3',3}) \right]
\\ \nonumber
&&\!\!\!\!\!\!\!\!\!\!\!\!
 \;\;\, +\,b^\nu_1 \Phi^\nu_1 \left[Z^\nu_{3',1} \Phi^\nu_{3',1}\!+ Z^\nu_{3',2} \Phi^\nu_{3',3}
\!+Z^\nu_{3',3} \Phi^\nu_{3',2} \right]
\\ \nonumber
&&\!\!\!\!\!\!\!\!\!\!\!\!
 \;\;\, +\,b^\nu_2 \left[ Z^\nu_{3',1} (\Phi^\nu_{2,1}\Phi^\nu_{3',2}\!+\Phi^\nu_{2,2}\Phi^\nu_{3',3})
+  Z^\nu_{3',2} (\Phi^\nu_{2,1}\Phi^\nu_{3',1}\!+\Phi^\nu_{2,2}\Phi^\nu_{3',2})
+  Z^\nu_{3',3} (\Phi^\nu_{2,1}\Phi^\nu_{3',3}\!+\Phi^\nu_{2,2}\Phi^\nu_{3',1})
\right]
\\ \nonumber
&&\!\!\!\!\!\!\!\!\!\!\!\!
 \;\;\, +\,b^\nu_3 \left[ Z^\nu_{3',1} ((\Phi^\nu_{3',1})^2\!-\Phi^\nu_{3',2}\Phi^\nu_{3',3})
+ Z^\nu_{3',2} ((\Phi^\nu_{3',2})^2\!-\Phi^\nu_{3',1}\Phi^\nu_{3',3})
+ Z^\nu_{3',3} ((\Phi^\nu_{3',3})^2\!-\Phi^\nu_{3',1}\Phi^\nu_{3',2})
\right]  ,
\eea
where the coefficients $a^\nu_i$ and $b^\nu_i$ are undetermined complex
parameters whose absolute values are of order one. The $F$-terms of the
components of $Y^\nu_2$ and $Z^\nu_{3'}$ vanish if the VEVs of $\Phi^\nu_1$,
$\Phi^\nu_2$ and $\Phi^\nu_{3'}$ take the following form\footnote{To be
  precise, we explicitly exclude  solutions in which the fields $\Phi^\nu_1$
  and  $\Phi^\nu_2$ acquire a VEV whereas $\Phi^\nu_{3'}$ does not.}
\be
\langle \Phi^\nu_{3'} \rangle = \varphi^\nu_{3'}
\begin{pmatrix}   \omega^{-p}\\\omega^{p}\\1 \end{pmatrix}  ,~\quad
\langle \Phi^\nu_{2} \rangle = \varphi^\nu_2
\begin{pmatrix}   1\\\omega^p \end{pmatrix} , ~\quad
\langle \Phi^\nu_{1} \rangle = \varphi^\nu_1 , ~\quad p=0,\pm 1\ ,
\label{nu-alignm}
\ee
with the scales, $\varphi^\nu_{3'}$, $\varphi^\nu_2$, $\varphi^\nu_1$, of the
VEVs related through
\be
\varphi^\nu_2 ~=~- \frac{b^\nu_1}{2 b^\nu_2} \omega^p \varphi^\nu_1 \ ,\qquad
({\varphi^\nu_{3'}})^2 ~=~ \frac{b^\nu_1}{6 b^\nu_2 a^\nu_3} \left(a^\nu_1 - \frac{b^\nu_1 a^\nu_2}{2 b^\nu_2} \right)
\omega^{2p} ({\varphi^\nu_1})^2 \ ,\label{nu-alignm-scale}
\ee
and $\varphi^\nu_1$ remaining undetermined. Thus, without assuming any
fine-tuning among the parameters $a^\nu_i$ and $b^\nu_i$ in the superpotential
the VEVs are expected to be of the same order of magnitude,
\be
\varphi^\nu_{3'} ~ \sim ~ \varphi^\nu_2 ~ \sim ~ \varphi^\nu_{1}  \ .
\label{vev-scale-nus}
\ee
This is a favourable situation as it ensures that all entries of the
right-handed neutrino mass matrix are  naturally of similar order of magnitude
so that a non-hierarchical light neutrino mass spectrum  is generated. In
Eq.~(\ref{vacuum-nu}) the
alignment with $p=0$ is given. Finally, we note that the alignment of the
flavons relevant to the neutrino sector at LO arises at $\m
O(\la^8)$. Thus, all combinations of flavons coupling to the driving fields
$Y^\nu_2$  and $Z^\nu_{3'}$ which might give a contribution to the alignment
of order $\gtrsim \la^8$ have to be absent.


Concerning the alignment of the VEV of $\wt\Phi^d_3$ we notice that for this
purpose two  driving fields are required, $\ol X^d_1$ and $X^{\nu
  d}_{1'}$. First, one aligns $\langle\wt\Phi^d_3\rangle$ through the
non-renormalisable operator
\bea\label{Xdbarterm}
&&\!\!\!\!\!\!\frac{1}{M}\ol X^d_1\Phi^d_2\Phi^d_3 \wt\Phi^d_3
\\ \nonumber
&&
=\frac{1}{M}\ol X^d_1 \left[
\Phi^d_{2,1} (\Phi^d_{3,1}\wt\Phi^d_{3,2}+\Phi^d_{3,2}\wt\Phi^d_{3,1}+\Phi^d_{3,3}\wt\Phi^d_{3,3})
+\Phi^d_{2,2} (\Phi^d_{3,1}\wt\Phi^d_{3,3}+\Phi^d_{3,2}\wt\Phi^d_{3,2}+\Phi^d_{3,3}\wt\Phi^d_{3,1})
\right]  .
\eea
Using the alignment of $\langle\Phi^d_2\rangle$ and $\langle\Phi^d_3\rangle$
as discussed above we can immediately infer from setting the $F$-term of $\ol
X^d_1$ to zero that
\be\label{alignphidt3_1}
\langle\wt\Phi^d_{3,1}\rangle=0 \; ,
\ee
so that only the second and third entry of $\langle\wt\Phi^d_3\rangle$ can
acquire a non-vanishing value. In order to correlate these entries we employ
the field $X^{\nu d}_{1'}$ which couples $\wt\Phi^d_3$ to $\Phi^\nu_{3'}$ through
\be
 X^{\nu d}_{1'}\wt\Phi^d_3\Phi^\nu_{3'}
= X^{\nu d}_{1'} (\wt\Phi^d_{3,1}\Phi^\nu_{3',1}+\wt\Phi^d_{3,2}\Phi^\nu_{3',3}+\wt\Phi^d_{3,3}\Phi^\nu_{3',2}) \; .
\ee
 For $\langle\Phi^\nu_{3'}\rangle$ being already aligned, the vanishing of
the $F$-term of $X^{\nu d}_{1'}$,
\be
\langle\wt\Phi^d_{3,1}\rangle+\langle\wt\Phi^d_{3,2}\rangle+\langle\wt\Phi^d_{3,3}\rangle =0  \; ,
\ee
together with Eq.~(\ref{alignphidt3_1}) shows that
$\langle\wt\Phi^d_{3,2}\rangle$ and $\langle\wt\Phi^d_{3,3}\rangle$ have to be
equal up to a relative sign. Thus, $\langle\wt\Phi^d_3\rangle$ is fully
aligned as
\be
\langle \wt \Phi^d_{3} \rangle \propto
\begin{pmatrix}   0\\-1\\1 \end{pmatrix}  \ .
\label{d3t-alignm}
\ee
Note that the operator from which $\langle\wt\Phi^d_{3,1}\rangle=0$ is
inferred arises at order $\la^6$ whereas the operator responsible for the
equality of $\langle\wt\Phi^d_{3,2}\rangle$ and
$\langle\wt\Phi^d_{3,3}\rangle$ is of order $\la^7$, using the orders of
magnitude shown in Eq.~(\ref{orderphid}) and Eq.~(\ref{orderphinu}).


Finally, the vacua of the flavons, $\Phi^u_{2}$ and $\wt\Phi^u_2$, responsible
for giving masses to up quarks at LO, can be aligned with the help
of two driving fields $Y^{du}_2$ and $X^u_1$. The field $Y^{du}_2$ allows
to couple the flavons $\Phi^u_2$ and $\Phi^d_2$ through the operator
\be
Y^{du}_2 \Phi^d_2  \Phi^u_2 = Y^{du}_{2,1} \Phi^d_{2,1}\Phi^u_{2,1} + Y^{du}_{2,2} \Phi^d_{2,2}\Phi^u_{2,2} \; .
\ee
Thus, from the vanishing of the $F$-term of $Y^{du}_{2,1}$ under the condition that $\langle\Phi^d_{2,1}\rangle\neq 0$ holds,
as discussed above, we immediately find that
\be
\langle  \Phi^u_2 \rangle ~\propto~
\begin{pmatrix}   0\\1 \end{pmatrix}  \ .\label{u2-align}
\ee
Similarly, the field $X^u_1$ couples the two fields $\Phi^u_{2}$ and $\wt \Phi^u_{2}$ through
\be
X^{u}_{1}  (\Phi^u_{2,1}\wt\Phi^u_{2,2}+\Phi^u_{2,2}\wt\Phi^u_{2,1}) \; .
\ee
Taking $\langle\Phi^u_2\rangle$ to be aligned as given in Eq.~(\ref{u2-align}), we derive from the vanishing $F$-term of
$X^{u}_{1}$ that $\langle\wt\Phi^u_2\rangle$ is aligned in the same way as $\langle\Phi^u_2\rangle$, i.e.
\be
\langle \wt \Phi^u_2 \rangle ~\propto~
\begin{pmatrix}   0\\1 \end{pmatrix}  \ .
\label{u2t-align}
\ee
These alignments are induced by operators that arise~-~according to
Eqs.~(\ref{orderphiu},\ref{orderphid})~- at the order $\la^5$
and $\la^8$, respectively.

Since some of the equations leading to the alignment of the flavon VEVs do not
have a unique solution, we arrive at a total 24 different degenerate
vacua (not counting the ones in which $\langle \Phi^\nu_{3'}\rangle=0$).
 We note that these 24 sets are related by $S_4$ transformations. 
Choosing one of the sets different from the one presented in 
Eqs~(\ref{vacuum-up},\ref{vacuum-down},\ref{vacuum-nu}) clearly leads to fermion mass matrices
which are of a different form from the one of those given in section 2. However, we have
checked explicitly that all these sets of different fermion mass matrices are related
to the one found in section 2 by $S_4$ transformations performed on the matter superfields
$T_3$, $T$, $F$ and $N$. We emphasise that the results for fermion mixings are not changed by
these transformations, because left-handed quarks as well as left-handed leptons
transform in the same way. Thus,
 our choice of the vacuum structure is a convention that can
be used without loss of generality.

Similar to the fact that the $F$-terms of the driving fields are the origin of
the alignment of the flavon VEVs, we can derive from the $F$-terms of the
latter fields the vacuum structure of the driving fields. As all terms in the
flavon superpotential are linear in the driving fields, the configuration in
which all these fields have vanishing VEVs is in any case a solution. However,
in our model we find that, plugging in the alignment of the flavon VEVs,
this is not the only possible solution satisfying the requirement that all
$F$-terms of the flavons vanish. In principle, the two fields $X^d_1$ and the
second component of $Y^{du}_2$ might have non-zero VEVs which fulfil a
non-trivial relation. The absolute size of these VEVs is not determined, however their
relative one. We note
that non-vanishing VEVs for driving fields could induce a $\mu$-term
for $H_{5}$  and $H_{\ol{5}}$ which, in our model, is forbidden by the $U(1)_R$
symmetry. In the following we will, however, assume that all VEVs of the driving
fields are zero.


\subsection{Discussion of dangerous operators in the flavon superpotential}


For specific choices of $x,y$ and $z$ additional operators which (can) spoil the
above alignment might be allowed by the $U(1)$ symmetry as well.
Thus, it is necessary to check each of the 26 possible choices of $U(1)$
charges $x$, $y$ and $z$ displayed in table~\ref{tab-U1s} for such unwanted
operators, using the vacuum alignments generated at the LO as shown
in the preceding section. We classify all operators as unwanted which
lead to contributions proportional to the same or to a lower power in $\la$ than the
LO terms  given above.

As an example of an unsuccessful case which is excluded
by our procedure, consider the $U(1)$ charge assignment $\# 1$ with
$(x,y,z)=(1,2,5)$. In this case the operator $Z^\nu_{3'} \Phi^d_3
(\Phi^d_2)^2/M$ is allowed by all the symmetries of the model. Inserting the desired
vacuum structure we arrive at a contribution of the form
$Z^\nu_{3',2} \varphi^d_3 (\varphi^d_2)^2/M$ being of order $\la^4$.
This has to be compared to the terms given in Eq.~(\ref{nu-potential}) leading to
the alignment of $\langle \Phi^\nu_{3',2,1} \rangle$ which are of order $\la^8$.
Thus, the additional operator $Z^\nu_{3'} \Phi^d_3
(\Phi^d_2)^2/M$ gives a contribution dominating even the assumed LO one,
so that the $U(1)$ charge assignment $\#1 $ has to be discarded.

Eventually, we are left with four potentially successful $U(1)$ charge assignments $(x,y,z)$
for which we do not find any operators that strongly perturb the LO
alignment if the flavons assume their LO VEVs. These are
\be
\label{U1sols4}
\# 10: ~ (4,5,2) \ , ~\quad
\# 13: ~ (5,4,1) \ , ~\quad
\# 21: ~ (3,-2,4) \ , ~\quad
\# 25: ~ (4,-1,5) \ .
\ee
We note that two of the solutions, namely $\# 10$ and $\# 21$,
allow for operators which could in principle strongly perturb the LO
result, $Y^d_2 \Phi^d_2 \Phi^u_2$, $X^u_1 \Phi^d_2 (\Phi^d_3)^2/M$ and
$Y^d_2 (\Phi^d_2)^3/M$, respectively.
However, inserting the LO structure of the flavon VEVs
we find that these operators give vanishing contributions.
Nevertheless, they might still perturb the vacuum alignment if corrections to
the LO vacua, caused by subleading terms, are taken into account
(see below). In contrast to this, the solutions $\# 13$ and $\# 25$ do not allow for any operator which can strongly perturb the
LO alignment, irrespective of the inserted vacua.
Finally, we remark that for the choice $\# 22$ of $U(1)$ charges, $(x,y,z)=(3,-2,5)$,
there is one operator $M_V \ol{X}^d_1 \Phi^\nu_1$ which, depending on the
size of the mass scale $M_V$, might or might not spoil the vacuum alignment achieved at LO.
Choosing $M_V \lesssim \la^3 M$ renders the associated contribution subdominant compared
to the one coming from the LO term, displayed in Eq.(\ref{Xdbarterm}). However, since we
would like to avoid the presence of such additional mass scales in the flavon superpotential
at this stage of the study, we discard case $\# 22$.


\subsection{Correlations among the flavon VEVs}
\label{4.3}


Having obtained the structure of the vacuum alignment, we now turn to the
question of relating the scales of the flavon VEVs. So far, the only
such relation is the one between the three flavons, relevant for right-handed neutrino masses, as stated in
Eq.~(\ref{vev-scale-nus}). Such a correlation of scales of more flavon
VEVs can be achieved by adding further driving fields. 

 Referring to the detailed analysis in appendix D for the four viable choices of $U(1)$
charges, $\# 10$, $\# 13$, $\# 21$ and $\# 25$, we find that only 
in case $\#13$ is it possible to find two (independent) further relations
among the flavon VEVs, if we introduce two further driving fields, transforming
as singlets under $S_4$. This result is achieved, if terms of a minimum size of order $\la^9$
are considered for scales of the flavon VEVs according to 
Eqs.~(\ref{orderphiu},\ref{orderphid},\ref{orderphinu}), and the possibility
of having couplings with positive mass dimension in the superpotential is 
not taken into account.

Explicitly we find
\be\label{correlnew12}
M \varphi^u_2 \sim  \varphi^d_2 \wt\varphi^d_3 \;\;\;
\mbox{and}
\;\;\;
M^2 \wt\varphi^u_2\wt\varphi^d_3 \sim  \varphi^d_2 (\varphi^d_3)^3 \; , \;\;\;\;
\ee
which comprise together with
Eq.~(\ref{vev-scale-nus}) the maximum set of
correlations that we can achieve in the context of our 26 possible $U(1)$ charge sets, see table \ref{tab-U1s}.
As a consequence, the eight flavon VEV scales face four constraints, thus leaving four parameters
undetermined.

In the following we shall choose the flavon VEVs
\be\label{scalefree}
\wt\varphi^u_2 \; , \;\; \varphi^d_3 \; , \;\; \varphi^d_2 \;\;\; \mbox{and}
\;\;\; \varphi^\nu_1 \ ,
\ee
by hand to have the following orders
\be
\wt\varphi^u_2/M \sim \la^4 \; , \;\; \varphi^d_3/M \sim \la^2 \; , \;\;
\varphi^d_2/M \sim \la \;\;\; \mbox{and} \;\;\; \varphi^\nu_1/M \sim \la^4\ .
\ee
Then, using the above correlations, we can deduce without further assumption
\be\label{scalefixed}
\varphi^u_2/M \sim \la^4 \; , \;\; \wt\varphi^d_3/M \sim \la^3 \; , \;\;
\varphi^\nu_2/M \sim \la^4 \;\;\; \mbox{and} \;\;\; \varphi^\nu_{3'}/M \sim \la^4\ .
\ee
 We find that the VEVs of the additional driving fields leading to the two further
correlations have to vanish. This is required by the $F$-term equations of
the flavons, if the LO alignments of
Eqs.~(\ref{vacuum-up},\ref{vacuum-down},\ref{vacuum-nu}) are applied.
As will be discussed in section 5.1 and in more detail in the second part of appendix D, the
 number of undetermined parameters among the flavon VEVs, see
 Eq.~(\ref{scalefree}), can be further reduced if we allow for couplings with
 positive mass dimension in the flavon superpotential.


\section{A specific model at NLO}
\cleqn


Fixing the $U(1)$ charges to take particular numerical values
may allow certain operators that are forbidden for a general set $(x,y,z)$ of $U(1)$ charges
so it is mandatory to study each model case by case.
In this section we discuss the full
results at NLO
for the particularly promising model \#13 where
the $U(1)$ charges are specified by $(x,y,z)=(5,4,1)$.
 We note that we checked that the results of the study of the messenger
  sector, relevant in order to properly generate the two operators $(F
  \wt\Phi^d_3)_1 ( T \Phi^d_2 )_1 H_{\ol{45}}/M^2$  and $(F \Phi^d_2
  \Phi^d_2)_3 ( T \wt\Phi^d_3 )_3 H_{\ol{5}}/M^3$, are not altered by this
  choice of $U(1)$ charges, especially no extra terms, not present
  in appendix~B, arise (at the renormalisable level).


\subsection{Flavon superpotential}


We first summarise the operator structures arising at LO in the flavon superpotential
\bea
&&X^d_1 (\Phi^d_2)^2 + \frac{1}{M^2} Y^d_2 (\Phi^d_2)^2 (\Phi^d_3)^2
\\ \nonumber
&&+ Y^\nu_2 \Phi^\nu_1 \Phi^\nu_2 + Y^\nu_2 (\Phi^\nu_2)^2 + Y^\nu_2 (\Phi^\nu_{3'})^2
+ Z^\nu_{3'} \Phi^\nu_1 \Phi^\nu_{3'} + Z^\nu_{3'} \Phi^\nu_2 \Phi^\nu_{3'} + Z^\nu_{3'} (\Phi^\nu_{3'})^2
\\ \nonumber
&&+ \frac{1}{M}\ol{X}^d_1 \Phi^d_2\Phi^d_3\wt\Phi^d_3 + X^{\nu d}_{1'}\wt\Phi^d_3\Phi^\nu_{3'}
+ Y^{du}_2 \Phi^d_2 \Phi^u_2 + X^u_1 \Phi^u_2  \wt\Phi^u_2
\\ \nonumber
&&+\frac{1}{M} X^{\mathrm{new}}_1 \Phi^u_2 (\Phi^d_3)^2
+ \frac{1}{M^2} X^{\mathrm{new}}_1 \Phi^d_2 \wt\Phi^d_3 (\Phi^d_3)^2
+ \frac{1}{M} \wt X^{\mathrm{new}}_{1'} \wt\Phi^u_2 \Phi^d_3 \wt\Phi^d_3
+ \frac{1}{M^3} \wt X^{\mathrm{new}}_{1'} \Phi^d_2 (\Phi^d_3)^4.
\eea
The effect of these operators has been discussed in detail in the preceding section
 and in appendix D.

In addition, for case \#13, we find several operators
which are subleading in the expansion in $\la$ relative to these, when the orders of the flavon VEVs
are chosen as in section 2 in order to reproduce in a satisfying way all fermion masses and mixings.
These subleading operators in general perturb the result for the vacuum alignment at the LO
in a particular way and thus affect the results for fermion masses and mixings.

In the following we consider
all subleading operators which can contribute at a level of up to and
including  order $\la^{12}$, for scales of VEVs as shown in Eqs.~(\ref{orderphiu},\ref{orderphid},\ref{orderphinu}).
Due to the four undetermined and the four
fixed VEV scales, see Eqs.~(\ref{scalefree},\ref{scalefixed}), we
have to parametrise the perturbed vacua of the flavons in the following way
\bea\label{shiftpara}
&& \langle \Phi^u_2 \rangle ~ = ~  \begin{pmatrix} \Delta^u_{2,1} \\ \varphi^u_2 + \Delta^u_{2,2}\end{pmatrix}
\ , \qquad
\langle \wt\Phi^u_2 \rangle ~=~
 \begin{pmatrix}\wt\Delta^u_{2,1}  \\ \wt\varphi^u_2\end{pmatrix} \ ,
\\ \nonumber
&& \langle \Phi^d_3 \rangle ~ = ~  \begin{pmatrix} \Delta^d_{3,1} \\ \varphi^d_3\\ \Delta^d_{3,3}\end{pmatrix}
\ , \qquad
\langle \wt\Phi^d_3 \rangle ~ =~
 \begin{pmatrix} \wt\Delta^d_{3,1}\\- (\wt\varphi^d_3 + \wt\Delta^d_{3,2})\\ \wt\varphi^d_3 + \wt\Delta^d_{3,3}\end{pmatrix}
\ , \qquad
\langle \Phi^d_2 \rangle ~=~  \begin{pmatrix} \varphi^d_2\\ \Delta^d_{2,2} \end{pmatrix} \ ,
\\ \nonumber
&&
\langle\Phi^\nu_{3'} \rangle~=~  \begin{pmatrix} \varphi^\nu_{3'} +\Delta^\nu_{3',1} \\ \varphi^\nu_{3'}
+\Delta^\nu_{3',2} \\ \varphi^\nu_{3'} + \Delta^\nu_{3',3}\end{pmatrix}
\  , \qquad
\langle\Phi^\nu_2\rangle~=~ \begin{pmatrix} \varphi^\nu_2 +\Delta^\nu_{2,1}\\ \varphi^\nu_2 +\Delta^\nu_{2,2}\end{pmatrix}
\;\;\; \mbox{and} \;\;\;
\langle \Phi^\nu_1 \rangle~=~ \varphi^\nu_1 \ .
\eea

Including all the above leading and subleading operators,
we solve the equations originating from the $F$-terms of the driving fields order by order in $\la$,
up to and including $\lambda^{12}$ in order to determine the size of all the shifts $\Delta^f_{i,j}$, $f=u,d,\nu$. We find as result
that the shifts are of the order in $\la$
\bea\label{shiftsize}
&& \Delta^u_{2,1}/M = \delta^u_{2,1} \la^8 \; , \;\; \Delta^u_{2,2}/M = \delta^u_{2,2} \la^6 \; , \;\;
\wt \Delta^u_{2,1}/M = \wt\delta^u_{2,1} \la^6 \; , \;\;
\Delta^d_{3,1}/M = \delta^d_{3,1} \la^6 \; , \;\;
\\ \nonumber
&& \Delta^d_{3,3}/M = \delta^d_{3,3} \la^6 \; , \;\; \wt\Delta^d_{3,1}/M = \wt\delta^d_{3,1} \la^7 \; , \;\;
\wt\Delta^d_{3,2}/M = \wt\delta^d_{3,2} \la^5 \; , \;\;
\wt\Delta^d_{3,3}/M = \wt\delta^d_{3,3} \la^5 \; , \;\;
\\ \nonumber
&& \Delta^d_{2,2}/M = \delta^d_{2,2} \la^7 \; , \;\; \Delta^\nu_{3',1}/M = \delta^\nu_{3',1} \la^8 \; , \;\;
\Delta^\nu_{3',2}/M = \delta^\nu_{3',2} \la^8 \; , \;\;
\Delta^\nu_{3',3}/M = \delta^\nu_{3',3} \la^8 \; , \;\;
\\ \nonumber
&& \Delta^\nu_{2,1}/M = \delta^\nu_{2,1} \la^8 \;\;
\mbox{and} \;\; \Delta^\nu_{2,2}/M = \delta^\nu_{2,2} \la^8 \; ,
\eea
where $\delta^f_{i,j}$, $f=u,d,\nu$, are complex numbers with absolute value of order one, determined by the couplings of the
superpotential. Notice that the shifts associated with the components of the
flavon $\Phi^\nu_{3'}$ are equal at this level, i.e.
\be
\Delta^\nu_{3',1} = \Delta^\nu_{3',2} = \Delta^\nu_{3',3} \ ,
\ee
so that the alignment, achieved at LO, is not perturbed up to the level $\la^8$. Since we are not interested in
the actual relation between $\varphi^\nu_{3'}$ and $\varphi^\nu_1$, we can absorb the shifts of the VEVs of the
components of the field $\Phi^\nu_{3'}$ into the LO VEV $\varphi^\nu_{3'}$. As we will see in section~5.2, this leads to the fact that tri-maximal mixing remains still preserved in the neutrino sector.

As one can see all ratios $\Delta^f_{i,j}/\varphi^f_{i}$ are small, at most of order $\la^2$, so that the
shifts relative to the LO alignment are small. Nevertheless these might lead to relevant corrections to
LO results for fermion masses and mixings, which is discussed in
section~5.2. Using the parametrisation in Eq.~(\ref{shiftpara}) and the results of the shifts given in Eq.~(\ref{shiftsize}), the
$F$-terms of all 15 driving fields vanish up to order $\la^{12}$, apart from
the one associated with the field $Y^{du}_{2,2}$ which has a contribution at
order $\la^{11}$ which can only vanish if either one
coupling of the flavon superpotential is tuned to cancel the term or
one of the involved flavon VEVs vanishes. This
tuning can be understood because the $F$-term equations of the 15 driving fields have to be fulfilled by solving for the
14 shifts $\Delta^f_{i,j}$. However, since the required tuning arises only at
 order $\la^{11}$ it has to be considered only a minor drawback in the
construction of the flavon superpotential.

 Finally, we briefly comment on how to ensure the spontaneous breaking of
  the family symmetry (i.e. avoid the trivial solution with all flavon VEVs being zero)
  by introducing a coupling with mass dimension two in the
  superpotential. This can be achieved by adding a driving field $V_0 \sim
  ({\bf 1},0)$ which is a total singlet under $S_4 \times U(1)$ so that the
  term $M_{V_0}^2 V_0$ is allowed. At the same time, a combination of the
  undetermined  VEVs, $\wt\varphi^u_2$, $\varphi^d_3$, $\varphi^d_2$ and
  $\varphi^\nu_1$, see  Eq.~(\ref{scalefree}), becomes fixed through $M_{V_0}$.
Furthermore, we find that introducing another driving field $V_2 \sim ({\bf
  2},-8)$ gives rise to a term $M_{V_2} V_2 \Phi^\nu_2$.
Considering operators resulting in contributions of order $\la^8$ or larger, the field $V_2$
leads to two additional constraints on the undetermined VEVs so that only one
free parameter remains. See second part of appendix~D for details.

We remark that one could fix the remaining undetermined parameter among the
flavon VEVs  through a Fayet-Iliopoulos term of an appropriate size provided
the $U(1)$ symmetry is gauged.  However, we do not pursue this possibility
further.


\subsection{Fermion masses and mixings}


In the following we study the effects of the subleading operators. We include
corrections caused by the shifted vacua as given above and the allowed
multi-flavon insertions with up to eight flavons.


\subsubsection{Quark  sector}


Including terms up to order $\la^8$ we find that the up quark
mass matrix remains nearly diagonal apart from the off-diagonal elements (23)
and (32) which are of order $\la^7$. This correction originates from the
operator structure $T T_3 (\Phi^d_2)^3 (\Phi^d_3)^2 H_5/M^5$. The diagonal
elements get corrected compared to the LO result: we find that
the (11) element does not only arise from the LO operator $T T \Phi^u_2
\wt\Phi^u_2 H_5/M^2$ but also from the operator $T T \Phi^u_2 H_5/M$ if the
non-zero shift $\Delta^u_{2,1}$ is taken into account. However, we include the
latter contribution into the former one. The (22) element receives a
correction of order $\la^6$ stemming from the insertion of the shifted vacuum
of $\Phi^u_2$ into the LO operator $T T \Phi^u_2 H_5/M$. All corrections which
might arise to the (33) element can be absorbed into the coupling of the LO
tree-level operator $T_3 T_3 H_5$. After taking into account possible
re-phasing of the right-handed fermion fields we find that $M_u$ can be
parametrised as
\be
M_u = \left( \begin{array}{ccc}
                y_u \la^8 & 0 & 0\\
		0 & y_c \la^4 & 0\\
		0 & 0 & y_t
\end{array}
\right) \; v_u  + \left( \begin{array}{ccc}
                0 & 0 & 0\\
		0 & z^u_1  e^{-i \alpha_{u,1}} \la^6 & z^u_2 e^{-i \alpha_{u,2}} \la^7\\
		0 & z^u_2 e^{-i \alpha_{u,3}} \la^7 & 0
\end{array}
\right) \; v_u \; .
\ee
Note that the parameters $y_{u,c,t}$ and $z^u_{1,2}$ are real and positive and the phases $\alpha_{u,i}$ are between $0$ and $2\pi$.
Here and in the following we display each mass matrix as the sum of the LO and the NLO result.
For the up quark masses we find
\be
\label{mups}
m_u = y_u \la^8 v_u \; , \;\;
m_c = \left( y_c \la^4 + \mathcal{O}(\la^6) \right) v_u \; , \;\;
m_t = \left( y_t + \mathcal{O}(\la^{14}) \right) v_u \; .
\ee
Thus, all corrections coming from NLO terms are small. In particular, all mixing angles in the up quark sector are negligible.

Similarly, we find the following parametrisation for the down quark mass matrix
\bea
\label{MdNLO}
M_d &=& \left( \begin{array}{ccc}
                0 & \wt x_2 \la^5 & - \wt x_2 e^{i \alpha_{d,2}} \la^5\\
		- \wt x_2 \la^5 & y_s e^{-i \alpha_{d,1}} \la^4 & (- y_s  e^{-i(\alpha_{d,1}-\alpha_{d,2})}
		                  \la^4 +  \wt x_2 e^{i \alpha_{d,2}} \la^5 )\\
		0 & 0 & y_b \la^2
\end{array}
\right) \; v_d
\\ \nonumber
&+&   \left( \begin{array}{ccc}
                z^d_1 e^{i \psi_{d,1}} \la^8 & - z^d_6  e^{i \psi_{d,6}} \la^7 & 0\\
		0 & z^d_4 e^{i \psi_{d,4}} \la^6   & - z^d_5  e^{i \psi_{d,5}} \la^6\\
		z^d_3 e^{i \psi_{d,3}} \la^6 & z^d_2 e^{i \psi_{d,2}} \la^6 & 0
\end{array}
\right) \; v_d \ ,
\eea
where we have only displayed the first subleading contribution to each of the
different matrix elements up to order $\la^8$. The parameters $y_s$ and $y_b$
are associated with the LO operators $(F \wt\Phi^d_3)_1 ( T \Phi^d_2 )_1
H_{\ol{45}}/M^2$  and $F T_3 \Phi^d_3 H_{\ol{5}}/M$, respectively. $\wt x_2$
coincides with the LO parameter $x_2$ as given in Eq.~(\ref{GSTrel}), up to
corrections of order $\la^2$ which are due to the shift
$\wt\Delta^d_{3,2}$. The $(11)$ element is of order $\la^8$ and originates
from several possible contractions of the operator structures $F T \wt\Phi^u_2 \Phi^d_2 \wt\Phi^d_3 H_{\ol{45}}/M^3$ and
 $F T (\Phi^d_2)^2 (\Phi^d_3)^3 H_{\ol{45}}/M^5$. The (32) element arises at order $\la^6$ from the following two
sources: through plugging the shifted vacuum of $\Phi^d_3$ in the LO operator $F T_3 \Phi^d_3 H_{\ol{5}}/M$
and through the subleading operator $F T_3 \wt\Phi^u_2 \Phi^d_3 H_{\ol{5}}/M^2$. Similarly, the (31) element of order $\la^6$ arises
from the LO term $F T_3 \Phi^d_3 H_{\ol{5}}/M$ and is proportional to the shift $\Delta^d_{3,1}$. The corrections of order
$\la^6$ in the (22) and the (23) elements, encoded in the parameters $z^d_4$ and $z^d_5$, originate from the LO operator
$(F \wt\Phi^d_3)_1 ( T \Phi^d_2 )_1 H_{\ol{45}}/M^2$ if the shifts of the vacuum alignment are included, and are proportional
to $\wt\Delta^d_{3,3}$ and to $\wt\Delta^d_{3,2}$, respectively. Finally, the correction to the (12) element is again the result of
the shifted vacuum of $\wt\Phi^d_3$, this time plugged into $(F \Phi^d_2 \Phi^d_2)_3 ( T \wt\Phi^d_3 )_3 H_{\ol{5}}/M^3$ and is generically
of order $\la^7$. We note that also in case of $M_d$ (and $M_e$, see below) all parameters,
$y_{b,s}$, $\wt x_2$ and $z^d_i$, are real and positive and that all appearing
phases, $\alpha_{d,i}$ and $\psi_{d,i}$, are within the interval $[0,2\pi)$.
The mass matrix in Eq.~(\ref{MdNLO}) leads to down quark masses of the form
\be
\label{mdowns}
m_d = \left( \frac{\wt x_2^2}{y_s} \la^6 + \mathcal{O}(\la^8) \right) v_d \; , \;\;
m_s = \left( y_s \la^4 + \mathcal{O}(\la^6) \right) v_d \; , \;\;
m_b = \left( y_b \la^2 + \mathcal{O}(\la^6) \right) v_d \; .
\ee

For the quark mixing angles we find
\be
\label{thqs}
\sin \theta_{13}^q = \frac{\wt x_2}{y_b} \la^3  \; , \;\;
\tan \theta_{12}^q = \frac{\wt x_2}{y_s} \la + \m{O}(\la^3) \;\;\; \mbox{and} \;\;\;
\tan \theta_{23}^q = \frac{y_s}{y_b} \la^2 + \m{O}(\la^3) \; ,
\ee
showing that the angles $\theta_{ij}^q$ are only determined by the LO results and all subleading corrections are very small.
The calculation of the Jarlskog invariant $J_{CP}$ yields
\be
\label{jcp}
J_{CP} = \frac{\wt x_2^3}{y_s y_b^2} \la^7 \sin \alpha_{d,1} + \m{O} (\la^8) \; ,
\ee
which turns out to be slightly below its expected size of
$\la^6$~\cite{PDG}. Eqs.~(\ref{mdowns},\ref{thqs}) confirm the achievement of the GST
relation \cite{gst} in our model, even after including corrections,
\be
\tan \theta_{12}^q \approx \sqrt{\frac{m_d}{m_s}} \; .
\ee
Due to the fact that only the parameters associated with the LO contributions
are relevant for the determination of masses, mixing angles and CP violation,
the model might turn out to be incapable of fitting the precise values for the
quantities determined from experiments~\cite{PDG}. However, we point out
that in our model these quantities are evaluated at a high energy scale and
any renormalisation group and threshold effects \cite{Ross:2007az}, which
among other things depend also on the actual value of $\tan\beta$, are not
taken into account in this analysis. 


\subsubsection{Lepton sector}


Coming to the lepton sector, we first note that
the structure of the charged lepton mass matrix is analogous to the one of $M_d$, apart from the GJ factor and
the slightly different positions of the phases, after re-phasing of all right-handed fields,
\bea
\label{MeNLO}
M_e &=& \left( \begin{array}{ccc}
                0 & - \wt x_2 \la^5 & 0\\
		\wt x_2 \la^5 & -3 y_s e^{-i \alpha_{d,1}} \la^4 & 0\\
		- \wt x_2 \la^5 & (3 y_s e^{-i \alpha_{d,1}} \la^4 + \wt x_2 \la^5) & y_b \la^2
\end{array}
\right) \; v_d
\\ \nonumber
&+& \left( \begin{array}{ccc}
                -3 z^d_1 e^{i \psi_{d,1}} \la^8 & 0 & z^d_3  e^{i (\alpha_{d,2}+ \psi_{d,3})} \la^6\\
		- z^d_6 e^{i \psi_{d,6}} \la^7 & -3 z^d_4 e^{i \psi_{d,4}} \la^6 & z^d_2  e^{i (\alpha_{d,2}+ \psi_{d,2})} \la^6\\
		0 & 3 z^d_5 e^{-i (\alpha_{d,2}-\psi_{d,5})} \la^6  & 0
\end{array}
\right) \; v_d \; .
\eea
The mass matrix given in Eq.~(\ref{MeNLO}) leads to charged lepton masses
\be
\label{mes}
m_e = \left( \frac{\wt x_2^2}{3 y_s} \la^6 + \mathcal{O}(\la^8) \right) v_d \; , \;\;
m_\mu = \left( 3 y_s \la^4 + \mathcal{O}(\la^6) \right) v_d \; , \;\;
m_\tau = \left( y_b \la^2 + \mathcal{O}(\la^6) \right) v_d \; , \;\;
\ee
which, similar to the quark masses, up to small corrections, are only
determined by the LO terms. The GJ relations \cite{GJ} are  confirmed by 
 Eqs.~(\ref{mdowns}) and (\ref{mes}). The charged lepton mixing
angles are of the form
\bea
&&\sin \theta_{13}^e = \frac{z^d_3}{y_b} \la^4 + \m{O}(\la^5) \; , \;\;
\tan \theta_{12}^e = \frac{\wt x_2}{3 y_s} \la + \m{O}(\la^3)  \; , \;\;
\\ \nonumber
&&\tan \theta_{23}^e = \left| 9 \left(\frac{y_s}{y_b}\right)^2 - e^{i (\alpha_{d,2}+\psi_{d,2})}
\left( \frac{z^d_2}{y_b} \right) \right| \la^4 + \m{O}(\la^5) \; , \;\;
\eea
coinciding with the estimate found in section 2.

The Dirac neutrino mass matrix elements also receive small corrections of
order $\la^4$ and $\la^6$, respectively. They read
\be
M_D = \left( \begin{array}{ccc}
                y_D & 0 & 0\\
                0 & 0  & y_D\\
		0 & y_D & 0
\end{array}
\right) \; v_u
+ \left( \begin{array}{ccc}
                2 z^D_3 \la^6 & z^D_2 \la^6 & z^D_1 \la^4\\
                z^D_2 \la^6  & z^D_1 \la^4  & - (z^D_3-z^D_4) \la^6\\
		z^D_1 \la^4 &  - (z^D_3+z^D_4) \la^6 & z^D_2 \la^6
\end{array}
\right) \; v_u \ ,
\ee
with $y_D$ being the coupling accompanying the LO tree-level term $F N H_5$. The corrections associated with the two parameters
$z^D_1$ and $z^D_2$ originate from the operator structure $F N \wt\Phi^u_2
H_5/M$. The one associated with $z^D_2$ is additionally suppressed
 by a factor of $\la^2$ because it is not proportional to the VEV
 $\wt\varphi^u_2$ but rather to the shift $\wt\Delta^u_{2,1}$.
The source of the two corrections to the (11), (23) and (32) elements of order $\la^6$, encoded in $z^D_3$ and $z^D_4$, is the operator
structure $F N (\Phi^d_2)^4 \Phi^d_3 H_5/M^5$. The two different parameters refer to two different possible contractions of the operator.

In the right-handed neutrino mass matrix corrections are of a relative order $\la^4$. They are encoded in two parameters, denoted by
$Z_1$ and $Z_2$ in the following. The general form of $M_R$ can be written as
\be
M_R = \left( \begin{array}{ccc}
                A+2 C &  B-C & B-C\\
                B-C & B+2 C & A-C\\
		B-C & A-C & B+2 C
\end{array}
\right) \la^4 \, M
+  \left( \begin{array}{ccc}
                0 &  Z_1 & Z_2\\
                Z_1 & Z_2 &  0\\
		Z_2 & 0 & Z_1
\end{array}
\right) \la^8 \, M
\; .
\ee
In the above equation we have also introduced the parameters $A$, $B$ and $C$, which (dominantly) originate from the LO terms
$\alpha N N \Phi^\nu_1$, $\beta N N \Phi^\nu_2$, $\gamma N N \Phi^\nu_{3'}$, respectively. Note that at the same time some of the
 subleading contributions are absorbed by re-defining $A$ as well as $C$. The latter incorporates then also the contribution
coming from the operator $N N \wt\Phi^u_2 \Phi^\nu_{3'}/M$.
One source of the contributions,
parametrised by $Z_1$ and $Z_2$, are the shifts $\Delta^\nu_{2,1}$ and
$\Delta^\nu_{2,2}$  if the LO term $N N \Phi^\nu_2$ is evaluated with the
shifted vacuum of $\Phi^\nu_2$.
Apart from that, the subleading
term $N N \Phi^\nu_2 \wt\Phi^u_2/M$ contributes to the correction associated with $Z_1$, while the two operator structures
$N N \Phi^\nu_1 \wt\Phi^u_2/M$ and $N N (\Phi^d_2)^8/M^7$ give a contribution to the (13) and (22) elements of $M_R$.
The effective light neutrino mass matrix which arises from the type I see-saw
mechanism can be arranged as
\bea
m_\nu^{eff} &=& \left( \begin{array}{ccc}
                B_\nu +C_\nu-A_\nu & A_\nu & A_\nu \\
                A_\nu & B_\nu & C_\nu \\
		A_\nu & C_\nu & B_\nu
\end{array}
\right) \left(\frac{v_{u}^{2}}{\la^4 M}\right)
\\ \nonumber
&+& \left( \begin{array}{ccc}
                z^\nu_1 & z^\nu_2 & z^\nu_3 \\
                z^\nu_2 & z^\nu_1 + z^\nu_3 - z^\nu_4 & z^\nu_4\\
		z^\nu_3 & z^\nu_4 & z^\nu_1 + z^\nu_2 - z^\nu_4
\end{array}
\right) \left(\frac{v_{u}^{2}}{M}\right) \; .
\eea
$A_\nu$, $B_\nu$ and $C_\nu$ parametrise the LO contributions and the four independent parameters $z^\nu_i$ the corrections to the
light neutrino mass matrix. In the following we will assume all these
parameters to be real since there is no experimental
evidence for CP violating phases in the lepton sector yet. As expected all
corrections to the light neutrino masses arise at a relative level
of $\la^4$.  Due to the fact that, up to the order $\la^8$, the shifted vacuum
of the flavon $\Phi^\nu_{3'}$ reveals the same alignment as the LO one, the tri-maximally mixed state remains an eigenstate of the light neutrino mass matrix
$m_{\nu}^{eff}$ (as well as of the right-handed neutrino mass matrix $M_R$).
 The neutrino mixing angles are thus still given by the TB mixing values up to corrections of $\mathcal{O}(\la^4)$.
Eventually, we find for the lepton mixing angles
\bea
\label{thls}
&&\sin \theta_{13}^l = \frac{\wt x_2}{3 \sqrt{2} y_s} \la +\m{O}(\la^3) \; , \;\;
\sin^2 \theta_{12}^l = \frac{1}{3} - \frac{2 \wt x_2}{9 y_s} \la \cos \alpha_{d,1} +\m{O}(\la^2)  \; , \;\; \\
\nonumber
&& \sin^2 \theta_{23}^l = \frac{1}{2} - \frac{\wt x_2^2}{36 y_s^2} \la^2  +\m{O}(\la^4)  \; , \;\;
\eea
coinciding with the estimates given in Eq.~(\ref{thl_approx}). Comparing the results for quark and lepton mixing angles, Eqs.~(\ref{thqs})
and (\ref{thls}), we see that our model incorporates the correlations \cite{King:2005bj}
\be
\label{rela1}
\sin^2 \theta_{12}^l \approx \frac{1}{3} - \frac{2}{9} \tan \theta_{12}^q \cos
\alpha_{d,1} \ ,
\ee
and
\be
\sin \theta_{13}^l \approx \tan \theta_{12}^q/(3 \sqrt{2}) \; ,
\ee
with $\alpha_{d,1}$ playing the role of the Dirac CP phase $\delta^l$ in the lepton sector, up to $\pi$
\be
\delta^l = \alpha_{d,1} + \pi \; .
\ee
This relation holds up to corrections of order $\la$. Thus, we can write Eq.~(\ref{rela1}) also as
\be
\label{rel_th12l_dl_th13l}
\sin^2 \theta_{12}^l \approx \frac{1}{3} \left(1 + 2 \sqrt{2} \sin \theta_{13}^l \cos \delta^l \right) \; .
\ee
Note that Eq.~(\ref{rel_th12l_dl_th13l}) holds without loss of generality,
although we have assumed
all parameters in the neutrino sector to be real since, as shown in \cite{King:2005bj},
the validity of this relation only depends on the fact that $\theta_{13}^e$, $\theta_{23}^e$ and $\theta_{13}^\nu$ are
(much) smaller than $\theta_{12}^e$.

It is convenient to define \cite{King:2007pr},
\be
\sin \theta^l_{13} = \frac{r}{\sqrt{2}}, \ \ \sin \theta^l_{12} = \frac{1}{\sqrt{3}}(1+s),
\ \ \sin \theta^l_{23} = \frac{1}{\sqrt{2}}(1+a),
\label{rsa}
\ee
where we have introduced the three real parameters
$r,s,a$ to describe the deviations of the reactor, solar and
atmospheric mixing angles from their TB values.
The present model predicts these deviation parameters to be,
\be
s\approx r\cos \delta^l, \ \  r\approx \lambda /3, \ \ a\approx -\lambda^2/36 \, ,
\ee
 up to renormalisation group effects and corrections associated with non-canonically normalised kinetic terms.
While the first equation above is the usual sum rule in terms of deviation parameters
\cite{King:2007pr}, we emphasise that the model predicts another new relation
\be
a\approx -r^2/4,
\ee
with $r\approx \lambda /3$, valid at the GUT scale.

\vspace{0.05in}
 In summary, all NLO corrections turn out to have a negligible effect on the results for fermion masses and mixings
achieved at LO and presented in section 2.


\section{\label{conclusion}Conclusions}
\cleqn


In this article we have constructed a model of fermion masses and mixings based on the combination of the minimal GUT $SU(5)$
and the family symmetry $S_4$. The latter is also minimal in the sense that it is the smallest non-Abelian finite group which contains all the symmetries necessary
to enforce TB neutrino mixing. At LO, the effective light neutrino mass matrix
arises from the type I see-saw mechanism where the TB mixing structure is imprinted
in the form of the Majorana mass matrix of the right-handed neutrinos. The
latter in turn originates from the vacuum alignment of three different flavon
fields $\Phi^\nu_{3'}$, $\Phi^\nu_{2}$, $\Phi^\nu_{1}$. As the right-handed neutrino mass matrix contains three independent parameters, our model can accommodate
all patterns for the neutrino masses; in particular we do not encounter the constraint of a neutrino mass sum rule as in the corresponding $A_4$ models.
At the same time the TB neutrino mixing is independent of the particular values of the neutrino masses and also stable under inclusion of NLO corrections.
Taking into account the corrections to the flavon alignments as well as higher-dimensional operators, we find that TB neutrino mixing remains exact up
to $\mathcal O(\lambda^4) \sim $  0.1\% at the GUT scale.

Regarding the masses of the charged fermions, we invoke additional $SU(5)$ singlet flavon fields. Their LO alignments give rise to acceptable quark
and charged lepton mass matrices, including the phenomenologically successful GJ and GST relations. The latter cannot be achieved in a generic effective
 theory, but require some specific set of messenger fields. Such a set has been explicitly constructed.
Having introduced eight flavon fields, it is necessary to study all allowed superpotential operators with two matter fields and an arbitrary number of flavons.
In order to forbid those terms which would spoil the LO results for the mass matrices, we introduce a new $U(1)$ symmetry, parametrised by three integers $(x,y,z)$.
Their specific values are determined when discussing how to obtain the
required vacuum alignment. In our model this originates  from the $F$-terms
of an additional set of fields, the driving fields, which cannot couple directly to the matter superfields. Solving the $F$-term equations of the driving fields
in the SUSY limit, we can obtain the desired flavon alignments at LO. Additional driving fields are then added to obtain further correlations between the
scales of the flavon VEVs. This study fixes our preferred choice of $U(1)$ charges, given by $(x,y,z)=(5,4,1)$. The number of undetermined parameters among
the flavon VEVs can be minimised by considering driving fields allowing for couplings with positive mass dimension in the flavon superpotential.
 In this way we can obtain by the (ad hoc) choice of the magnitude of two mass parameters and one flavon VEV, which remains undetermined, the 
correct size of the VEVs of all flavons coupling to the superfields $T_3$, $T$, $F$ and $N$, as required to achieve the observed fermion mass and mixing
patterns in the quark and lepton sector.
In the final part of this work we have scrutinised the NLO effects on the flavon alignments as well as the fermion mass matrices.
Our results reveal that the NLO corrections have a negligible effect on quark and lepton masses and mixings, thus confirming the stability of the original
LO structure of the model.
 Since the main purpose of this work is the study of fermion masses and mixings, we have not discussed the GUT Higgs sector and the corresponding
(super-)potential necessary in order to correctly break $SU(5)$ to the SM gauge group.

In conclusion we have constructed a SUSY GUT of Flavour
based on $S_4\times SU(5)$, together with an additional (global or local) Abelian symmetry,
and studied it to NLO accuracy.
We have specified the complete effective theory for general $U(1)$ charges, valid just
below the GUT scale, relevant for fermion masses and mixings, 
and performed a full operator analysis
taking into account all relevant higher order terms with several
insertions of flavons.
The model includes a successful description of quark masses and
mixing angles at LO incorporating the GST relation. In addition,
at LO, charged lepton and down quark masses fulfil GJ relations.
Our predictions apply just below the GUT scale, and the determination of
the fermion masses and mixings at the electroweak scale would require a
detailed investigation of renormalisation group and threshold effects which is
beyond the scope of this paper.
We have studied the vacuum alignment arising from $F$-terms to NLO and
the resulting corrections have been shown to not affect the LO predictions significantly
for specific choices of $U(1)$ charges. A specific model evaluated to NLO predicts
TB mixing in the neutrino sector very accurately up to corrections of order 0.1\%.
Including charged lepton mixing corrections leads to small deviations from TB
lepton mixing described by a precise sum rule, with
 accurately maximal atmospheric mixing and a reactor mixing angle close to three degrees.


\section*{Acknowledgments}

We thank Ferruccio Feruglio, Marco Serone and Robert Ziegler for discussions.
CH thanks the Galileo Galilei Institute for Theoretical Physics for
hospitality and the INFN for partial support during the completion of this
work. SFK and CL acknowledge support from the STFC Rolling Grant ST/G000557/1.
SFK is grateful to the Royal Society for a Leverhulme Trust Senior Research
Fellowship.


\appendix

\section*{Appendix}


\section{Group theory of $\bs{S_4}$}
\cleqn


The group $S_4$ is the permutation group of four distinct objects and is
isomorphic to the symmetry group $O$ of a regular octahedron. Its order is 24
and it contains five real irreducible representations: ${\bf 1}$, ${\bf
  1^\prime}$, ${\bf 2}$, ${\bf 3}$ and ${\bf 3^\prime}$. Only the two triplet
representations are faithful.
A decisive feature among the two triplets ${\bf 3}$ and ${\bf 3^\prime}$ is
that only ${\bf 3}$ can be identified with the fundamental representation of
the continuous groups $SO(3)$ and $SU(3)$. The three generators $S$, $T$ and
$U$ are of the following form for the five different representations
\begin{center}
\begin{math}
\begin{array}{llll}
{\bf 1}:       & S=1 \; , & T=1 \; ,  & U=1 \ ,\\[2mm]
{\bf 1^\prime}:    & S=1  \; , & T=1 \; ,  & U=-1\ ,\\[2mm]
{\bf 2}: & S= \left( \begin{array}{cc}
    1&0 \\
    0&1
    \end{array} \right) \; ,
    & T= \left( \begin{array}{cc}
    \omega&0 \\
    0&\omega^2
    \end{array} \right) \; ,
    & U=  \left( \begin{array}{cc}
    0&1 \\
    1&0
    \end{array} \right)\ ,\\[2mm]
{\bf 3}: & S= \frac{1}{3} \left(\begin{array}{ccc}
    -1& 2  & 2  \\
    2  & -1  & 2 \\
    2 & 2 & -1
    \end{array}\right) \; ,
    & T= \left( \begin{array}{ccc}
    1 & 0 & 0 \\
    0 & \omega^{2} & 0 \\
    0 & 0 & \omega
    \end{array}\right) \; ,
    & U= - \left( \begin{array}{ccc}
    1 & 0 & 0 \\
    0 & 0 & 1 \\
    0 & 1 & 0
    \end{array}\right)\ ,\\[2mm]
{\bf 3^\prime}: & S= \frac{1}{3} \left(\begin{array}{ccc}
    -1& 2  & 2  \\
    2  & -1  & 2 \\
    2 & 2 & -1
    \end{array}\right) \; ,
    & T= \left( \begin{array}{ccc}
    1 & 0 & 0 \\
    0 & \omega^{2} & 0 \\
    0 & 0 & \omega
    \end{array}\right) \; ,
    & U= \left( \begin{array}{ccc}
    1 & 0 & 0 \\
    0 & 0 & 1 \\
    0 & 1 & 0
    \end{array}\right)\ ,
\end{array}
\end{math}
\end{center}
with $\omega = e^{2\pi i/3}$.

The generators fulfil the relations
\begin{eqnarray}\nonumber
&& S^2 = \mathds{1} \; , \;\; T^3 = \mathds{1} \; , \;\; U^2= \mathds{1} \; ,\\[0mm]
\nonumber
&&  (S T)^3 = \mathds{1} \; , \;\; (S U)^2 = \mathds{1} \; , \;\; (T U)^2 = \mathds{1} \; , \;\;
  (S T U)^4 = \mathds{1} \; .
\end{eqnarray}
Note that the minimal number of generators necessary to define $S_4$ is
actually only two, compare e.g.
\cite{S4-group}. However, in order to emphasise the correlation between the
groups $A_4$ and $S_4$ it is advantageous to choose the set $S$, $T$ and $U$, since then one easily sees that $S$ and $T$ alone generate
the group $A_4$, see fifth reference in \cite{A4-L}. Notice that similarly, the two generators $T$ and $U$
alone generate the group $S_3$ \cite{Lomont}.
The character table is given in table \ref{characters}.
\begin{table}
\begin{center}
\begin{tabular}{|c|ccccc|}
\hline
&\multicolumn{5}{|c|}{Classes}                                                 \\ \cline{2-6}
&$\mathcal{C}_{1}$&$\mathcal{C}_{2}$&$\mathcal{C}_{3}$&$\mathcal{C}_{4}$&$\mathcal{C}_{5}$\\
\cline{1-6}
\rule[0.15in]{0cm}{0cm} $G$         &$\mathds{1}$&$S$&$U$ &$T$ &$S T U$\\
\cline{1-6}
$n_i$                  &1      &3      &6      &8      &6\\
\cline{1-6}
$h_i$                     &1      &2      &2      &3      &4\\
\hline
${\bf 1}$                                &1      &1      &1      &1      &1              \\[0.1cm]
${\bf 1^\prime}$                                &1      &1      &-1     &1      &-1             \\[0.1cm]
${\bf 2}$                                       &2      &2      &0      &-1     &0              \\[0.1cm]
${\bf 3}$                                &3      &-1     &-1     &0      &1              \\[0.1cm]
${\bf 3^\prime}$                                &3      &-1     &1      &0      &-1             \\[0.1cm]
\hline
\end{tabular}
\end{center}
\begin{center}
\caption[]{Character table of the group
  $S_4$. $\mathcal{C}_i$ denote the five classes of $S_4$, $n_i$ the number of distinct elements in the classes $\mathcal{C}_i$ and $h_i$
the order of the elements contained in class $\mathcal{C}_i$. For each of the
classes we give a representative $G$ in terms of the generators $S$, $T$ and~$U$. \label{characters}}
\end{center}
\end{table}
The Kronecker products are of the form
\begin{eqnarray}\nonumber
&&\bf 1 \times {\bs \mu} = {\bs \mu} \;\; \forall \;\; {\bs \mu} \; , \;\; 1^\prime \times 1^\prime =1 \; , \;\; 1^\prime \times 2 = 2 \; ,\\[0mm]
\nonumber
&&\bf 1^\prime \times 3 = 3^\prime \; , \;\; 1^\prime \times 3^\prime = 3 \; ,\\[0mm]
\nonumber
&&\bf 2 \times 2 = 1 + 1^\prime + 2 \; , \;\; 2 \times 3 = 2 \times 3^\prime = 3 + 3^\prime\; ,\\[0mm]
\nonumber
&&\bf 3 \times 3 = 3^\prime \times 3^\prime = 1 + 2 + 3 + 3^\prime \; , \;\;
3 \times 3^\prime = 1^\prime + 2 + 3 + 3^\prime \; .
\end{eqnarray}
In the following we list the Clebsch-Gordan coefficients using the notation $a
\sim {\bf   1}$, $a^\prime \sim {\bf 1^\prime}$,
$( b_1 , b_2 )^t, ( \tilde b_1, \tilde b_2 )^t \sim~{\bf 2}$,
$( c_1 ,c_2 , c_3 )^t, ( \tilde c_1 , \tilde c_2 , \tilde c_3 )^t \sim {\bf 3}$,
$( c'_1 , c'_2 , c'_3 )^t, ( \tilde c'_1 , \tilde c'_2 , \tilde c'_3 )^t \sim
{\bf 3'}$.

\noindent For a singlet multiplied with a doublet or a triplet
\begin{eqnarray}\nonumber
{\bf 1^{(\prime)}\times 2}: && (a b_1 , a b_2 )^t \sim {\bf 2}  \; , \;\; (a' b_1 , -a' b_2 )^t \sim {\bf 2} \; ,\\[2mm]
\nonumber
{\bf 1^{(\prime)} \times 3}: && (a c_1 , a c_2 , a c_3 )^t \sim {\bf 3} \; , \;\; (a' c_1 , a' c_2 , a' c_3 )^t \sim {\bf 3'} \; ,\\[2mm]
\nonumber
{\bf 1^{(\prime)} \times 3'}: \hspace{-1mm} && (a c'_1 , a c'_2 , a c'_3 )^t \sim {\bf 3'} \; , \;\; (a' c'_1 , a' c'_2 , a' c'_3 )^t \sim {\bf 3} \; .
\end{eqnarray}
For a doublet coupled to a doublet
\begin{equation}\nonumber
{\bf 2 \times 2}:\;\;\; b_1 \tilde b_2 + b_2 \tilde b_1 \sim {\bf 1} \; , \;\;
b_1 \tilde b_2 - b_2 \tilde b_1 \sim {\bf 1'} \; , \;\;
(b_2 \tilde b_2 , b_1 \tilde b_1)^t \sim {\bf 2} \; .
\end{equation}
For a doublet multiplied with a triplet
\begin{equation}\nonumber
{\bf 2 \times 3}: \;\;\,
(b_1 c_2 + b_2 c_3 , b_1 c_3 + b_2 c_1 , b_1 c_1 + b_2 c_2 )^t \sim {\bf 3} \; , \;\;
(b_1 c_2 - b_2 c_3 , b_1 c_3 - b_2 c_1 , b_1 c_1 - b_2 c_2 )^t \sim {\bf 3'} \; ,
\end{equation}
and
\begin{equation}\nonumber
{\bf 2 \times 3'}: \;\;\,
(b_1 c'_2 - b_2 c'_3 , b_1 c'_3 - b_2 c'_1 , b_1 c'_1 - b_2 c'_2 )^t \sim {\bf 3} \; , \;\;
(b_1 c'_2 + b_2 c'_3 , b_1 c'_3 + b_2 c'_1 , b_1 c'_1 + b_2 c'_2 )^t \sim {\bf
  3'}  \, .
\end{equation}
For the product ${\bf 3 \times 3}$
\begin{eqnarray}\nonumber
&& c_1 \tilde c_1 + c_2 \tilde c_3 +  c_3 \tilde c_2 \sim {\bf 1} \; , \;\;
(c_1 \tilde c_3 + c_2 \tilde c_2 +  c_3 \tilde c_1 , c_1 \tilde c_2 + c_2 \tilde c_1 +  c_3 \tilde c_3)^t \sim {\bf 2} \; ,\\[2mm]
\nonumber
&& (c_2 \tilde c_3 - c_3 \tilde c_2 , c_1 \tilde c_2 - c_2 \tilde c_1, c_3 \tilde c_1 - c_1 \tilde c_3)^t \sim {\bf 3} \; ,\\[2mm]
\nonumber
&& (2 c_1 \tilde c_1 - c_2 \tilde c_3 - c_3 \tilde c_2 , 2 c_3 \tilde c_3 - c_1 \tilde c_2 - c_2 \tilde c_1,
2 c_2 \tilde c_2 - c_1 \tilde c_3 - c_3 \tilde c_1)^t \sim {\bf 3'} \; ,
\end{eqnarray}
as well as for the product ${\bf 3' \times 3'}$
\begin{eqnarray}\nonumber
&& c'_1 \tilde c'_1 + c'_2 \tilde c'_3 +  c'_3 \tilde c'_2 \sim {\bf 1} \; , \;\;
(c'_1 \tilde c'_3 + c'_2 \tilde c'_2 +  c'_3 \tilde c'_1 , c'_1 \tilde c'_2 + c'_2 \tilde c'_1 +  c'_3 \tilde c'_3)^t \sim {\bf 2} \; ,\\[2mm]
\nonumber
&& (c'_2 \tilde c'_3 - c'_3 \tilde c'_2 , c'_1 \tilde c'_2 - c'_2 \tilde c'_1, c'_3 \tilde c'_1 - c'_1 \tilde c'_3)^t \sim {\bf 3} \; ,\\[2mm]
\nonumber
&& (2 c'_1 \tilde c'_1 - c'_2 \tilde c'_3 - c'_3 \tilde c'_2 , 2 c'_3 \tilde c'_3 - c'_1 \tilde c'_2 - c'_2 \tilde c'_1,
2 c'_2 \tilde c'_2 - c'_1 \tilde c'_3 - c'_3 \tilde c'_1)^t \sim {\bf 3'} \; ,
\end{eqnarray}
and finally for the product ${\bf 3 \times 3'}$
\begin{eqnarray}\nonumber
&& c_1 c'_1 + c_2 c'_3 +  c_3 c'_2 \sim {\bf 1'} \; , \;\;
(c_1 c'_3 + c_2 c'_2 +  c_3 c'_1 ,-( c_1 c'_2 + c_2 c'_1 +  c_3 c'_3))^t \sim {\bf 2} \; ,\\[2mm]
\nonumber
&& (2 c_1 c'_1 - c_2 c'_3 - c_3 c'_2 , 2 c_3 c'_3 - c_1 c'_2 - c_2 c'_1,
2 c_2 c'_2 - c_1 c'_3 - c_3 c'_1)^t \sim {\bf 3} \; ,\\[2mm]
\nonumber
&& (c_2 c'_3 - c_3 c'_2 , c_1 c'_2 - c_2 c'_1, c_3 c'_1 - c_1 c'_3)^t \sim {\bf 3'} \; .
\end{eqnarray}

\noindent These results are in accordance with \cite{S4-group}.

\noindent Note that due to the choice of $T$ being complex for the
real representations ${\bf 2}$, ${\bf 3}$ and ${\bf 3'}$ for fields which
transform
as $(\phi_1,\phi_2)^t\sim {\bf 2}$, $(\psi_1,\psi_2,\psi_3)^t\sim {\bf 3}$ and
$(\psi'_1,\psi'_2,\psi'_3)^t\sim {\bf 3'}$ their conjugates,
$(\phi_1^\star,\phi_2^\star)^t$,
$(\psi_1^\star,\psi_2^\star,\psi_3^\star)^t$ and
$((\psi'_1)^\star,(\psi'_2)^\star, (\psi'_3)^\star)^t$ are in
${\bf 2^\star}$, ${\bf 3^\star}$ and ${\bf ( 3')^\star}$, respectively, and
only $
(\phi_2^\star,\phi_1^\star)^t\sim {\bf 2}$,
$(\psi_1^\star,\psi_3^\star,\psi_2^\star)^t\sim {\bf 3}$ and
$((\psi'_1)^\star,(\psi'_3)^\star,(\psi'_2)^\star)^t\sim {\bf 3'}$
holds.

\noindent Eventually, we display the embedding of $S_4$ into the continuous
groups $SO(3)$ and $SU(3)$ as well as its breaking to the discrete groups
$A_4$ and $S_3$ \cite{Luhn:2007yr}. The smallest representations of $SO(3)$
and $SU(3)$ are decomposed into $S_4$ representations, respectively
\begin{center}
\parbox{2.5in}{
\begin{eqnarray} \nonumber
&\underline{SO(3)}& \;\;\; \rightarrow \;\;\; \underline{S_4}\\[0.0mm] \nonumber
&\bf 1& \;\;\; \rightarrow \;\;\;\bf  1 \\[0.0mm] \nonumber
&\bf 3& \;\;\; \rightarrow \;\;\;\bf  3 \\[0.0mm] \nonumber
&\bf 5& \;\;\; \rightarrow \;\;\;\bf  2 + 3^\prime \\[0.0mm] \nonumber
&\bf 7& \;\;\; \rightarrow \;\;\;\bf  1^\prime + 3 + 3^\prime \\[0.0mm] \nonumber
&\bf 9& \;\;\; \rightarrow \;\;\;\bf  1 + 2 + 3 + 3^\prime
\end{eqnarray}}
\parbox{2.5in}{
\begin{eqnarray} \nonumber
&\underline{SU(3)}& \;\;\; \rightarrow \;\;\; \underline{S_4}\\[0.0mm] \nonumber
&\bf 1& \;\;\; \rightarrow \;\;\; \bf 1 \\[0.0mm] \nonumber
&\bf 3& \;\;\; \rightarrow \;\;\; \bf 3 \\[0.0mm] \nonumber
&\bf 6& \;\;\; \rightarrow \;\;\; \bf 1 + 2 + 3^\prime\\[0.0mm] \nonumber
&\bf 8& \;\;\; \rightarrow \;\;\; \bf 2 + 3 + 3^\prime \\[0.0mm] \nonumber
&\bf 10& \;\;\; \rightarrow \;\;\; \bf 1^\prime + 3 + 3 + 3^\prime
\end{eqnarray}}
\end{center}
The decomposition of the irreducible representations of $S_4$ into
those of the groups $A_4$ and $S_3$ leads to
\vspace{-6.8mm}
\begin{center}
\parbox{2.5in}{
\begin{eqnarray} \nonumber
&\underline{S_4}& \;\;\; \rightarrow \;\;\; \underline{A_4}\\ \nonumber
&\bf 1& \;\;\; \rightarrow \;\;\; \bf 1 \\ \nonumber
&\bf 1^\prime& \;\;\; \rightarrow \;\;\; \bf 1 \\ \nonumber
&\bf 2& \;\;\; \rightarrow \;\;\; \bf 1^\prime + 1^{\prime\prime} \\ \nonumber
&\bf 3& \;\;\; \rightarrow \;\;\; \bf 3 \\ \nonumber
&\bf 3^\prime& \;\;\; \rightarrow \;\;\; \bf 3
\end{eqnarray}}
\parbox{2.5in}{
\begin{eqnarray} \nonumber
&\underline{S_4}& \;\;\; \rightarrow \;\;\; \underline{S_3}\\ \nonumber
&\bf 1& \;\;\; \rightarrow \;\;\; \bf 1 \\ \nonumber
&\bf 1^\prime& \;\;\; \rightarrow \;\;\; \bf 1^\prime \\ \nonumber
&\bf 2& \;\;\; \rightarrow \;\;\; \bf 2\\ \nonumber
&\bf 3& \;\;\; \rightarrow \;\;\; \bf 1^\prime + 2 \\ \nonumber
&\bf 3^\prime& \;\;\; \rightarrow \;\;\; \bf 1 + 2
\end{eqnarray}}
\end{center}
\vspace{-3.0mm}
Note that due to the choice of $S$, $T$ and $U$ the decomposition of $S_4$ representations
into those of $A_4$ and $S_3$ can be nicely read off from the generators.


\section{Messenger sector}
\cleqn


As already discussed in section 2, in order to generate the operators  $(F\wt \Phi^d_3)_1 (T \Phi^d_2)_1 H_{\overline{45}}/M^2$ and
$(F \Phi^d_2 \Phi^d_2)_3 (T \wt\Phi^d_3)_3 H_{\ol{5}}/M^3$ we have to
require a specific choice of mediators to exist in a high energy completion of our effective theory.
This is necessary in order to correctly achieve the GJ relations among the down quark and charged lepton masses
as well as in order to ensure the validity of the GST relation.
To this end, we add five pairs of heavy fields $\left\{\Sigma,\Sigma^c \right\}$, $\left\{\Delta,
\Delta^c \right\}$, $\left\{\Upsilon, \Upsilon^c \right\}$, $\left\{\Omega,
\Omega^c \right\}$ and $\left\{\Theta, \Theta^c \right\}$ which are vector-like under
$SU(5) \times U(1)$. Apart from the pair $\left\{\Sigma,\Sigma^c \right\}$ all fields
transform as $5$-plets under $SU(5)$.
Similarly to the supermultiplets containing the SM particles, they
carry a charge +1 under the $U(1)_R$ symmetry.
Their transformation properties under the family symmetry
$S_4 \times U(1)$ as well as under the $SU(5)$ gauge group can be found in
table \ref{heavyfields}. The $U(1)$ charges are given in terms of the general
parameters $(x,y,z)$ as well as for the specific case $\# 13$ where $(x,y,z)=(5,4,1)$,
as chosen in section~4.3.
\begin{table}
\begin{center}
\begin{tabular}{|c||c|c|c|c|c|c|c|c|c|c|}
\hline
\rule[0.16in]{0cm}{0cm} \!Particle\phantom{\Big|}\!\!\! & $\Sigma$ & $\Sigma^c$ & $\Delta$ & $\Delta^c$ & $\Upsilon$ & $\Upsilon^c$
& $\Omega$ & $\Omega^c$ & $\Theta$ & $\Theta^c$\\
\hline
\rule[0.16in]{0cm}{0cm}$SU(5)$\phantom{\Big|}\!\! & $\bf \overline{10}$ & $\bf 10$ &  $\bf 5$ & $\bf \overline{5}$ & $\bf 5$ & $\bf \overline{5}$  &  $\bf \overline{5}$ & $\bf 5$ & $\bf \overline{5}$ & $\bf 5$ \\
\hline
$S_4$\phantom{\Big|}\!\! & $\bf 1$ & $\bf 1$ & $\bf 1$ & $\bf 1$ & $\bf 2$ & $\bf 2$ & $\bf 3$ & $\bf 3$ & $\bf 3$ & $\bf 3$\\
\hline
$U(1)$\phantom{\Big|}\!\! & \!$-x-z$\! & \!$x+z$\! & \!$x+2 z$\! & \!$-x-2 z$\! & $x$ & $-x$ & \!$y+2 z$\! & \!$-y-2 z$\! & \!$y+z$\! & \!$-y-z$\!\\
\hline
\hline
$U(1)$\phantom{\Big|}\!\! & $-6$ & $6$ & $7$ & $-7$ & $5$ & $-5$ & $6$ & $-6$ & $5$ & $-5$\\
\hline
\end{tabular}\end{center}
\begin{center}
\caption[]{Heavy fields necessary to generate the diagrams given in figure \ref{diagram}.
We list their $U(1)$ charges
in terms of the parameters $(x,y,z)$ as well as for the specific case $\#13$
where $(x,y,z)=(5,4,1)$, as chosen in section~4.3. All fields carry a $U(1)_R$ charge
$+1$. \label{heavyfields}}
\end{center}
\end{table}
The relevant terms in the superpotential which give rise to the first diagram
of figure~\ref{diagram}, and thus to $(F\wt \Phi^d_3)_1 (T \Phi^d_2)_1 H_{\overline{45}}/M^2$, are
\begin{eqnarray}
w_{\mathrm{heavy}}&\supset&\alpha_1 T \Phi^d_2 \Sigma + \alpha_2 H_{\overline{45}}\Delta^c\Sigma^c
+ \alpha_3 \Delta F \wt\Phi^d_3
\\[0mm] \nonumber
&&+ M_{\Sigma} \Sigma^c \Sigma + M_{\Delta} \Delta^c \Delta \; .
\end{eqnarray}
The second diagram of figure~\ref{diagram},
corresponding to the operator $(F \Phi^d_2 \Phi^d_2)_3 (T
\wt\Phi^d_3)_3 H_{\ol{5}}/M^3$, is generated from the terms
\begin{eqnarray}
w_{\mathrm{heavy}}&\supset&\beta_1 T \Upsilon^c H_{\ol{5}}+ \beta_2 \Upsilon \Omega \wt\Phi^d_3  + \beta_3 \Omega^c \Theta \Phi^d_2
+ \beta_4 F \Theta^c \Phi^d_2
\\[0mm] \nonumber
&& + M_\Upsilon \Upsilon^c\Upsilon + M_\Omega \Omega^c\Omega +
M_\Theta \Theta^c\Theta  \ .
\end{eqnarray}
Note that we omit $SU(5)$ indices throughout this calculation.
According to \cite{integrateout} we can integrate out the heavy degrees of freedom
$\left\{\Sigma, \Sigma^c\right\}$, $\left\{\Delta,
\Delta^c\right\}$, $\left\{\Upsilon, \Upsilon^c\right\}$, $\left\{\Omega,
\Omega^c\right\}$  and
$\left\{\Theta, \Theta^c\right\}$  by computing the derivatives of $w_{\rm heavy}$ with respect to
these fields, setting them to zero and plugging the result for the heavy fields
back into the superpotential $w_{\rm heavy}$ as well as the
K\"{a}hler potential. The K\"{a}hler potential for the heavy fields is canonical in the lowest order
in the expansion of flavon fields.
We find for the derivatives of $w_{\rm heavy}$

\small
\hspace{-10.65mm}
\parbox{3.638in}{\begin{eqnarray}
\frac{\partial w_{\mathrm{heavy}}}{\partial\Sigma} \!\!&\!\!=\!\!&\! M_\Sigma \Sigma^c + \alpha_1 (T_1 \Phi^d_{2,2} + T_2 \Phi^d_{2,1})  ,
\nonumber \\[0mm]
\frac{\partial w_{\mathrm{heavy}}}{\partial\Sigma^c} \!\!&\!\!=\!\!&\! M_\Sigma \Sigma + \alpha_2 H_{\overline{45}}\Delta^c   ,
\nonumber \\[0mm]
\frac{\partial w_{\mathrm{heavy}}}{\partial\Delta} \!\!&\!\!=\!\!&\! M_\Delta \Delta^c + \alpha_3 (F_1 \wt\Phi^d_{3,1} + F_2 \wt\Phi^d_{3,3} +F_3 \wt\Phi^d_{3,2})   ,
\nonumber \\[0mm]
\frac{\partial w_{\mathrm{heavy}}}{\partial\Delta^c} \!\!&\!\!=\!\!&\! M_\Delta \Delta + \alpha_2  H_{\overline{45}}\Sigma^c   ,
\nonumber \\[0mm]
\frac{\partial w_{\mathrm{heavy}}}{\partial\Upsilon_1} \!\!&\!\!=\!\!&\! M_\Upsilon \Upsilon^c_2
+ \beta_2 (\Omega_1 \wt\Phi^d_{3,2} + \Omega_2 \wt\Phi^d_{3,1} +\Omega_3 \wt\Phi^d_{3,3})   ,
\nonumber \\[0mm]
\frac{\partial w_{\mathrm{heavy}}}{\partial\Upsilon_2} \!\!&\!\!=\!\!&\! M_\Upsilon \Upsilon^c_1
+ \beta_2 (\Omega_1 \wt\Phi^d_{3,3} + \Omega_2 \wt\Phi^d_{3,2} +\Omega_3 \wt\Phi^d_{3,1})   ,
\nonumber \\[0mm]
\frac{\partial w_{\mathrm{heavy}}}{\partial\Upsilon_1^c} \!\!&\!\!=\!\!&\! M_\Upsilon \Upsilon_2 + \beta_1 H_{\ol{5}} T_2    ,
\nonumber \\[0mm]
\frac{\partial w_{\mathrm{heavy}}}{\partial\Upsilon_2^c} \!\!&\!\!=\!\!&\! M_\Upsilon \Upsilon_1 + \beta_1 H_{\ol{5}} T_1    ,
\nonumber \\[0mm]
\frac{\partial w_{\mathrm{heavy}}}{\partial\Omega_1} \!\!&\!\!=\!\!&\! M_\Omega \Omega^c_1
+ \beta_2 (\Upsilon_1 \wt\Phi^d_{3,2} + \Upsilon_2 \wt\Phi^d_{3,3})     ,
\nonumber \\[0mm]
\frac{\partial w_{\mathrm{heavy}}}{\partial\Omega_2} \!\!&\!\!=\!\!&\! M_\Omega \Omega^c_3
+ \beta_2 (\Upsilon_1 \wt\Phi^d_{3,1} + \Upsilon_2 \wt\Phi^d_{3,2})     ,
\nonumber
\end{eqnarray}}
\parbox{2.77in}{
\begin{eqnarray}
\frac{\partial w_{\mathrm{heavy}}}{\partial\Omega_3} \!\!&\!\!=\!\!&\! M_\Omega \Omega^c_2
+ \beta_2 (\Upsilon_1 \wt\Phi^d_{3,3} + \Upsilon_2 \wt\Phi^d_{3,1})    ,
\nonumber \\[0mm]
\frac{\partial w_{\mathrm{heavy}}}{\partial\Omega^c_1} \!\!&\!\!=\!\!&\! M_\Omega \Omega_1
+ \beta_3 (\Theta_2 \Phi^d_{2,1} + \Theta_3 \Phi^d_{2,2})    ,
\nonumber \\[0mm]
\frac{\partial w_{\mathrm{heavy}}}{\partial\Omega^c_2} \!\!&\!\!=\!\!&\! M_\Omega \Omega_3
+ \beta_3 (\Theta_2 \Phi^d_{2,2} + \Theta_1 \Phi^d_{2,1})    ,
\nonumber \\[0mm]
\frac{\partial w_{\mathrm{heavy}}}{\partial\Omega^c_3} \!\!&\!\!=\!\!&\! M_\Omega \Omega_2
+ \beta_3 (\Theta_1 \Phi^d_{2,2} + \Theta_3 \Phi^d_{2,1})   ,
\nonumber \\[0mm]
\frac{\partial w_{\mathrm{heavy}}}{\partial\Theta_1} \!\!&\!\!=\!\!&\! M_\Theta \Theta^c_1
+ \beta_3 (\Omega^c_2 \Phi^d_{2,1} + \Omega^c_3 \Phi^d_{2,2})   ,
\nonumber \\[0mm]
\frac{\partial w_{\mathrm{heavy}}}{\partial\Theta_2} \!\!&\!\!=\!\!&\! M_\Theta \Theta^c_3
+ \beta_3 (\Omega^c_1 \Phi^d_{2,1} + \Omega^c_2 \Phi^d_{2,2})   ,
\nonumber \\[0mm]
\frac{\partial w_{\mathrm{heavy}}}{\partial\Theta_3} \!\!&\!\!=\!\!&\! M_\Theta \Theta^c_2
+ \beta_3 (\Omega^c_1 \Phi^d_{2,2} + \Omega^c_3 \Phi^d_{2,1})    ,
\nonumber \\[0mm]
\frac{\partial w_{\mathrm{heavy}}}{\partial\Theta^c_1} \!\!&\!\!=\!\!&\! M_\Theta \Theta_1
+ \beta_4 (F_2 \Phi^d_{2,1} + F_3 \Phi^d_{2,2})    ,
\nonumber \\[0mm]
\frac{\partial w_{\mathrm{heavy}}}{\partial\Theta^c_2} \!\!&\!\!=\!\!&\! M_\Theta \Theta_3
+ \beta_4 (F_1 \Phi^d_{2,1} + F_2 \Phi^d_{2,2})  ,
\nonumber \\[0mm]
\frac{\partial w_{\mathrm{heavy}}}{\partial\Theta^c_3} \!\!&\!\!=\!\!&\! M_\Theta \Theta_2
+ \beta_4 (F_1 \Phi^d_{2,2} + F_3 \Phi^d_{2,1})  .
\nonumber\\[0mm]
\end{eqnarray}}

\normalsize
Plugging the solution for the heavy fields back into $w_{\rm heavy}$ and
using the vacuum structure of $\Phi^d_2$ and $\wt \Phi^d_3$,
as shown in Eq.~(\ref{vacuum-down}), we arrive at
\be
 \frac{\alpha_1 \alpha_2 \alpha_3}{M_\Delta M_\Sigma} (F_2 T_2 -F_3 T_2) H_{\overline{45}} \varphi^d_2 \wt\varphi^d_3 \; ,
\ee
\be \frac{\beta_1 \beta_2 \beta_3 \beta_4}{M_\Omega M_\Theta M_\Upsilon} (-F_2 T_1  + F_1 T_2 + F_3 T_1 - F_3 T_2) H_{\ol{5}} \varphi^d_2
\varphi^d_2 \wt\varphi^d_3 \ ,
\ee
which shows that terms of exactly the required form are generated and no further ones, compare to Eqs.~(\ref{GJfact},\ref{GSTrel},\ref{MdLO},\ref{MeLO}).

There are additional terms\footnote{One can easily check that this set of
additional renormalisable operators is exhaustive even if we consider the
specific $U(1)$ charges of case $\# 13$, see table~\ref{heavyfields}.} which
also arise at the renormalisable level involving the messengers
\begin{equation}
\gamma_1 \wt\Phi^u_2 \Upsilon \Upsilon^c
+ \gamma_2 \wt\Phi^u_2 \Omega \Omega^c
+ \gamma_3 \wt\Phi^u_2 \Theta \Theta^c \; .
\end{equation}
These terms are expected to give small corrections to the LO mass terms of the messengers which are
$M_\Upsilon$, $M_\Omega$ and $M_\Theta$. The terms involving $\wt\Phi^u_2$ should be small compared to those,
since $\langle \wt\Phi^u_2 \rangle \approx \lambda^4 M$ with $M$ being the generic messenger mass. These terms can also be taken
into account when integrating out the heavy fields and give a subleading contribution to the fermion mass matrices
which is suppressed by $\lambda^4$ compared to the LO one. The flavour structure deviates from the one of the LO
term. However, all such corrections have a small effect on the mass spectrum of the fermions and their mixings.
Plugging the solution for the heavy fields into their K\"ahler potential shows that the non-canonical terms generated
for the supermultiplets containing SM fermions are small and thus do not considerably affect our assumption of
a canonical K\"ahler potential for all fields.


\section{\label{app-list}List of unwanted terms with up to three flavons}
\cleqn


Here we present all operators with up to three flavons which are classified either as dangerous or as marginal.
As done in section 3 we only consider the case of $k=1$.
Apart from the operator we show in this table also the $\la$-suppression of the contribution(s) due to this operator as well as
the entries of the mass matrices which are in conflict with the LO
setup. Entries for which the operator is marginal in the above sense are
marked with square brackets, whereas for all other ones the operator is dangerous.
Note that we only give one of the two entries $(ij)$ and
$(ji)$ in the case of symmetric or symmetrised terms, $TTH^{}_{5}$, $T_3TH^{}_{5}$,
$NN$. The three operators denoted with a prime ($43'$, $48'$, $54'$) differ from the
LO terms of the down quark sector in
Eq.~(\ref{downLO}) only by the exchange of $H_{\ol{5}}$ and
$H_{\ol{45}}$. All other terms given for the down quark sector must be forbidden for both
 Higgs fields, $H_{\ol{5}}$ as well as $H_{\ol{45}}$.
 Note that in this calculation we assumed the vacuum
alignment of the flavons as given in Eqs.~(\ref{vacuum-up},\ref{vacuum-down},\ref{vacuum-nu}),
 apart from the fact that we allow $\langle\wt\Phi^d_3\rangle$ to be aligned as $(0, \kappa, 1)^t$ with $|\kappa|=1$,
$\kappa$ complex, instead of using $(0, -1, 1)^t$ as shown in Eq.~(\ref{vacuum-down}).
\footnote{Note that in the discussion of the flavon superpotential in section 4 we present a setup
of driving fields which only leads to the alignment $\langle \wt\Phi^d_3\rangle \propto (0,-1,1)^t$.}
Albeit these alignments lead to the same LO results for fermion masses and mixings, the results for the
classification of dangerous and marginal operators are slightly different:
the operator $\# 18$ which is dangerous becomes irrelevant for the specific
alignment $\langle\wt\Phi^d_3\rangle \propto (0, -1, 1)^t$ as does the
marginal operator $\# 32$.
Note that, for notational simplicity, the appropriate powers of the messenger
scale $M$, necessary to give
the correct mass dimension of the operators, are omitted in the following table. For the operator $F T H_{\ol{5},\ol{45}} \Phi^\nu_{3'}/M$
$(i1)$ indicates that the contributions to the $(11)$, $(21)$ as well as $(31)$ elements are classified as dangerous.
\begin{center}
{\footnotesize{

\begin{tabular}{cc}
~\\[0mm]
\begin{tabular}{c}
$
\begin{array}{|c|c|c|c|}\hline
\multicolumn{4}{|c|}{TTH^{}_{5}}\\\hline
\# & \text{Operator}
& \text{Structure}
& \m O(\lambda)  \\ \hline

1 & TTH^{}_{5} \Phi^d_2
& (11)
& \lambda \\ \hline

2 & TTH^{}_{5} (\Phi^d_2)^2
& (22)
& \lambda^2 \\ \hline

3 & TTH^{}_{5} (\Phi^d_2)^3
&(12)
& \lambda^3 \\ \hline

4 & T T H^{}_{5} \Phi^\nu_2
&(11)[(22)]
& \la^{4} \\ \hline

5  & T T H^{}_{5} \Phi^\nu_1
& (12)
& \la^{4} \\ \hline

6  & T T H^{}_{5} (\Phi^d_3)^2
& (11)
& \la^{4} \\\hline

7 & TTH^{}_{5} \Phi^d_3 \wt\Phi^d_3
& (11),(12)
& \lambda^{5}  \\ \hline

8 & TTH^{}_{5} \Phi^u_2 \Phi^d_2
& (12)
& \lambda^{5}   \\ \hline

9 & T T H^{}_{5} \wt\Phi^u_2 \Phi^d_2
& (12)
& \la^{5} \\\hline

10  & T T H^{}_{5} \Phi^d_2 \Phi^\nu_2
& (12)
& \la^{5} \\\hline

11   & T T H^{}_{5} \Phi^d_2 \Phi^\nu_1
& (11)
& \la^{5} \\\hline

12 & TTH^{}_{5} \Phi^d_3 \Phi^\nu_{3'}
&(11)
& \lambda^{6} \\ \hline

13 & T T H^{}_{5} (\wt\Phi^d_3)^2
& (11)[(12)]
& \la^{6} \\\hline

14 & TTH^{}_{5}  \Phi^d_3 \wt\Phi^d_{3} \Phi^d_2
&(11)
& \lambda^{6}   \\ \hline

15 & TTH^{}_{5} (\Phi^d_2)^2 \Phi^\nu_{2}
& (11)[(12)]
& \lambda^{6}  \\ \hline

16  & T T H^{}_{5} \Phi^u_2 (\Phi^d_2)^2
& (11)
&\la^{6} \\\hline

17  & T T H^{}_{5} \wt\Phi^u_2(\Phi^d_2)^2
& (11)
& \la^{6} \\\hline

18 & TTH^{}_{5} \wt\Phi^d_3 \Phi^\nu_{3'}
& (11)
& \lambda^{7} \\ \hline

19 & TTH^{}_{5} \Phi^d_3 \Phi^d_2 \Phi^\nu_{3'}
& (11)
& \lambda^{7} \\ \hline

20 & TTH^{}_{5}  (\wt\Phi^d_3)^2 \Phi^d_2
& (11)
& \lambda^{7}   \\ \hline

21 & T T H^{}_{5} \wt\Phi^u_2
& [(22)]
& \la^{4} \\\hline

22 & T T H^{}_{5} (\Phi^u_2)^2
& [(11)]
& \la^8 \\\hline

23 & T T H^{}_{5} (\wt\Phi^u_2)^2
& [(11)]
& \la^8 \\\hline

24 & T T H^{}_{5} \Phi^u_2 \Phi^\nu_2
& [(11)]
& \la^{8} \\\hline

25 & T T H^{}_{5} \wt\Phi^u_2 \Phi^\nu_2
& [(11)]
& \la^{8} \\\hline

26 & T T H^{}_{5} (\Phi^\nu_{3'})^2
&[(11)]
& \la^{8} \\\hline

27 & T T H^{}_{5} (\Phi^\nu_2)^2
& [(11)]
& \la^{8} \\\hline

28 & T T H^{}_{5} \Phi^\nu_2 \Phi^\nu_1
& [(11)]
& \la^{8} \\\hline

29 & TTH^{}_{5}  (\Phi^d_3)^2 \Phi^\nu_{3'}
&[(11)]
& \lambda^{8}  \\ \hline

30  & T T H^{}_{5} (\Phi^d_3)^2 \Phi^\nu_1
& [(11)]
& \la^{8} \\\hline

31  & T T H^{}_{5} \Phi^d_3 (\wt\Phi^d_3)^2
& [(11)]
& \la^{8} \\\hline

32  & T T H^{}_{5} \wt\Phi^d_3 \Phi^d_2 \Phi^\nu_{3'}
& [(11)]
& \la^{8} \\\hline

\end{array}
$

\\~\\[-2mm]

$
\begin{array}{|c|c|c|c|}\hline
\multicolumn{4}{|c|}{T_3TH^{}_{5}}\\\hline
\# &\text{Operator}
& \text{Structure}
& \m O(\lambda) \\ \hline

33 & T_3 T H^{}_{5} \Phi^d_2
& (32)
& \la \\ \hline

34 & T_3TH^{}_{5} (\Phi^d_2)^2
& (31)
&\lambda^{2} \\ \hline

35 & T_3 T H^{}_{5} \Phi^u_2
& [(31)]
& \la^4 \\\hline

36 & T_3 T H^{}_{5} \wt\Phi^u_2
& [(31)]
& \la^4 \\ \hline

37 & T_3 T H^{}_{5} \Phi^\nu_2
& [(31)]
& \la^{4} \\ \hline

\end{array}
$

\end{tabular}

&\hspace{-6mm}

\begin{tabular}{c}

$
\begin{array}{|c|c|c|c|}\hline
\multicolumn{4}{|c|}{FTH_{\ol{5},\ol{45}}}\\\hline
\# &\text{Operator}
& \text{Structure}
& \m O(\lambda)  \\ \hline

38 & FTH_{\ol{5},\ol{45}} \Phi^d_3
& (11),(22)
& \lambda^{2}  \\ \hline

39 & FTH_{\ol{5},\ol{45}} \wt\Phi^d_3
&(11),(12),(22),(31)
& \lambda^{3}   \\ \hline

40 & F T H_{\ol{5},\ol{45}} \Phi^d_3 \Phi^d_2
& (21),(32)
& \la^{3} \\\hline

41 & F T H_{\ol{5},\ol{45}} \Phi^\nu_{3'}
& (i1),(12)[(22),(32)]
& \la^{4} \\\hline

42 & FTH_{\ol{5},\ol{45}} (\Phi^d_3)^2
& (12),(31)
& \lambda^{4}   \\ \hline

43' & F T H_{\ol{5}} \wt\Phi^d_3 \Phi^d_2
& (11),(21)[(22),(32)]
& \la^{4} \\\hline

44 & F T H_{\ol{5},\ol{45}} \Phi^d_3 (\Phi^d_2)^2
& (12),(31)
& \la^{4} \\\hline

45 & F T H_{\ol{5},\ol{45}} \Phi^d_2 \Phi^\nu_{3'}
& (11)[(12),(21),(31)]
& \la^{5} \\\hline

46 & F T H_{\ol{5},\ol{45}} (\Phi^d_3)^2 \Phi^d_2
& (11)
& \la^{5} \\\hline

47 & FTH_{\ol{5},\ol{45}} \Phi^d_3 \wt\Phi^d_3
& [(12),(21),(31)]
& \lambda^{5}   \\ \hline

48' & FTH_{\ol{45}} \wt \Phi^d_3 (\Phi^d_2)^2
& [(12),(21),(31)]
& \lambda^{5}   \\ \hline

49 & F T H_{\ol{5},\ol{45}} (\wt\Phi^d_3)^2
& [(11)]
& \la^{6} \\\hline

50 & F T H_{\ol{5},\ol{45}} \Phi^d_3 \Phi^\nu_{3'}
& [(11)]
& \la^{6} \\ \hline

51 & F T H_{\ol{5},\ol{45}} \Phi^d_3 \Phi^\nu_1
& [(11)]
& \la^{6} \\ \hline

52 & F T H_{\ol{5},\ol{45}} (\Phi^d_2)^2 \Phi^\nu_{3'}
& [(11)]
& \la^{6} \\\hline

53 & FTH_{\ol{5},\ol{45}} \Phi^d_3 \wt\Phi^d_3 \Phi^d_2
& [(11)]
& \lambda^{6} \\ \hline

\end{array}
$

\\~\\[-2mm]

$
\begin{array}{|c|c|c|c|}\hline
\multicolumn{4}{|c|}{FT_3H_{\ol{5},\ol{45}}}\\\hline
\# &\text{Operator}
& \text{Structure}
& \m O(\lambda)  \\ \hline

54' & F T_3 H_{\ol{45}} \Phi^d_3
& [(33)]
& \la^{2} \\\hline

55 & F T_3 H_{\ol{5},\ol{45}} \wt\Phi^d_3
& (23)
& \la^{3} \\\hline

56 & F T_3 H_{\ol{5},\ol{45}} \Phi^d_3\Phi^d_2
& (13)
& \la^{3} \\\hline

57 & F T_3 H_{\ol{5},\ol{45}} \wt\Phi^d_3\Phi^d_2
& (13)
& \la^{4}  \\\hline

58 & F T_3 H_{\ol{5},\ol{45}} \Phi^d_3 (\Phi^d_2)^2
& [(23)]
& \la^{4} \\\hline

59 & FT_3H_{\ol{5},\ol{45}}  \Phi^d_{3}\wt\Phi^d_{3}
& [(13)]
& \lambda^{5} \\ \hline

60 & F T_3 H_{\ol{5},\ol{45}} \Phi^d_2 \Phi^\nu_{3'}
& [(13)]
& \la^{5} \\\hline

61 & F T_3 H_{\ol{5},\ol{45}} \wt\Phi^d_3 (\Phi^d_2)^2
& [(13)]
& \la^{5} \\\hline

\end{array}
$

\\~\\[-2mm]

$
\begin{array}{|c|c|c|c|}\hline
\multicolumn{4}{|c|}{NN}\\\hline
\# &\text{Operator}
& \text{Structure}
& \m O(\lambda)  \\ \hline

62 & NN \Phi^d_2
& (12),(33)
& \lambda  \\ \hline

63 & NN (\Phi^d_2)^2
& (13),(22)
& \lambda^2   \\ \hline
64 & NN  \Phi^d_3 \Phi^d_2
& (11),(23)
& \lambda^{3}   \\ \hline

65 & N N (\Phi^d_2)^3
& (11),(23)
& \la^{3}  \\ \hline

66 & NN \Phi^u_2
& (13),(22)
& \lambda^4   \\ \hline

67 & NN \wt\Phi^u_2
& (13),(22)
& \lambda^4   \\ \hline

68 & NN  (\Phi^d_3)^2
& (12),(33)
& \lambda^{4}   \\ \hline

69 & NN  \wt\Phi^d_3 \Phi^d_2
& (11),(13),(22),(23)
& \lambda^{4}  \\ \hline

70 & NN \Phi^d_3 (\Phi^d_2)^2
& (12),(33)
 & \lambda^{4}   \\ \hline

\end{array}
$

\end{tabular}

\end{tabular}
}}\end{center}

\normalsize


\section{Relations of flavon VEVs}
\cleqn


\subsection{Correlations among flavon VEVs}
\cleqn


In this part of appendix D we detail the calculations which lead to the results given in section 4.3. If 
we want to couple additional driving fields, not already present in table 3, to the flavons, in order to 
correlate the flavon VEVs further such fields obviously have
 to couple to at least two operator structures with different flavon content.
For this to work, the latter operators have to:
$(i)$ have identical $U(1)$ charges, $(ii)$ transform identically under $S_4$
and $(iii)$ obviously have the same overall $\lambda$-suppression if we insert the assumed
  suppression of the occurring flavon scales as given in
  Eqs.~(\ref{orderphiu},\ref{orderphid},\ref{orderphinu}). Furthermore, as already mentioned
at length above, we avoid introducing new mass scales into the flavon
superpotential at this stage (with the exception of
case $\# 10$, see below and table 6).

In the case that the additional driving field furnishes a doublet or a triplet
representation of $S_4$, we have to
ensure that the $F$-terms of {\it all} the components vanish for the
LO vacuum structure of the flavons. The following example
illustrates this issue: let us consider the $U(1)$ charge assignment
$\# 10$ and a driving field $Z^{\mathrm{new}}_{3}$ being a triplet {\bf 3}
under $S_4$. For $Z^{\mathrm{new}}_{3}$ having the $U(1)$ charge
$+13$ we find two flavon combinations $\wt \Phi^u_2 \wt \Phi^d_3$ and $(\Phi^d_3)^3
\Phi^d_2/M^2$ which can couple to $Z^{\mathrm{new}}_{3}$ in order to form
an invariant under $S_4 \times U(1)$. Furthermore, these combinations
reveal the same $\la$-suppression ($\la^7$).
 However, inserting the vacuum alignment of Eqs.~(\ref{vacuum-up},\ref{vacuum-down},\ref{vacuum-nu}), we
find that for $\langle\wt \Phi^u_2 \wt \Phi^d_3\rangle$ the first as well as the third component
of the triplet {\bf 3} are non-zero, whereas only the third component of {\bf 3}
is non-zero for $\langle(\Phi^d_3)^3\Phi^d_2/M^2 \rangle$. Thus, we cannot
satisfy the requirement of vanishing $F$-terms for all components of
$Z^{\mathrm{new}}_{3}$ unless we set some of the flavon VEVs to
zero.\footnote{Moreover, a detailed analysis shows that there exist two
  operators, $M_{Z^{\mathrm{new}}_{3}} Z^{\mathrm{new}}_{3} \wt\Phi^d_3$, with
  $M_{Z^{\mathrm{new}}_{3}} \sim \lambda^{x^{}_M}_{} M$ being an explicit mass
  scale, and $Z^{\mathrm{new}}_{3} \Phi^d_3\Phi^u_2$, arising respectively at
  $\la^{3+x_M}$ and $\la^6$, that also
  give non-vanishing contributions if the flavon vacuum structure at LO
  is employed. These contributions would perturb any
  possible correlation among the VEVs $\wt \varphi^u_2$,
  $\varphi^d_2$,$\varphi^d_3$ and $\wt \varphi^d_3$.}
This discussion shows that the constraint of the
 vanishing of the $F$-terms of all components considerably reduces
 the possibilities of introducing additional driving fields, which give rise to correlations between
the scales of flavon VEVs.
Moreover, also for any new driving field which couples consistently to
at least two terms with identical $\lambda$-suppression, we have to check that the same
driving field does not couple to other terms which are less
suppressed and thus would strongly perturb the desired correlation.
\begin{table}
{\small
\begin{center}
$$
\begin{array}{|l|c|c|c|c|c|c|c|} \hline
U(1)~\text{symmetry}\phantom{\Big|}\!\!\! & \# 10 & \multicolumn{3}{|c|}{\# 13} & \# 21 & \multicolumn{2}{|c|}{\# 25} \\\hline \text{Driving~field}\phantom{\Big|} & X^{\mathrm{new}}_{1\,\mathrm{or}\,1'} & \wt X^{\mathrm{new}}_{1'} & X^{\mathrm{new}}_{1\,\mathrm{or}\,1'} & X^{\mathrm{new}}_{1\,\mathrm{or}\,1'} & X^{\mathrm{new}}_{1\,\mathrm{or}\,1'}~\mathrm{or}~
Z^{\mathrm{new}}_{3\,\mathrm{or}\,3'} &
X^{\mathrm{new}}_{1\,\mathrm{or}\,1'} &
X^{\mathrm{new}}_{1\,\mathrm{or}\,1'}
\\\hline
U(1)~\text{charge}\phantom{\Big|} & 0
& 15
& 18
& 25
& 10
& 6
& 20 \\\hline
\text{Order~in~} \la \phantom{\Big|} & \la^8 & \la^9 & \la^8 & \la^9 & \la^8 & \la^8 & \la^9 \\\hline \text{Correlation}\phantom{\Big|} & C_{\#10} & C_{\#13} &  \multicolumn{2}{|c|}{C'_{\#13}} & C_{\#21} & \multicolumn{2}{|c|}{C_{\#25}}\\\hline
\end{array}
$$
\end{center} }
\caption{\label{correlations}The driving fields and the resulting
  correlations for the four successful charge assignments.
  The correlations are as follows,
$C_{\#10}: M^2 (\varphi^d_3)^2 \varphi^\nu_1 \sim  (\varphi^d_2)^4
\varphi^u_2
\big[ + M_{X_1^{\mathrm{new}}}^2 M^3 \big]$; %
$C_{\#13}: M^2 \wt\varphi^u_2\wt\varphi^d_3 \sim  \varphi^d_2 (\varphi^d_3)^3$; %
$C'_{\#13}: M \varphi^u_2 \sim  \varphi^d_2 \wt\varphi^d_3$; %
$C_{\#21}: M^2 \varphi^u_2 \varphi^\nu_1 \sim  (\varphi^d_2)^2 (\wt\varphi^d_3)^2$ and
$C_{\#25}: M \varphi^u_2 \sim  \varphi^d_2 \wt\varphi^d_3$.
Note that for case
  $\#10$ additionally an explicit mass term for the field
  $X^{\mathrm{new}}_{1}$ is allowed as it is a (trivial) singlet under $S_4$ and uncharged with respect
  to the $U(1)$ symmetry. The size of $M_{X_1^{\mathrm{new}}}$ has to be chosen as $\la^4 M$.
}
\end{table}

Restricting ourselves, for practical purposes, to operators whose
 order in $\la$ is $\leq \la^{9}$ if the scales of the flavons according to
Eqs.~(\ref{orderphiu},\ref{orderphid},\ref{orderphinu}) are plugged
in, the number of possible new driving fields which correlate the different
scales $\varphi^u_2$, $\wt\varphi^u_2$, $\varphi^d_3$, $\wt\varphi^d_3$, $\varphi^d_2$
and $\varphi^\nu_1$ narrows down to only a few. In particular, for our four $U(1)$ symmetries
we find the correlations in table~\ref{correlations}.
A few aspects are interesting to observe: first of all, notice that for solution
$\# 10$ the additional driving field(s) has (have) to be neutral under the $U(1)$ symmetry. For this
reason a coupling with mass dimension two is allowed, if the field $X^{\mathrm{new}}_1$
is used. Furthermore, we find only a limited number of
possible correlations among the VEVs. Especially for a given choice of $U(1)$ charges
$x$, $y$ and $z$ we find at most two distinct relations among the VEVs. In the case in
which a certain relation can be reproduced through several different driving fields,
e.g. in the case $\# 21$ in which $M^2 \varphi^u_2 \varphi^\nu_1 \sim  (\varphi^d_2)^2 (\wt\varphi^d_3)^2$
can be achieved through four different driving fields, $X^{\mathrm{new}}_{1\,\mathrm{or}\,1'}$ or
$Z^{\mathrm{new}}_{3\,\mathrm{or}\,3'}$,
obviously only one of these can be added to the model because otherwise
we would have to require ad hoc relations among the parameters in the superpotential
to reconcile the results. For a similar reason
it is also not possible to introduce a driving field $Y^{\mathrm{new}}_2$
in the case of solution $\# 21$ instead of $X^{\mathrm{new}}_{1\,\mathrm{or}\,1'}$
or $Z^{\mathrm{new}}_{3\,\mathrm{or}\,3'}$. Thirdly, notice that only for the
$U(1)$ charge assignment $\# 13$ we find two distinct additional relations among
the scales of the flavon VEVs. One might argue that we can also
achieve two non-trivial relations if,  in the case $\# 10$, we make use of
both fields, $X^{\mathrm{new}}_1$ and $X^{\mathrm{new}}_{1'}$. (This is
possible since the consequential conditions for the flavon VEVs differ by the
term related to the mass scale $M_{X_1^{\mathrm{new}}}^2$.) However, as stated
in section 4.1 we avoid the introduction of such terms into the flavon
superpotential at this stage. Therefore we focus on scenario $\# 13$ in the
 phenomenological analysis presented in section 5.

From table~\ref{correlations} we see that scenario $\# 13$ can lead to the relation
\be
M \varphi^u_2 \sim  \varphi^d_2 \wt\varphi^d_3 \ ,
\ee
through a new driving field transforming as ${\bf 1}$ or ${\bf 1'}$ under $S_4$ carrying
either charge $+18$ or charge $+25$ under the $U(1)$. Since the relation
arises at order $\la^8$ in the former case we will include - without
loss of generality - the $S_4$ singlet driving field $X^{\mathrm{new}}_{1}$
with charge $+18$ in our model. Additionally, we add
the field $\wt X^{\mathrm{new}}_{1'}$ with $U(1)$ charge $+15$ giving rise to
the correlation
\be
M^2 \wt\varphi^u_2\wt\varphi^d_3 \sim  \varphi^d_2 (\varphi^d_3)^3 \; . \;\;\;\;
\ee
The additional terms in the superpotential read
\bea
&&\!\!\!\!\!\!\!\frac{1}{M}\, X^{\mathrm{new}}_1 \Phi^u_2 (\Phi^d_3)^2
+ \frac{1}{M^2} \,X^{\mathrm{new}}_1 \Phi^d_2 \wt\Phi^d_3 (\Phi^d_3)^2
\\ \nonumber
&&\!\!  =
\frac{1}{M} \, X^{\mathrm{new}}_1  \left[ \Phi^u_{2,1} ((\Phi^d_{3,3})^2 + 2\Phi^d_{3,1}\Phi^d_{3,2})
+ \Phi^u_{2,2} ((\Phi^d_{3,2})^2 + 2\Phi^d_{3,1}\Phi^d_{3,3})  \right]
\\ \nonumber
&&\!\! + \,\frac{1}{M^2} X^{\mathrm{new}}_1  \left[ (\Phi^d_{2,1}\wt\Phi^d_{3,2}-\Phi^d_{2,2}\wt\Phi^d_{3,3})((\Phi^d_{3,1})^2-\Phi^d_{3,2}\Phi^d_{3,3})
\right.
\\ \nonumber
&&\!\!
+ \left.(\Phi^d_{2,1}\wt\Phi^d_{3,3}-\Phi^d_{2,2}\wt\Phi^d_{3,1})((\Phi^d_{3,2})^2-\Phi^d_{3,1}\Phi^d_{3,3})
+ (\Phi^d_{2,1}\wt\Phi^d_{3,1}-\Phi^d_{2,2}\wt\Phi^d_{3,2})((\Phi^d_{3,3})^2-\Phi^d_{3,1}\Phi^d_{3,2})
\right],
\eea
and
\bea
&&\!\!\!\!\!\!\!\!\frac{1}{M} \,\wt X^{\mathrm{new}}_{1'} \wt\Phi^u_2 \Phi^d_3 \wt\Phi^d_3
+ \frac{1}{M^3} \,\wt X^{\mathrm{new}}_{1'} \Phi^d_2 (\Phi^d_3)^4
\\ \nonumber
&& \!\!\!=
 \frac{1}{M} \,\wt X^{\mathrm{new}}_{1'} \left[ \wt\Phi^u_{2,1} (\Phi^d_{3,1}\wt\Phi^d_{3,2} +\Phi^d_{3,2}\wt\Phi^d_{3,1}+ \Phi^d_{3,3}\wt\Phi^d_{3,3})
- \wt\Phi^u_{2,2} (\Phi^d_{3,1}\wt\Phi^d_{3,3} +\Phi^d_{3,2}\wt\Phi^d_{3,2}+ \Phi^d_{3,3}\wt\Phi^d_{3,1})
\right]
\\ \nonumber
&&\!\!\!
+ \,\frac{a^{\mathrm{new}}_1}{M^3} \,\wt X^{\mathrm{new}}_{1'} ((\Phi^d_{3,1})^2 + 2\Phi^d_{3,2}\Phi^d_{3,3})\left[
\Phi^d_{2,1} ((\Phi^d_{3,3})^2 + 2\Phi^d_{3,1}\Phi^d_{3,2}) -\Phi^d_{2,2} ((\Phi^d_{3,2})^2 + 2\Phi^d_{3,1}\Phi^d_{3,3})
\right]
\\ \nonumber
&&\!\!\!
+ \,\frac{a^{\mathrm{new}}_2}{M^3} \,\wt X^{\mathrm{new}}_{1'} \left[
\Phi^d_{2,1}((\Phi^d_{3,2})^2+ 2\Phi^d_{3,1}\Phi^d_{3,3})^2 - \Phi^d_{2,2}((\Phi^d_{3,3})^2+ 2\Phi^d_{3,1}\Phi^d_{3,2})^2
\right]  .
\eea
Finally, one can check that the VEVs of the driving fields $X^{\mathrm{new}}_1$ and
$\wt X^{\mathrm{new}}_{1'}$, which are determined by the $F$-term equations of
the flavons, vanish if the LO alignments of
Eqs.~(\ref{vacuum-up},\ref{vacuum-down},\ref{vacuum-nu}) are applied.


\subsection{Mass scales in the flavon superpotential}
\cleqn


 Here we introduce further driving fields which allow for additional
mass scales in the flavon superpotential. Thus the trivial vacuum in which all
flavon VEVs vanish can be destabilised. In addition, the number of free
parameters among the undetermined VEVs, $\wt\varphi^u_2$, $\varphi^d_3$, $\varphi^d_2$, and $\varphi^\nu_1$, can be reduced to a minimum of only one parameter.
 We consider only the specific case $\# 13$ which we have singled out in section 4.3 and in the first part of this appendix.

Invoking a driving field $V_0$ which is neutral under the family symmetry $S_4
\times U(1)$ allows for the following couplings up to order $\la^8$ (if we
already take into account the phenomenologically determined sizes of the
different flavon VEVs)
\be
V_0 M_{V_0}^2 + V_0 (\wt\Phi^u_2)^2 + V_0 (\Phi^d_3)^2 \Phi^\nu_1/M + V_0 (\Phi^d_3)^2 \Phi^\nu_2/M
+ V_0 (\Phi^d_3)^2 \Phi^\nu_{3'}/M \; .
\ee
Studying the equation derived from the $F$-term of $V_0$ and using the LO form of the flavon VEVs,
we find the relation
\be
\label{V0F}
M_{V_0}^2 + (\varphi^d_3)^2 \varphi^\nu_2/M + (\varphi^d_3)^2 \varphi^\nu_{3'}/M =0 \; .
\ee
As the VEVs $\varphi^\nu_2$ and $\varphi^\nu_{3'}$ are already related to
$\varphi^\nu_1$ through the $F$-terms of the driving fields $Z^\nu_{3'}$ and
$Y^\nu_2$, see Eq.~(\ref{vev-scale-nus}), we find that $\varphi^d_3$  can be
expressed through $M_{V_0}$ and $\varphi^\nu_1$. In order to achieve the
correct order of magnitude of the VEV $\varphi^d_3$ we demand that $M_{V_0}
\sim \la^4 M$. More importantly, in order to fulfil Eq.~(\ref{V0F})
the flavon VEVs $\varphi^\nu_1$ and $\varphi^d_3$ must be non-zero.
This excludes the trivial solution with only vanishing VEVs which cannot be
avoided if only the driving fields listed in the main text are
present. Therefore $V_0$ ensures that the family symmetry actually gets broken.

In a similar way we can introduce a field $V_2 \sim ({\bf 2},-8)$ which allows, up to the order $\la^8$, for the
following couplings
\be
M_{V_2} V_2 \Phi^\nu_2 + V_2 \wt\Phi^u_2 \Phi^\nu_1 + V_2 \wt\Phi^u_2 \Phi^\nu_2 + V_2 (\Phi^d_2)^8/M^6 \; .
\ee
From the $F$-terms of the two components of $V_2$, $V_{2,1}$ and $V_{2,2}$, we find the relations
\bea
&& M_{V_2} \varphi^\nu_2 + \wt\varphi^u_2 \varphi^\nu_1 + (\varphi^d_2)^8/M^6
=0 \; , \label{ddd555}
\\
&& M_{V_2} \varphi^\nu_2 + \wt\varphi^u_2 \varphi^\nu_2 = 0 \; ,
\eea
if we apply the LO results for the flavon VEVs. As one can see,
we can relate the VEV $\wt\varphi^u_2$ to the mass scale
$M_{V_2}$. In order to end up with the correct order of magnitude for
$\wt\varphi^u_2$, we have to set $M_{V_2} \sim \la^4 M$. Using $\varphi^\nu_2 \sim \varphi^\nu_1$,
Eq.~(\ref{ddd555}) additionally leads to a determination
of $\varphi^d_2$ in terms of $\varphi^\nu_1$ and $M_{V_2}$ and consistently leads to $\varphi^d_2 \sim \la M$.
Notice further that the inclusion of the driving field $V_2$ is essential
  for giving non-zero VEVs to all flavon fields. (Here we are still assuming that a solution with
$\langle\Phi^\nu_{3'}\rangle\neq 0$ is chosen, as shown in Eqs.(\ref{nu-alignm},\ref{nu-alignm-scale}).)

In summary, by adding the two further driving fields $V_0$ and $V_2$ we can enforce the breaking of the family symmetry,
 eliminate three of the four undetermined parameters among the flavon VEVs and ensure that all these VEVs have to be
non-vanishing. The explicit mass scales,
$M_{V_0}$ and $M_{V_2}$, as well as the free parameter $\varphi^\nu_1$ all have to be of the order of $\la^4 M$
in order to generate the sizes of the flavon VEVs as invoked in the discussion
of fermion masses and mixings in section 2. 
 Obviously, we have to choose these values by hand.
In order to fully include these fields into the model presented in section 5, a careful study of
the subleading corrections arising from higher-dimensional operators as well as a re-calculation of the
shifts in the flavon VEVs would have to be performed.

We have also studied the effect of other possible driving fields $V$ allowing
for terms of the form $M_V V \Phi$ with $\Phi$ being a flavon. However,
several of these $(i)$ cannot be consistently introduced, $(ii)$ lead to some
parameter fine-tuning if considered in a setup together with $V_0$ or $(iii)$
lead to redundant results only. Therefore we conclude that the presented
choice of fields, $V_0$ and $V_2$, is the most favourable one.

Obviously, such fields could also be considered for the choices of $U(1)$ charges which we have discarded in section 4, see
Eq.~(\ref{U1sols4}). We have checked that a consistent introduction of such
fields is generally possible, however, it does not lead to a
scenario with less parameters than the one presented in the paper.



\begin{thebibliography}{10}


\bibitem{HPS}
P.~F.~Harrison, D.~H.~Perkins and W.~G.~Scott,
  Phys.\ Lett.\ B \textbf{530} (2002) 167 [hep-ph/0202074];~\\
P.~F.~Harrison and W.~G.~Scott,
  Phys.\ Lett.\ B \textbf{535} (2002) 163 [hep-ph/0203209];~\\
P.~F.~Harrison and W.~G.~Scott,
  Phys.\ Lett.\ B \textbf{557} (2003) 76 [hep-ph/0302025];~\\
C.~I.~Low and R.~R.~Volkas,
  Phys.\ Rev.\  D {\bf 68} (2003) 033007 [hep-ph/0305243].


\bibitem{Lam}
C.~S.~Lam,
 Phys.\ Lett.\  B {\bf 656} (2007) 193 [arXiv:0708.3665];~\\
C.~S.~Lam,
  Phys.\ Rev.\ Lett.\  {\bf 101} (2008) 121602 [arXiv:0804.2622];~\\
C.~S.~Lam,
  Phys.\ Rev.\  D {\bf 78} (2008) 073015 [arXiv:0809.1185].


\bibitem{King:2009ap}
  S.~F.~King and C.~Luhn,
  JHEP {\bf 0910} (2009) 093
  [arXiv:0908.1897].


\bibitem{S3-L}
W.~Grimus and L.~Lavoura,
  JHEP {\bf 0508} (2005) 013 [hep-ph/0504153];~\\
W.~Grimus and L.~Lavoura,
  JHEP {\bf 0601} (2006) 018  [hep-ph/0509239];~\\
R.~N.~Mohapatra, S.~Nasri and H.~B.~Yu,
  Phys.\ Lett.\  B {\bf 639} (2006) 318 [hep-ph/0605020];~\\
Y.~Koide,
  Eur.\ Phys.\ J.\  C {\bf 50} (2007) 809 [hep-ph/0612058];~\\
M.~Mitra and S.~Choubey,
  Phys.\ Rev.\  D {\bf 78} (2008) 115014 [arXiv:0806.3254].


\bibitem{Dn-L}
 W.~Grimus and L.~Lavoura,
  Phys.\ Lett.\  B {\bf 572} (2003) 189
  [hep-ph/0305046];~\\
A.~Adulpravitchai, A.~Blum and C.~Hagedorn,
  JHEP {\bf 0903} (2009) 046  [arXiv:0812.3799].


\bibitem{A4-L}
E.~Ma and G.~Rajasekaran,
   Phys.\ Rev.\  D {\bf 64} (2001) 113012 [hep-ph/0106291];~\\
E.~Ma,
  Phys.\ Rev.\  D {\bf 73} (2006) 057304 [hep-ph/0511133];~\\
G.~Altarelli and F.~Feruglio,
  Nucl.\ Phys.\  B {\bf 720} (2005) 64
  [hep-ph/0504165];~\\
K.~S.~Babu and X.~G.~He,
  hep-ph/0507217;~\\
G.~Altarelli and F.~Feruglio,
  Nucl.\ Phys.\  B {\bf 741} (2006) 215
  [hep-ph/0512103];~\\
G.~Altarelli, F.~Feruglio and Y.~Lin,
  Nucl.\ Phys.\  B {\bf 775} (2007) 31
  [hep-ph/0610165];~\\
M.~Hirsch, A.~S.~Joshipura, S.~Kaneko and J.~W.~F.~Valle,
  Phys.\ Rev.\ Lett.\  {\bf 99} (2007) 151802 [hep-ph/0703046];~\\
M.~Honda and M.~Tanimoto,
  Prog.\ Theor.\ Phys.\  {\bf 119} (2008) 583 [arXiv:0801.0181];~\\
Y.~Lin,
  Nucl.\ Phys.\  B {\bf 813} (2009) 91  [arXiv:0804.2867];~\\
M.~C.~Chen and S.~F.~King,
  JHEP {\bf 0906} (2009) 072
  [arXiv:0903.0125];~\\
G.~Altarelli and D.~Meloni,
  J.\ Phys.\ G {\bf 36} (2009) 085005
  [arXiv:0905.0620].


\bibitem{S4-L}
H.~Zhang,
  Phys.\ Lett.\  B {\bf 655} (2007) 132
  [hep-ph/0612214];~\\
Y.~Koide,
  JHEP {\bf 0708} (2007) 086 [arXiv:0705.2275].


\bibitem{delta54-L}
H.~Ishimori, T.~Kobayashi, H.~Okada, Y.~Shimizu and M.~Tanimoto,
  JHEP {\bf 0904} (2009) 011  [arXiv:0811.4683].


\bibitem{S3-LQ}
J.~Kubo, A.~Mondragon, M.~Mondragon and E.~Rodriguez-Jauregui,
   Prog.\ Theor.\ Phys.\  {\bf 109} (2003) 795,
   Erratum-ibid.\  {\bf 114} (2005) 287 [hep-ph/0302196];~\\
S.~L.~Chen, M.~Frigerio and E.~Ma,
   Phys.\ Rev.\  D {\bf 70} (2004) 073008
  [Erratum-ibid.\  D {\bf 70} (2004) 079905]
  [hep-ph/0404084];~\\
F.~Feruglio and Y.~Lin,
  Nucl.\ Phys.\  B {\bf 800} (2008) 77  [arXiv:0712.1528].


\bibitem{Dn-LQ}
A.~Blum, C.~Hagedorn and M.~Lindner,
  Phys.\ Rev.\  D {\bf 77} (2008) 076004 [arXiv:0709.3450];~\\
A.~Blum, C.~Hagedorn and A.~Hohenegger,
  JHEP {\bf 0803} (2008) 070  [arXiv:0710.5061];~\\
A.~Blum and C.~Hagedorn,
  Nucl.\ Phys.\  B {\bf 821} (2009) 327
  [arXiv:0902.4885].


\bibitem{Q6-LQ}
M.~Frigerio, S.~Kaneko, E.~Ma and M.~Tanimoto,
  Phys.\ Rev.\  D {\bf 71} (2005) 011901  [hep-ph/0409187];~\\
K.~S.~Babu and J.~Kubo,
  Phys.\ Rev.\  D {\bf 71} (2005) 056006 [hep-ph/0411226];~\\
Y.~Kajiyama, E.~Itou and J.~Kubo,
  Nucl.\ Phys.\  B {\bf 743} (2006) 74 [hep-ph/0511268].


\bibitem{A4-LQ}
K.~S.~Babu, E.~Ma and J.~W.~F.~Valle,
  Phys.\ Lett.\  B {\bf 552} (2003) 207  [hep-ph/0206292];~\\
F.~Bazzocchi, S.~Morisi and M.~Picariello,
  Phys.\ Lett.\  B {\bf 659} (2008) 628 [arXiv:0710.2928].


\bibitem{A4-LQ:King}
S.~F.~King and M.~Malinsky,
  Phys.\ Lett.\  B {\bf 645} (2007) 351 [hep-ph/0610250].


\bibitem{doubleA4-LQ}
A.~Aranda, C.~D.~Carone and R.~F.~Lebed,
  Phys.\ Rev.\  D {\bf 62} (2000) 016009  [hep-ph/0002044];~\\
F.~Feruglio, C.~Hagedorn, Y.~Lin and L.~Merlo,
  Nucl.\ Phys.\  B {\bf 775} (2007) 120   [hep-ph/0702194];~\\
P.~H.~Frampton and T.~W.~Kephart,
  JHEP {\bf 0709} (2007) 110  [arXiv:0706.1186];~\\
A.~Aranda,
  Phys.\ Rev.\  D {\bf 76} (2007) 111301  [arXiv:0707.3661];~\\
P.~H.~Frampton and S.~Matsuzaki,
  Phys.\ Lett.\  B {\bf 679} (2009) 347
  [arXiv:0902.1140].


\bibitem{S4-LQ:Hagedorn}
K.~S.~Babu, T.~Enkhbat and I.~Gogoladze,
  Phys.\ Lett.\  B {\bf 555} (2003) 238
  [hep-ph/0204246];~\\
C.~Hagedorn, M.~Lindner and R.~N.~Mohapatra,
  JHEP {\bf 0606} (2006) 042 [hep-ph/0602244];~\\
G.~J.~Ding,
  Nucl.\ Phys.\  B {\bf 827} (2010) 82
  [arXiv:0909.2210];~\\
D.~Meloni,
  arXiv:0911.3591.



\bibitem{S4-group}
F.~Bazzocchi, L.~Merlo and S.~Morisi,
  Nucl.\ Phys.\  B {\bf 816} (2009) 204
  [arXiv:0901.2086].


\bibitem{A4-SU5}
G.~Altarelli, F.~Feruglio and C.~Hagedorn,
  JHEP {\bf 0803} (2008) 052
  [arXiv:0802.0090];~\\
P.~Ciafaloni, M.~Picariello, E.~Torrente-Lujan and A.~Urbano,
  Phys.\ Rev.\  D {\bf 79} (2009) 116010
  [arXiv:0901.2236];~\\
T.~J.~Burrows and S.~F.~King,
  arXiv:0909.1433.


\bibitem{Tpr-SU5}
M.~C.~Chen and K.~T.~Mahanthappa,
  Phys.\ Lett.\  B {\bf 652} (2007) 34 [arXiv:0705.0714].


\bibitem{A4-LQ:Morisi}
S.~Morisi, M.~Picariello and E.~Torrente-Lujan,
  Phys.\ Rev.\  D {\bf 75} (2007) 075015  [hep-ph/0702034];~\\
F.~Bazzocchi, M.~Frigerio and S.~Morisi,
  Phys.\ Rev.\  D {\bf 78} (2008) 116018  [arXiv:0809.3573].


\bibitem{S4-LQ}
H.~Ishimori, Y.~Shimizu and M.~Tanimoto,
  Prog.\ Theor.\ Phys.\  {\bf 121} (2009) 769
  [arXiv:0812.5031];~\\
H.~Ishimori, T.~Kobayashi, H.~Ohki, H.~Okada, Y.~Shimizu and M.~Tanimoto,
  arXiv:1003.3552.




\bibitem{Mohapatra:2003tw}
  D.~G.~Lee and R.~N.~Mohapatra,
  Phys.\ Lett.\  B {\bf 329} (1994) 463
  [hep-ph/9403201];~\\
  R.~N.~Mohapatra, M.~K.~Parida and G.~Rajasekaran,
  Phys.\ Rev.\  D {\bf 69} (2004) 053007
  [hep-ph/0301234];~\\
  Y.~Cai and H.~B.~Yu,
  Phys.\ Rev.\  D {\bf 74} (2006) 115005
  [hep-ph/0608022];~\\
  M.~K.~Parida,
  Phys.\ Rev.\  D {\bf 78} (2008) 053004
  [arXiv:0804.4571];~\\
  B.~Dutta, Y.~Mimura and R.~N.~Mohapatra,
  arXiv:0911.2242.


\bibitem{King:2009mk}
S.~F.~King and C.~Luhn,
  Nucl.\ Phys.\  B {\bf 820} (2009) 269
  [arXiv:0905.1686];~\\
S.~F.~King and C.~Luhn,
  Nucl.\ Phys.\  B {\bf 832} (2010) 414
 [arXiv:0912.1344].



\bibitem{Z7Z3-LQ}
C.~Luhn, S.~Nasri and P.~Ramond,
  Phys.\ Lett.\  B {\bf 652} (2007) 27 [arXiv:0706.2341].

\bibitem{T7-LQ}
C.~Hagedorn, M.~A.~Schmidt and A.~Y.~Smirnov,
  Phys.\ Rev.\  D {\bf 79} (2009) 036002 [arXiv:0811.2955].



\bibitem{delta27-LQ:King}
I.~de Medeiros Varzielas, S.~F.~King and G.~G.~Ross,
   Phys.\ Lett.\  B {\bf 644} (2007) 153
  [hep-ph/0512313];~\\
I.~de Medeiros Varzielas, S.~F.~King and G.~G.~Ross,
  Phys.\ Lett.\  B {\bf 648} (2007) 201  [hep-ph/0607045];~\\
F.~Bazzocchi and I.~de Medeiros Varzielas,
  Phys.\ Rev.\  D {\bf 79} (2009) 093001
  [arXiv:0902.3250].


\bibitem{SO(3)-LQ:King}
S.~F.~King and M.~Malinsky,
  JHEP {\bf 0611} (2006) 071 [hep-ph/0608021].


\bibitem{SU(3)-LQ:Ross}
S.~F.~King and G.~G.~Ross,
  Phys.\ Lett.\ B \textbf{520} (2001) 243 [hep-ph/0108112];~\\
S.~F.~King and G.~G.~Ross,
  Phys.\ Lett.\ B \textbf{574} (2003) 239 [hep-ph/0307190];~\\
I.~de Medeiros Varzielas and G.~G.~Ross,
   Nucl.\ Phys.\  B {\bf 733} (2006) 31 [hep-ph/0507176].


\bibitem{Reviews}
S.~F.~King,
  Rept.\ Prog.\ Phys.\ \textbf{67} (2004) 107 [hep-ph/0310204];~\\
R.~N.~Mohapatra {\it et al.},
  Rept.\ Prog.\ Phys.\  {\bf 70} (2007) 1757 [hep-ph/0510213];~\\
R.~N.~Mohapatra and A.~Y.~Smirnov,
  Ann.\ Rev.\ Nucl.\ Part.\ Sci.\  {\bf 56} (2006) 569 [hep-ph/0603118];~\\
C.~H.~Albright,
  arXiv:0905.0146;~\\
G.~Altarelli and F.~Feruglio,
  arXiv:1002.0211.


\bibitem{gst}
R.~Gatto, G.~Sartori and M.~Tonin,
  Phys.\ Lett.\ B {\bf 28} (1968) 128.


\bibitem{GJ}
H.~Georgi and C.~Jarlskog,
  Phys.\ Lett.\ B {\bf 86} (1979) 297.


\bibitem{King:2005bj}
  S.~F.~King,
  JHEP {\bf 0508} (2005) 105
  [hep-ph/0506297];~\\
I.~Masina,
  Phys.\ Lett.\  B {\bf 633} (2006) 134
  [hep-ph/0508031];~\\
S.~Antusch and S.~F.~King,
  Phys.\ Lett.\  B {\bf 631} (2005) 42
  [hep-ph/0508044];~\\
S.~Antusch, P.~Huber, S.~F.~King and T.~Schwetz,
  JHEP {\bf 0704} (2007) 060
  [hep-ph/0702286];~\\
S.~Antusch, S.~F.~King and M.~Malinsky,
  Phys.\ Lett.\  B {\bf 671} (2009) 263
  [arXiv:0711.4727];~\\
S.~Antusch, S.~F.~King and M.~Malinsky,
  JHEP {\bf 0805} (2008) 066
  [arXiv:0712.3759];~\\
S.~Boudjemaa and S.~F.~King,
  Phys.\ Rev.\  D {\bf 79} (2009) 033001
  [arXiv:0808.2782];~\\
S.~Antusch, S.~F.~King and M.~Malinsky,
  Nucl.\ Phys.\  B {\bf 820} (2009) 32
  [arXiv:0810.3863].


\bibitem{SU5-EDs}
E.~Witten,
  Nucl.\ Phys.\  B {\bf 258} (1985) 75;~\\
Y.~Kawamura,
  Prog.\ Theor.\ Phys.\  {\bf 105} (2001) 999
  [hep-ph/0012125];~\\
A.~E.~Faraggi,
  Phys.\ Lett.\  B {\bf 520} (2001) 337
  [arXiv:hep-ph/0107094]~\\
and references therein.



\bibitem{SU5DTsplit}
H.~Georgi,
  Phys.\ Lett.\  B {\bf 108} (1982) 283;~\\
A.~Masiero, D.~V.~Nanopoulos, K.~Tamvakis and T.~Yanagida,
  Phys.\ Lett.\  B {\bf 115} (1982) 380;~\\
B.~Grinstein,
  Nucl.\ Phys.\  B {\bf 206} (1982) 387;~\\
E.~Witten,
  Phys.\ Lett.\  B {\bf 105} (1981) 267;~\\
D.~V.~Nanopoulos and K.~Tamvakis,
  Phys.\ Lett.\  B {\bf 113} (1982) 151;~\\
S.~Dimopoulos and F.~Wilczek, NSF-ITP-82-07;~\\
M.~Srednicki,
  Nucl.\ Phys.\  B {\bf 202} (1982) 327.


\bibitem{noncanonkin}
 S.~F.~King and I.~N.~R.~Peddie,
   Phys.\ Lett.\  B {\bf 586} (2004) 83
  [hep-ph/0312237];~\\
 J.~R.~Espinosa and A.~Ibarra,
  JHEP {\bf 0408} (2004) 010
  [hep-ph/0405095].


\bibitem{Wolfpara}
 L.~Wolfenstein,
  Phys.\ Rev.\ Lett.\  {\bf 51} (1983) 1945.



\bibitem{GfbreakED}
N.~Haba, A.~Watanabe and K.~Yoshioka,
  Phys.\ Rev.\ Lett.\  {\bf 97} (2006) 041601
  [hep-ph/0603116];~\\
T.~Kobayashi, Y.~Omura and K.~Yoshioka,
  Phys.\ Rev.\  D {\bf 78} (2008) 115006
  [arXiv:0809.3064];~\\
A.~Adulpravitchai and M.~A.~Schmidt,
  arXiv:1001.3172.


\bibitem{PDG}
 C.~Amsler {\it et al.}  [Particle Data Group],
  Phys.\ Lett.\  B {\bf 667} (2008) 1.


\bibitem{Ross:2007az}
  G.~Ross and M.~Serna,
  Phys.\ Lett.\  B {\bf 664} (2008) 97
  [arXiv:0704.1248].


\bibitem{King:2007pr}
  S.~F.~King,
  Phys.\ Lett.\  B {\bf 659} (2008) 244
  [arXiv:0710.0530].


\bibitem{Lomont}
J. ~Lomont, {\it Applications of Finite Groups}, Acad. Press (1959) 346 p..


\bibitem{Luhn:2007yr}
C.~Luhn, S.~Nasri and P.~Ramond,
  J.\ Math.\ Phys.\  {\bf 48} (2007) 123519
  [arXiv:0709.1447];~\\
C.~Luhn and P.~Ramond,
  JHEP {\bf 0807} (2008) 085
  [arXiv:0805.1736].


\bibitem{integrateout}
I.~Affleck, M.~Dine and N.~Seiberg,
  Nucl.\ Phys.\  B {\bf 256} (1985) 557;~\\
D.~Gallego and M.~Serone,
  JHEP {\bf 0901} (2009) 056
  [arXiv:0812.0369];~\\
D.~Gallego and M.~Serone,
  JHEP {\bf 0906} (2009) 057
  [arXiv:0904.2537].


\end{thebibliography}
\end{document}